\documentclass[12pt]{article}

\usepackage{amssymb}
\usepackage{amsfonts}
\usepackage{amsmath}
\usepackage{bbm}
\usepackage{graphicx}
\usepackage{color}
\usepackage{rotating}
\usepackage{lscape}
\usepackage{lscape}
\usepackage{wasysym}

\def\Z{\mathbb Z}
\def\C{\mathbb C}
\def\R{\mathbb R}
\def\NN{\mathbb N}

\newcommand{\al}{\alpha}

\newcommand{\ga}{\gamma}
\newcommand{\de}{\delta}
\newcommand{\eps}{\epsilon}

\renewcommand{\th}{\theta}
\newcommand{\si}{\sigma}

\newcommand{\De}{{\Delta\vphantom{\big|}}}
\newcommand{\Ga}{\Gamma}

\newcommand{\wb}{{\bar{w}}}
\newcommand{\rb}{{\bar{\rho}}}
\newcommand{\cD}{{\cal D}}
\newcommand{\cH}{{\cal H}}
\newcommand{\cL}{{\cal L}}
\newcommand{\cM}{{\cal M}}

\newcommand{\s}{{\mathrm{s}}}
\renewcommand{\c}{{\mathrm{c}}}

\newcommand{\sfrac}[2]{{\textstyle\frac{#1}{#2}}}
\newcommand{\half}{{\sfrac12}}
\newcommand{\pa}{\partial}

\newcommand{\im}{{\mathrm{i}}}
\newcommand{\ep}{{\mathrm{e}}}
\newcommand{\ph}{\phantom{-}}
\newcommand{\unity}{\mathbbm{1}}
\newcommand{\rf}{\sqrt{5}}

\newcommand{\beq}{\begin{equation}}
\newcommand{\eeq}{\end{equation}}
\newcommand{\eq}{\end{equation}}
\newcommand{\bea}{\begin{eqnarray}}
\newcommand{\eea}{\end{eqnarray}}
\newcommand{\with}{{\quad{\rm with}\quad}}
\newcommand{\for}{{\quad{\rm for}\quad}}
\renewcommand{\and}{{\quad{\rm and}\quad}}
\newcommand{\und}{{\qquad{\rm and}\qquad}}
\renewcommand{\=}{\ =\ }

\advance \textheight by \topskip
\makeatletter
\@addtoreset{equation}{section}

\makeatother

\begin{document}

\begin{titlepage}
\setcounter{page}{0}

\phantom{.}
\vskip 1.5cm

\begin{center}

{\LARGE\bf
${\cal PT}$ deformation of angular Calogero models
}

\vspace{12mm}
{\Large Francisco Correa${}^{+}$ and \
Olaf Lechtenfeld${}^{\times}$ 
}
\\[8mm]

\noindent ${}^+${\em 
Instituto de Ciencias F\'isicas y Matem\'aticas\\
Universidad Austral de Chile, Casilla 567, Valdivia, Chile}\\
{Email: francisco.correa@uach.cl}\\[6mm]

\noindent ${}^\times${\em
Institut f\"ur Theoretische Physik and Riemann Center for Geometry and Physics\\
Leibniz Universit\"at Hannover \\
Appelstra\ss{}e 2, 30167 Hannover, Germany }\\
{Email: olaf.lechtenfeld@itp.uni-hannover.de}\\[6mm]

\vspace{12mm}

\begin{abstract}
\noindent
The rational Calogero model based on an arbitrary rank-$n$ Coxeter root system
is spherically reduced to a superintegrable angular model of a particle moving on $S^{n-1}$
subject to a very particular potential singular at the reflection hyperplanes.
It is outlined how to find conserved charges and to construct intertwining operators.
We deform these models in a ${\cal PT}$-symmetric manner by judicious complex 
coordinate transformations, which render the potential less singular. The ${\cal PT}$
deformation does not change the energy eigenvalues but in some cases adds a previously 
unphysical tower of states. For integral couplings the new and old energy levels coincide, 
which roughly doubles the previous degeneracy and allows for a conserved nonlinear 
supersymmetry charge. We present the details for the generic rank-two ($A_2$, $G_2$)
and all rank-three Coxeter systems ($AD_3$, $BC_3$ and $H_3$), including a reducible 
case ($A_1^{\otimes 3}$).
\end{abstract}

\end{center}

\end{titlepage}

\section{Introduction and summary}

\noindent
The rational Calogero model (for a review, see \cite{Poly06-rev}) generalizes to any root system of a
(finite-dimensional) Lie algebra or, better, to any Coxeter root system.
Given such a system of rank~$n$, it describes a conformal particle moving in~$\R^n$ under
the influence of a very special potential. Since this potential has a universal
inverse-square radial dependence and otherwise depends only on the angular coordinates
(of $S^{n-1}$), a spherical reduction to its angular subsystem, the angular Calogero
model, is natural. Like the full model on $\R^n$, the reduced dynamics on $S^{n-1}$ is
superintegrable, so that it enjoys $2n{-}3$ integrals of motion, which are however
not in involution. Recently, the angular models have been analyzed in some detail,
both classically and quantum mechanically
\cite{Feigin03,HNY,HKLN,LNY,HLNS,HLN,HLNSY,FeLePo13,CoLePl13,FeHa14,CoLe15}.

It has been known for a long time that hermiticity is not an essential feature of
a Hamiltonian for its spectrum to be real. For instance, it suffices that
the Hamiltonian commutes with an antilinear involution (one example is provided by the $\cal{PT}$ operator where ${\cal P}$ correspond to the parity operator and ${\cal T}$ the time reversal operator)
which also leaves the eigenfunctions invariant (``unbroken $\cal{PT}$ symmetry'')~\cite{BeBo98}.
Such a non-hermitian Hamiltonian is related to a hermitian one by a (non-unitary)
similarity transformation, which may be impossibly complicated.
Often, however, there exists a family $H_\eps$ of non-hermitian $\cal{PT}$-invariant
Hamiltonians representing a smooth deformation of a hermitian~$H_0$.
In this case we speak of a ``$\cal{PT}$ deformation'', with
the parameter~$\eps$ measuring the deviation from hermiticity.
For rational Calogero models, a particularly nice set of $\cal{PT}$ deformations
can be generated by a specific {\it complex\/} orthogonal deformation of
the coordinates in the expression for the Hamiltonian.
If such a $\cal{PT}$ deformation is in accordance with the Coxeter reflection symmetry 
of the system, integrability will be preserved.
This kind of $\cal{PT}$ deformation has been applied to the full rational Calogero model
about ten years ago by Fring and Znojil~\cite{FrZn08}, and corresponding complex root systems
were constructed by Fring and Smith thereafter~\cite{fringsmith1,fringsmith2,fringsmith3}.
For a review of $\cal{PT}$ deformations of integrable models, see~\cite{Fring12-rev}.

It is worth recalling the relevant part (for this paper) of the Calogero model's long history:
\begin{itemize}
\addtolength{\itemsep}{-5pt}
\item 1971 \quad  Calogero \cite{Cal71}:\\ 
Solution of the one-dim'l N-body problem with $\ldots$ inversely
quadratic pair potentials
\item 1981 \quad  Olshanetsky \&\ Perelomov
\cite{OlshaPere81-rev,OlshaPere83-rev}:\\ 
Classical integrable finite-dimensional systems related to Lie algebras
(1983: quantum)
\item 1983 \quad  Wojciechowski \cite{Woj}:\\ 
Superintegrability of the Calogero--Moser system
\item 1989 \quad  Dunkl \cite{Dunkl89}:\\ 
Differential-difference operators associated to reflection groups
\item 1990 \quad  Chalykh \&\ Veselov \cite{ChaVes90}:\\ 
Commutative rings of partial differential operators and Lie algebras,
supercompleteness
\item 1991 \quad  Heckman \cite{Heckman}:\\ 
Elementary construction for commuting charges and intertwiners (shift
operators)
\item 2003 \quad  M. Feigin \cite{Feigin03}:\\ 
Intertwining relations for the spherical parts of generalized Calogero
operators
\item 2008 \quad  A. Fring, M. Znojil
\cite{FrZn08}:\\ 
${\cal PT}$-symmetric deformations of Calogero models
\item 2008 \quad  Hakobyan, Nersessian, Yeghikyan \cite{HNY}:\\
The cuboctahedric Higgs oscillator from the rational Calogero model
(classical)
\item 2010 \quad  A. Fring, M. Smith
\cite{fringsmith1,fringsmith2,fringsmith3}:\\ 
Complex root systems in the Calogero model
\item 2013 \quad  M. Feigin, Lechtenfeld, Polychronakos
\cite{FeLePo13}:\\ 
The quantum angular Calogero--Moser model (spectra, eigenstates)
\item 2013 \quad  Correa, Lechtenfeld, Plyushchay
\cite{CoLePl13}:\\
Nonlinear supersymmetry in the quantum Calogero model
\item 2014 \quad  M. Feigin, Hakobyan \cite{FeHa14}:\\ 
On the algebra of Dunkl angular momentum operators
\item 2015 \quad  Correa, Lechtenfeld \cite{CoLe15}:\\
The tetrahexahedric angular Calogero model
\end{itemize}

The present paper describes the superintegrable spherical reduction of the rational
quantum Calogero model for any Coxeter root system (Section~2) and some of its complex 
${\cal PT}$ deformations (Section~3). The emphasis is on the Weyl-singlet energy spectrum 
including degeneracy and eigenstates, and on the conserved charges and intertwiners, 
in particular for a coupling strength $g(g{-}1)$ with $g\in\Z$.  
We discuss all features in some detail for the rank-two cases of $A_2$ and $G_2$ 
(Sections 4 and~5) and for all rank-three cases, i.e.~$AD_3$, $BC_3$ and $H_3$ 
(Sections 6, 7 and~9), as well as for $A_1^3$ as a reducible example (Section~8). 
Tables of low-lying states are collected in the Appendix.

Our results generalize those of \cite{CoLe15} to general Coxeter root systems,
in particular to the non-simply-laced case, where two independent couplings wrongly suggest
the existence of long-root and short-root intertwiners. Instead, we find that all intertwiners
respecting the reflection symmetry either shift both couplings or only one of them,
so not all states with integral couplings can be connected. 
We identify a geometric condition for complex orthogonal coordinate transformations
to yield a ${\cal PT}$ deformation (with ${\cal P}$ given by a Coxeter element) and
display the simplest solutions. It turns out that such deformations reduce the singularities
of the angular Calogero potential from codimension one to codimension two. 
We also present a nonlinear ${\cal PT}$ deformation which may completely remove those
singularities (it does so for rank three). 
In such a situation, the non-normalizable eigenstates (formally given by sending
$g\mapsto1{-}g$ for $g\in\NN$) become normalizable and have to be added to the spectrum.
Not only does this roughly double the state degeneracy, but it also gives rise to
new `odd' conserved charges, which connect the old and the new states.
We display these effects for the generic rank-two and all rank-three Coxeter systems.

\newpage

\section{The angular rational Calogero model}

\noindent
The well known rational Calogero model describing $n$ interacting identical particles 
moving on~$\R$ can be formulated for any finite reflection group~$W$, with the 
multi-particle potential encoded in the associated Coxeter root system~${\cal R}\subset\R^n$.
Since this interaction is not translation invariant\footnote{
The $A_n$ model decribes the {\sl relative\/} coordinates of $n{+}1$
particles after decoupling the center of mass.}
it is more natural to view such systems as a single particle moving in~$\R^n$
under the influence of a rather particular external potential determined by~$\cal R$. 
As the Hamiltonian is homogeneous under a common coordinate rescaling (the couplings
are dimensionless) the model may be reduced over the $(n{-}1)$-sphere. 
The result is what we have named the {\sl angular\/} Calogero model, since 
it describes a particle moving on~$S^{n-1}$, parametrized by angular coordinates~$\vec{\th}$ only.
Because hyperspherical coordinates are rather unwieldy however, we prefer to
employ the homogeneous $\R^n$ coordinates $x=(x^i)$ with $i=1,\ldots,n$ and define
\beq
\sum_{i=1}^n (x^i)^2\ =:\ r^2\ .
\eeq
In terms of the latter, the angular Calogero Hamiltonian takes the form
\beq \label{H}
H \= \half L^2 + U \with
L^2\= -\sum_{i<j}(x^i\pa_j-x^j\pa_i)^2 \and
U \= r^2\!\sum_{\alpha\in{\cal R}_+} \frac{g_\al(g_\al{-}1)\,\al{\cdot}\al}{2\,(\al\cdot x)^2}
\eeq
where ${\cal R}_+$ is the positive half of~$\cal R$, $g_\al\in\R$ are the couplings,
and $\cdot$ is the standard scalar product in~$\R^n$.
Due to the invariance of the Hamiltonian under $g_\al{+}1\leftrightarrow-g_\al$, 
it suffices to consider $g_\al\ge\frac12$, but we shall not impose this restriction
because intermediate results do not reflect this symmetry.
Each positive root~$\al$ contributes a term of the form $\cos^{-2}\phi_\al$,
where $\phi_\al$ is the geodesic distance to $\hat\al$. 
This so-called Higgs oscillator potential~\cite{higgs,leemon} 
is singular on a great $S^{n-2}$, where the hyperplane orthogonal to $\al$ 
cuts our $(n{-}1)$-sphere into two hemispheres. Taken together,
these singular loci of codimension one tessalate the $(n{-}1)$-sphere,
and our particle is confined to a given Weyl chamber, with its wave function
vanishing at the walls (except for $g{=}0$ and~$g{=}1$).
The potential breaks the SO($n$) invariance of~$L^2$ to its discrete subgroup~$W$,
so the energy eigenstates fall into $W$~representations. 
Motivated by the physical interpretation, we admit only singlet states, i.e.~wave functions 
are either totally symmetric or totally antisymmetric under Coxeter reflections.

The Weyl-invariant spectrum of $H$ has been derived in~\cite{FeLePo13} 
(see also the appendices of~\cite{Cal71}),
\beq
H\,v_{\{\ell\}} \,=\, E_\ell\,v_{\{\ell\}} \with
\{\ell\}=(\ell_3,\ell_4,\ldots,\ell_{n+1}) \and 
\ell=d_3\ell_3+d_4\ell_4+\cdots+d_{n+1}\ell_{n+1}
\eeq
where $d_2{=}2$, $d_3, \ldots, d_{n+1}$ are the degrees of the basic homogeneous $W$-invariant polynomials
$\si_2=\sum_i(x^i)^2$, $\si_3, \ldots, \si_{n+1}$ and the quantum numbers $\ell_3, \ell_4, \ldots, \ell_{n+1}$ 
are nonnegative integers.\footnote{
The unconventional labelling is chosen to match with the standard choice for the $A_1\oplus A_n$ model.}    
Note that $\si_2$ does not contribute because $\ell_2$ labels the radial excitations.
The energy depends only on the `deformed angular momentum'~$q$,
\beq \label{Eell}
E_\ell \= \half\,q\,(q+n-2) \quad\with\quad q\=\ell+\sum_{\alpha\in{\cal R}_+}g_\al\ .
\eeq
For vanishing couplings, $H=\sfrac12L^2$, and $q=\ell$ is the familiar
total angular momentum for a free particle on $S^{n-1}$. 
Nevertheless, the degeneracy of $E_\ell$ is greatly reduced by $W$-invariance to the
number of partitions of~$\ell$ into integers from the set~$\{d_3,\ldots,d_{n+1}\}$.

The angular wave function $v^{(g)}_{\{\ell\}}$ for couplings $g=\{g_\al\}$ 
can be constructed in the following way~\cite{FeLePo13}.
First, we split off a suitable power of $r$ and a `Vandermonde factor',
\beq \label{vellg}
v^{(g)}_{\{\ell\}} \= r^{-q}\,\De^g\,h^{(g)}_{\{\ell\}}(x) \quad\with\quad
\De^g = \prod_{\alpha\in{\cal R}_+} (\al\cdot x)^{g_\al}
\eeq
and obtain a homogenous polynomial $h^{(g)}_{\{\ell\}}$ of degree~$\ell$ in~$x$.
Second, the latter is a $W$-invariant Dunkl-deformed harmonic function given by
\beq
h^{(g)}_{\{\ell\}}(x) \= r^{n-2+2q}\,
\biggl(\prod_{\mu=3}^{n+1} \sigma_\mu\bigl(\{\widetilde{\cD_i}\}\bigr)^{\ell_\mu} \biggr)\, r^{2-n-2(q-\ell)}\ ,
\eeq
where 
\beq
\widetilde{\cD}_i 
\= \pa_i \ +\ \sum_{\alpha\in{\cal R}_+} \frac{g_\al\,\alpha_i}{\alpha\cdot x}\,(1-s_\alpha)
\= \De^{-g}\,\cD_i\,\De^g
\eeq
denotes the Dunkl differential-reflection operator~\cite{Dunkl89,DunklXu}, which 
involves the Coxeter reflections~$s_\al$ about the hyperplane $\al\cdot x=0$.
The tilde signifies the so-called potential-free frame, which is related to the `potential frame'
by a similarity transformation with $\De^g$,
\beq
\cD_i \= \De^g\,\widetilde{\cD}_i\,\De^{-g}
\= \pa_i \ -\ \sum_{\alpha\in{\cal R}_+} \frac{g_\al\,\alpha_i}{\alpha\cdot x}\,s_\alpha\ .
\eeq
In particular, for the ground state one has
\beq
h_{\{0\}}^{(g)} \= 1 \qquad\Longrightarrow\qquad 
v_{\{0\}}^{(g)} \= r^{-q} \De^g \= \prod_{\al\in{\cal R}_+} \bigl(\al\cdot\sfrac{x}{r}\bigr)^{g_\al}\ ,
\eeq
and hence the full ground-state wave function is totally symmetric (antisymmetric)
under Coxeter reflections for even (odd) integer values of~$g_\al$.
Since all other ingredients besides $\De^g$ in~(\ref{vellg}) are completely symmetric,
this symmetry property of the integer-$g_\al$ ground state extends to all excited states above it.
The degeneracy of the energy levels decreases with growing values of~$g_\al$.
Furthermore, the reflection symmetry $g_\al{+}1\leftrightarrow{-}g_\al$ of the Hamiltonian~(\ref{H})
is broken since one tower of states is Weyl symmetric while the other one is antisymmetric.
However, due to singularities at $\al\cdot x=0$ coming from the Vandermonde factor in~(\ref{vellg}), 
for $g_\al<0$ the formal eigenstates are not normalizable (i.e.~not in $L_2(S^{n-1})$) and thus unphysical. 
In other words, the singularities in the potential~$U$ enforce boundary conditions, which admit
only one of the two symmetry types. The free case is an exception, because then those boundary
conditions are absent, and so both values $g_\al=0$ and $g_\al=1$ contribute to the same spectrum,
leading to a rough doubling of the states.

Our Hamiltonian and other conserved quantities are conveniently constructed from the
algebra of Dunkl-deformed angular momenta,
\beq
\cL_{ij} \= x^i\cD_j-x^j\cD_i\ ,
\eeq
which yields
\beq
-\half\sum_{i<j}\cL_{ij}^2 \= \cH\ -\ \sfrac12\,S\,(S+n{-}2)
\eeq
with
\beq \label{calH}
\cH \= \half L^2 + r^2\!\sum_{\alpha\in{\cal R}_+} 
\frac{\al{\cdot}\al/2}{(\al\cdot x)^2} g_\al(g_\al{-}s_\al)
\und 
S \= \sum_\al g_\al s_\al\ .
\eeq
The restriction `res' to $W$-symmetric functions provides the Hamiltonian,
\beq 
-\half\textrm{res}\Bigl(\sum_{i<j}\cL_{ij}^2\Bigr) \= 
\textrm{res}(\cH)\ -\ \half\sum_\al g_\al\Bigl(\sum_\al g_\al+n{-}2\Bigr)
\= H\ -\ E_0 \ .
\eeq
As was shown in \cite{Feigin03}, the center of the algebra generated by $\{\cL_{ij}\}$
is spanned by $\cH$ and the constants. Therefore, any polynomial~${\cal C}$ 
built from the $\cL_{ij}$ will commute with~$\cH$. If such a polynomial is Weyl invariant,
then its restriction yields a conserved quantity,
\beq
{\cal C}\quad\textrm{Weyl invariant}\qquad\Longrightarrow\qquad
[\,C\,,\,H\,] \= 0 \quad\for\quad C = \textrm{res}({\cal C})\ .
\eeq
It is not clear whether some combinations of these are in involution or how to classify them.

It is actually more fruitful to investigate Weyl antiinvariant polynomials in $\cL_{ij}$,
since they give rise to intertwiners (shift operators) which connect Hamiltonians and eigenspaces
differing by unit values in the couplings.
To be more precise, let us split the set of positive roots into Weyl orbits,
\beq
{\cal R}_+ \= {\cal R}' \cup {\cal R}''\ ,
\eeq
where one of the following four situations occurs:
\begin{center}
\begin{tabular}{|c|cccc|}
\hline
case & A & B & C & D \\
\hline
${\cal R}'$ & all +ve roots & long +ve roots & short +ve roots & empty \\
${\cal R}''$ & empty & short +ve roots & long +ve roots & all +ve roots \\
\hline
\end{tabular}
\end{center}
Because all couplings $g_\al$ in a given Weyl orbit must coincide, we can have at most two
different values, $g'$ and $g''$. 
The objects of interest are polynomials $\cM$ in $\cL_{ij}$ which are Weyl antiinvariant under
${\cal R}'$ reflections but Weyl invariant under ${\cal R}''$ reflections.
Because the structure of (\ref{calH}) implies that
\beq
\textrm{res}\bigl(\cM\,\cH^{(g',g'')}\bigr) \= M\,H^{(g',g'')}
\qquad\textrm{but}\qquad
\textrm{res}\bigl(\cH^{(g',g'')}\,\cM\bigr) \= H^{(g'+1,g'')}\,M
\eeq
the commutation of $\cM$ and $\cH$ qualifies $M=\textrm{res}(\cM)$ as an intertwiner,
\beq
{\cal M}\quad\begin{cases}
{\cal R}'\ \textrm{antiinvariant}\\ {\cal R}''\ \textrm{invariant}
\end{cases}\!\!\!\!\!\!\Bigg\}\qquad\Longrightarrow\qquad
M\,H^{(g',g'')} \= H^{(g'+1,g'')}\,M\ .
\eeq
Note that $\cM$ and $M$ depend on~$(g',g'')$, which we have suppressed.
This operator relation may be applied to $W$-noninvariant states.
Hence, $M$ maps $H^{(g',g'')}$ eigenstates of energy $E_\ell^{(g',g'')}$ to
$H^{(g'+1,g'')}$ eigenstates of (the same) energy $E_{\ell'}^{(g'+1,g'')}$ with
$\ell'=\ell-|{\cal R}'|$ (see (\ref{Eell})). In particular,
\beq
M\,v_{\{\ell\}}^{(g',g'')} \= 
\smash{\sum_{\{\ell'\}\atop\ell'=\ell-|{\cal R}'|}} c^{\{\ell\}}_{\{\ell'\}} \ v_{\{\ell'\}}^{(g'+1,g'')}
\eeq
$\ph$\\
with some coefficients $c^{\{\ell\}}_{\{\ell'\}}\in\R$.
Generically, such a map~$M$ has a nonempty kernel.
The action on the deformed harmonic polynomials~$h_{\{\ell\}}^{(g)}$ is obtained by passing
to the potential-free frame,
\beq
\widetilde{M}\,h_{\{\ell\}}^{(g',g'')} \= 
\smash{\sum_{\{\ell'\}\atop\ell'=\ell-|{\cal R}'|}} 
c^{\{\ell\}}_{\{\ell'\}} \ \De\,h_{\{\ell'\}}^{(g'+1,g'')}
\quad\with \widetilde{M}\=\De^{-g} M\,\De^g \ .
\eeq
$\ph$\\
It is a nontrivial problem for a given Coxeter group~$W$ to identify a complete set of intertwiners,
their algebra and its generators.
We remark that case~D does not shift any coupling and describes the constants of motion~$C$
mentioned above, while case~A pertains to the simply-laced Coxeter groups.
When both couplings $g'$ and $g''$ are integer, repeated intertwining may relate all quantities with their
analogs in the free theory, which allows one to generate analytic expressions for all wave functions.

\section{${\cal PT}$-symmetric complex coordinate deformations}

\noindent
We implement a complex deformation of the (angular) coordinates~$\vec{\th}$ through a family of complex linear maps
\beq
\Ga(\eps) : \R^n\ \to\ \C^n  \quad\with \Ga(0)\=\textrm{id}
\eeq
which respect the standard scalar product of~$\R^n$, so 
\beq
\Ga(\eps)^\top \= \Ga(\eps)^{-1} \ .
\eeq
Hence, $\Ga(\eps)\in\textrm{SO}(n,\C)$, but because real coordinate rotations are inessential
our family is parametrized by the coset $\textrm{SO}(n,\C)/\textrm{SO}(n,\R)$ of real dimension
$\frac12n(n{-}1)$,
\beq \label{Gdef}
\Ga(\eps)\= \exp\Bigl\{ \sum_{i<j} \eps_{ij} G_{ij} \Bigr\}
\quad\with\quad
G_{ij} :  x^k\ \mapsto\ \im \bigl(\de^{kj}x^i - \de^{ki}x^j \bigr)\ ,
\eeq
and thus we also have
\beq
\Ga(\eps)^* \= \Ga(\eps)^\top \= \Ga(-\eps)\ .
\eeq
A coordinate change effected by~$\Ga(\eps)$,
\beq
(x^1,x^2,\ldots,x^n)^\top \= x \ \mapsto\ \Ga(\eps)\,x \ =:\ x(\eps)\ ,
\eeq
leaves $r^2$ and the kinetic term~$\half L^2$ invariant but generates a complex
deformation $U\mapsto U(\eps)$ of the angular potential~(\ref{H}), via
\beq
\al\cdot x \ \mapsto\ \al\cdot \Ga(\eps)\,x \= \Ga(\eps)^\top\al\cdot x\ ,
\eeq
which may also be interpreted as a complex (dual) deformation of the roots~$\al$.
Formally, the deformed Hamiltonian~$H(\eps)$ is isospectral to~$H=H(0)$, 
and its $W$-invariant eigenfunctions
are simply given by
\beq \label{vdef}
v_{\{\ell\}\,\eps}^{(g)} \= r^{-q}\,\De_\eps^g\,h_{\{\ell\}\,\eps}^{(g)} 
\quad\with\quad
\De_\eps^g=\prod_{\al\in{\cal R}_+} \bigl(\al\cdot x(\eps)\bigr)^{g_\al}
\and h_{\{\ell\}\,\eps}^{(g)}(x) = h_{\{\ell\}}^{(g)}\bigl(x(\eps)\bigr)\ .
\eeq

Our Hamiltonian is ${\cal PT}$ symmetric if there exist two involutions, one linear (${\cal P}$) and 
one antilinear (${\cal T}$), under whose combined action it is invariant.
For ${\cal T}$ we take the conventional choice of complex conjugation.
In the context of Calogero models, a natural ${\cal P}$ transformation is provided by some
element~$s$ of order~2 in the Coxeter group~$W$.
The kinetic term~$\half L^2$ is separately invariant under ${\cal P}$ and~${\cal T}$
but, in order for $U(\eps)$ to be ${\cal PT}$ invariant, the action of the involutive Coxeter element~$s$
on the deformed coordinate $x(\eps)$ has to be undone by complex conjugation, implying
\beq
{\cal P}\,\Ga(\eps) \= s\,\Ga(\eps)\,s \ \buildrel{!}\over{=}\ 
\Ga(-\eps) \= \Ga(\eps)^* \= {\cal T}\,\Ga(\eps)\ .
\eeq
On the Lie-algebra level this condition reads
\beq \label{PTsymcond}
\{ s\, ,\,\eps{:}G \} = 0 
\qquad\Leftrightarrow\qquad
s\,(\eps{:}G)\,s = -\eps{:}G
\qquad\Leftrightarrow\qquad
P_\pm\,(\eps{:}G)\, P_\pm = 0 
\eeq
with $\eps{:}G=\sum_{i<j}\eps_{ij}G_{ij}$ and projectors
\beq
P_-\=\half(1-s) \und P_+\=\half(1+s)
\eeq
on the $-1$ and $+1$ eigenspaces of~$s$, repectively.
It means that $\eps{:}G$ intertwines between those two eigenspaces,
and so 
\beq
\textrm{rank}(\eps{:}G) \= 
\textrm{min}\bigl( 2\,\textrm{rank}(P_-),2\,\textrm{rank}(P_+)\bigr)\ .
\eeq

If $s$ is just a Coxeter reflection~$s_\ga$ pertaining to some (positive) root~$\ga$,
then we can say a bit more.
Since in this case $P_-$ is of rank~one, it follows that $\eps{:}G$ is of rank~two only
and parallel to~$\ga$,
\beq \label{Ggamma}
\eps{:}G \= -\im\eps\,\hat\ga \wedge\hat\eta \ \in su(1,1)
\quad\with (\eps{:}G)\,\ga\ \sim\ \eta\ \perp\ \ga
\eeq
for some real vector~$\eta$, carrying $n{-}1$ parameters.
The hats denote unit vectors, 
and the overall scale has been absorbed into a single parameter~$\eps$.
For this situation, the infinitesimal transformation can be integrated explicitly to
\beq \label{Gamma2rank}
\Ga(\eps) \= \exp\bigl\{ -\im\eps\,\hat\ga \wedge\hat\eta \bigr\} \=
P^\perp_{\ga\wedge\eta}\ -\ P_{\ga\wedge\eta} 
\bigl( \cosh(\eps)-\im\sinh(\eps)\,\hat\ga \wedge\hat\eta\bigr)\ ,
\eeq
with the help of projectors onto the plane spanned by $\ga$ and~$\eta$ and orthogonal to it.
This is just a complex rotation (boost) in the plane determined by $\ga$ and~$\eta$.
A similar analysis applies in the co-rank-one case, i.e.\ when $P_+$ is of rank one.
In adapted coordinates,
\beq
\Ga(\eps) \= \ep^{\eps{:}G} \=
\begin{pmatrix}
\cosh(\eps) & -\im\sinh(\eps) & 0\cdots0 \\[4pt]
\im\sinh(\eps) & \ph\cosh(\eps) & 0\cdots0 \\[4pt]
0 & 0 & \\[-8pt] \vdots & \vdots & \unity_{n-2} \\[-4pt]
0 & 0 & \end{pmatrix}\ .
\eeq

The complex deformation 
greatly improves the singularities of~$U$ by generically increasing 
their codimension from one to two. The singularity relation $\al\cdot \Ga(\eps)\,x=0$ decomposes
into a real and imaginary part giving two conditions, 
\beq
\al\cdot x \=0 \und \al\cdot(\eps{:}G)\cdot x \=0 \qquad\textrm{mod $O(\eps^2)$}\ ,
\eeq
leaving an $S^{n-3}$ plus its antipode
as the singular locus for each positive root~$\al$ contributing to~$U$. 
Specializing to ${\cal PT}$-symmetric deformations~(\ref{PTsymcond}), 
the second condition may be empty if $\al$ lies in the kernel of $\eps{:}G$. 
However, such a situation can be avoided by a slight change in the parameters~$\eps_{ij}$.
For the case of $s=s_\ga$, the singular loci appear at
\beq \label{singcond}
\al\cdot\bigl( P_{\ga\wedge\eta}^\perp\,x + \cosh(\eps)\,P_{\ga\wedge\eta}\bigr)\,x \= 0 
\und \al\cdot(\ga\wedge\eta)\,P_{\ga\wedge\eta}\,x \= 0\ .
\eeq
The second condition gets lost if $\al$ lies in the kernel of $P_{\ga\wedge\eta}$, i.e.~if
\beq
\al\cdot(\ga\wedge\eta) \= 0\ ,
\eeq
However, by a suitable (generic) choice of~$\eta$ one can tilt the plane 
spanned by $\ga$ and~$\eta$ such as to avoid any roots and so evade this degenerate situation.

The deformation also ameliorates the singularities in the unphysical wave functions for
negative values of the couplings. 
{}From the form of~(\ref{vdef}) it is clear that $\De_\eps$ vanishes at antipodal pairs $(x_\al,-x_\al)$
obeying~(\ref{singcond}), for each $\al\in{\cal R}_+$. Hence, on a collection of $(n{-}3)$-spheres
in $S^{n-1}$ our wave functions have nodes for positive values of~$g_\al$, but they still
blow up for negative couplings when $n>2$. Hence, for rank~3 and larger, the formal energy eigenstates
at $g_\al<0$ remain non-normalizable under the linear deformation~(\ref{Gdef}). 
Passing to the deformed metric under which $H$ becomes hermitian unfortunately does not change this,
and so the ${\cal PT}$ deformation in general does not enlarge the degeneracy of the energy spectrum.
An exception occurs for $n{=}2$, which will be outlined below.

The conserved quantities and intertwiners naturally carry over to the deformed situation, 
\beq
C_\eps \= \textrm{res}({\cal C}_\eps) \und  M_\eps \= \textrm{res}(\cM_\eps) \ ,
\eeq
built from `doubly deformed' angular momenta $\cL_{ij}^{\eps}$ made from $x(\eps)$ and
\beq
\cD_i^\eps \= \bigl(\Gamma(\eps)\,\pa\bigr)_i \ -\ 
\sum_{\al\in{\cal R}_+} \frac{g_\al\,\al^i}{\al\cdot\Gamma(\eps)\,x}\,s_\al^\eps
\quad\with\quad s_\al^\eps \= \Gamma(\eps)\,s_\al\,\Gamma(-\eps)
\eeq
in the case of a linear deformation.
Therefore, the superintegrability of the model is unchanged.

One may consider also nonlinear complex deformations of the coordinates.
A particular one consists in a complex shift of each angle in a hyperspherical parametrization,
\beq \label{nonlinearPT}
\begin{aligned}
x^1(\eps) &\= r\,\cos(\phi_1{+}\im\eps_1) \ ,\\
x^2(\eps) &\= r\,\sin(\phi_1{+}\im\eps_1)\,\cos(\phi_2{+}\im\eps_2) \ ,\\
x^3(\eps) &\= r\,\sin(\phi_1{+}\im\eps_1)\,\sin(\phi_2{+}\im\eps_2)\,\cos(\phi_3{+}\im\eps_3) \ ,\\
&\cdots \\
x^{n-1}(\eps) &\= r\,\sin(\phi_1{+}\im\eps_1)\,\sin(\phi_2{+}\im\eps_2)\,
\cdots\,\sin(\phi_{n-2}{+}\im\eps_{n-2})\,\cos(\phi_{n-1}{+}\im\eps_{n-1}) \ ,\\
x^n(\eps) &\= r\,\sin(\phi_1{+}\im\eps_1)\,\sin(\phi_2{+}\im\eps_2)\,
\cdots\,\sin(\phi_{n-2}{+}\im\eps_{n-2})\,\sin(\phi_{n-1}{+}\im\eps_{n-1}) \ .
\end{aligned}
\eeq
Such a deformation will (for $n{>}2$) also modify the kinetic term $\half L^2$.
The obvious choice for ${\cal P}$ is 
\beq
\phi_i\ \mapsto\ -\phi_i \qquad\Leftrightarrow\qquad
x^i\ \mapsto\ (-1)^{i+1}x^i\ .
\eeq
The correspondingly deformed Hamiltonian is ${\cal PT}$ invariant if this transformation
is a symmetry of the root system, i.e.\ if it is contained in the Coxeter group extended by 
the outer automorphisms (symmetries of the Dynkin diagram).

The advantage of such a deformation is that the singular locus of the potential $U(\eps)$
and thus the zero set of the Vandermonde $\De_\eps$ may be empty. This renders the formal energy eigenstates
for $g_\al<0$ normalizable and, hence, produces new towers of physical states for negative couplings.
Due to $H(-g_\al)=H(g_\al{+}1)$, these new states enlarge the state space for $g_\al>1$.
For integral~$g_\al$ we can connect the two towers by a string of intertwiners.\footnote{
This is also possible for odd half-integral couplings but does not yield an independent charge.}
In the enlarged state space then acts an additional, `odd' conserved charge,
\beq
\begin{aligned}
& Q_\eps^{(g',g'')} \= M_\eps^{(g'-1,g'')} M_\eps^{(g'-2,g'')} \cdots M_\eps^{(1-g',g'')} \\
\qquad\Rightarrow\qquad
& Q_\eps^{(g',g'')} H_\eps^{(g',g'')} \=Q_\eps^{(g',g'')} H_\eps^{(1-g',g'')} 
\= H_\eps^{(g',g'')} Q_\eps^{(g',g'')}\ ,
\end{aligned}
\eeq
which intertwines between the $g'>0$ and $g'\le0$ towers.
In the potential-free frame,
\beq
\widetilde{Q}_\eps^{(g',g'')} \= \De_\eps^{-g',-g''} Q_\eps\,\De_\eps^{1-g',g''} : \ 
h_{\{\ell\}\,\eps}^{(1-g',g'')}\ \mapsto\ h_{\{\ell'\}\,\eps}^{(g',g'')}
\eeq
relates the two Dunkl- and ${\cal PT}$-deformed harmonic polynomials to each other.
Note that in contrast to $Q_\eps^{(g',g'')}$, the potential-free intertwiner 
$\widetilde{Q}_\eps^{(g',g'')}$ is not conserved.
The new odd charge squares to a polynomial in the conserved `even' charges~$C$
and extends the algebra of conserved quantities to a nonlinear supersymmetric one.
Due to the ${\cal PT}$ regularization of the negative-coupling states, 
$Q_\eps^{(g',g'')}$~now has a regular action in the state space.
In general there exist more than one intertwiner, giving rise to various such odd charges.

\section{$A_2$ model}

\noindent
The simplest case to consider is the $A_2$ model, 
which is based on the roots
\beq
{\cal R}_+ \= \bigl\{ e_1\,,\,\sfrac12(e_1+\sqrt{3}e_2)\,,\,\sfrac12(-e_1+\sqrt{3}e_2) \bigr\}\ ,
\eeq
yielding the Coxeter reflections
\beq
\Bigl(\begin{smallmatrix} -1 & \ph 0 \\[4pt] \ph 0 & \ph 1 & \end{smallmatrix}\Bigr) \ ,\ 
\sfrac12 \Bigl(\begin{smallmatrix} \ph 1 & -\sqrt{3} \\[2pt] -\sqrt{3} & -1 \end{smallmatrix}\Bigr) \ ,\
\sfrac12 \Bigl(\begin{smallmatrix} 1 & \sqrt{3} \\[2pt] \sqrt{3} & -1 \end{smallmatrix}\Bigr) \ .
\eeq
Its spherical reduction yields the
P\"oschl-Teller model, which describes a particle on~$S^1$ in the potential
\beq
U\= \sfrac92\,g(g{-}1)\,r^6\,\bigl(x^1\bigr)^{-2}\bigl((x^1)^2-3(x^2)^2\bigr)^{-2}
\= {g(g{-}1)}\sfrac{18\,(w\wb)^3}{(w^3+\wb^3)^2}
\= \sfrac92 {g(g{-}1)}\,\cos^{-2}(3\phi)\ ,
\eeq
where we introduced a complex homogeneous $\R^2$ coordinate~$w$ and polar coordinates $(r,\phi)$,
\beq
w\ :=\ x^1+\im x^2 \= r\,\ep^{\im\phi} 
\qquad\Leftrightarrow\qquad
x^1 \= r\,\cos\phi \and x^2 \= r\,\sin\phi\ .
\eeq
Since $A_2$ is simply-laced, all couplings must coincide, $g_\al=g$.
The two basic homogeneous polynomials invariant under $W=S_3$ are
\beq
\si_2=(x^1)^2{+}(x^2)^2= w\,\wb = r^2 \und 
\si_3=3(x^1)^2x^2{-}(x^2)^3 \sim w^3{-}\wb^3 \sim r^3\sin(3\phi)\ .
\eeq
Hence, $d_3=3$, $\{\ell\}=\ell_3$ and $\ell=3\ell_3$, and we have
the $S_3$-invariant spectrum
\beq
E_\ell \= \half\,q^2 \quad\with\quad q\=\ell+3g\=3(\ell_3{+}g)
\und \textrm{deg}(E_\ell)=1\ .
\eeq
For $g>0$ this implies $E_{\textrm{min}}=\frac92g^2$, 
but for $g<0$ the spectrum goes down to zero energy.
The Vandermonde factor takes the simple form
\beq
\De\ \sim\ (x^1)^3-3x^1(x^2)^2\ \sim\ w^3+\wb^3 \ \sim\ r^3\,\cos(3\phi)\ ,
\eeq
and the Dunkl operator in the potential-free frame reads ($\rho=\ep^{2\pi\im/3}$)
\beq
\widetilde{\cD}_w \= \pa_w\ +\ \frac{3\,g\,w^2}{w^3+\wb^3} \ -\ g\,\Bigl\{
\frac{1}{w+\wb}\,s_0 + \frac{\rho}{\rho w+\rb\wb}\,s_+ + \frac{\rb}{\rb w+\rho\wb}\,s_- \Bigr\}
\eeq
with the Coxeter reflections
\beq
s_0: \ w\ \mapsto\ -\wb\quad,\qquad
s_+: \ w\ \mapsto\ -\rho\wb\quad,\qquad
s_-: \ w\ \mapsto\ -\bar{\rho}\wb\quad.
\eeq
Thus, the $S_3$-invariant wave functions in the potential-free frame are (with $r^0\to\ln r$)
\beq \label{hanalytic}
\begin{aligned}
h_\ell^{(g)} &\= r^{2\ell+6g} \bigl( \widetilde{\cD}_w^3-\widetilde{\cD}_\wb^3 \bigr)^{\ell_3} r^{-6g} \\
&\ \sim\ \sum_{k=0}^{\ell_3} \frac{\Gamma(1{+}\ell_3)\,\Gamma(g{+}k)\,\Gamma(g{+}\ell_3{-}k)}
{\Gamma(2g{+}\ell_3)\,\Gamma(g)\,\Gamma(1{+}k)\,\Gamma(1{+}\ell_3{-}k)} \ w^{\ell-3k}(-\wb)^{3k} \\[4pt]
&\=\sfrac{\Gamma(g{+}\ell_3)}{\Gamma(2g{+}\ell_3)}\ 
{}_2F_1\bigl(g,{-}\ell_3;1{-}g{-}\ell_3;(-\sfrac{\wb}{w})^3\bigr)\ w^\ell \\[4pt]
&\=\sfrac{\Gamma(g)\,\Gamma(\ell_3{+}1)}{\Gamma(2g{+}\ell_3)}\ 
P_{\ell_3}^{(-g-\ell_3,2g-1)}\bigl(1{+}2(\sfrac{\wb}{w})^3\bigr)\ (-w)^\ell\ ,
\end{aligned}
\eeq
expressed in terms of the hypergeometric function ${}_2F_1$ or the Jacobi polynomials~$P_n^{(\alpha,\beta)}$.
The gamma-function prefactors are irrelevant for $g>0$ but are chosen such as to enable
an analytic continuation to $g<0$, which will become relevant in a while.
A table of states for small values of~$\ell$ can be found in Appendix~A.1.

The Dunklized angular momentum is given by
\beq
\cL\ \equiv\ \cL_{12}\= x^1\cD_2-x^2\cD_1 \= \im\bigl(w\cD_w-\wb\cD_\wb\bigr)
\eeq
with $\cD_w=\widetilde{\cD}_w-\frac{3\,g\,w^2}{w^3+\wb^3}$. 
{}From this we can build only one algebraically independent $S_3$-symmetric polynomial (case~D),
\beq
{\cal C}_2 \= \cL^2 \= -2\,\cH\ +\ g^2(s_0{+}s_+{+}s_-)^2\ ,
\eeq
whose restriction~$C_2$ to $S_3$-symmetric functions provides the P\"oschl-Teller Hamiltonian
minus its ground-state energy.
The single basic $S_3$-antiinvariant polynomial (case~A) is $\cL$ itself, from which we get
\beq
\cM_1\=\cL \qquad\Rightarrow\qquad
M_1\ \equiv\ \textrm{res}(\cL) \= 
\im \bigl( w\pa_w-\wb\pa_\wb \bigr) - 3\im g\,\frac{w^3{-}\wb^3}{w^3{+}\wb^3} \=
\pa_\phi + 3g\,\tan(3\phi)\ ,
\eeq
which obeys
\beq \label{intertwine}
M_1^{(g)}\,H^{(g)} \= H^{(g+1)}\,M_1^{(g)} \und
M_1^{(1-g)}\,H^{(g)} \= H^{(g-1)}\,M_1^{(1-g)} \ .
\eeq

Because $\cM_1$ is linear in~$\cL$, in this case it is also true that
\beq
\widetilde{M}_1\ \equiv\ \De^{-g}M_1\,\De^g \= 
\textrm{res}\bigl(\widetilde{\cL}\bigr) \=
\im\,\textrm{res}\bigl(w\widetilde{\cD}_w{-}\wb\widetilde{\cD}_\wb\bigr) \= 
L \= \im \bigl( w\pa_w-\wb\pa_\wb \bigr) \= \pa_\phi\ ,
\eeq
which exceptionally does not depend on~$g$.
The ladder relation for the deformed harmonic polynomials (remember deg$(E_\ell)=1$),
\beq \label{Mtildeladder}
\widetilde{M}_1\ h_\ell^{(g)} \= \De\,h_{\ell-3}^{(g+1)}
\qquad\Leftrightarrow\qquad
\partial_\phi \ h_\ell^{(g)} \= r^3\cos(3\phi)\,h_{\ell-3}^{(g+1)}\ ,
\eeq
may for positive integer~$g$ be iterated to generate them from the free ($g{=}0$) ones,
\beq \label{hiterated}
\begin{aligned}
h_\ell^{(g>0)} &\= \bigl(\De^{-1}\widetilde{M}_1\bigr)^g\,h_{\ell+3g}^{(0)} 
\= r^{-3g}\bigl(\cos^{-1}\!(3\phi)\,\partial_\phi\bigr)^g\,h_{\ell+3g}^{(0)} \\[2pt]
&\ \sim\ \bigl( (w^3{+}\wb^3)^{-1} (w\pa_w{-}\wb\pa_\wb) \bigr)^g 
\bigl(w^{\ell+3g}+(-\wb)^{\ell+3g}\bigr)\ ,
\end{aligned}
\eeq
which reproduces the analytic expression~(\ref{hanalytic}).
Eventually, the iteration hits the kernel of $\widetilde{M}_1$, i.e.~$h_0^{(g)}=1$ corresponding to
the ground state, where it ceases.

The $g<0$ states can as well be obtained directly from (\ref{intertwine}), which also implies that
\beq
\widetilde{M}_1\,\De^{2g-1} h_\ell^{(g)} \= \De^{2g-2} h_{\ell+3}^{(g-1)}
\qquad\Leftrightarrow\qquad
\bigl(\partial_\phi-3(2g{-}1)\tan(3\phi)\bigr) h_\ell^{(g)} 
\= r^{-3}\cos^{-1}\!(3\phi)\,h_{\ell+3}^{(g-1)}\;.
\eeq
Its iteration for negative integer~$g$ produces
\beq
\begin{aligned}
h_\ell^{(g<0)} &\= \De^{-2g}\,\bigl(\widetilde{M}_1\,\De^{-1}\bigr)^{-g}\,h_{\ell+3g}^{(0)} 
\= r^{-3g}\cos^{-2g}\!(3\phi)\bigl(\partial_\phi\cos^{-1}\!(3\phi)\bigr)^{-g}\,h_{\ell+3g}^{(0)}\\[2pt]
&\ \sim\ (w^3{+}\wb^3)^{-2g}\,\bigl( (w\pa_w{-}\wb\pa_\wb)\,(w^3{+}\wb^3)^{-1} \bigr)^{-g}
\bigl(w^{\ell+3g}+(-\wb)^{\ell+3g}\bigr)\ ,
\end{aligned}
\eeq
which may be checked to reproduce the analytic continuation of~(\ref{hanalytic}) to $g<0$.
However, without ${\cal PT}$ deformation the full wave functions $v_{\ell}^{(g<0)}$ are
not normalizable.
For illustration, in Appendix A.1 we display the polynomials 
$h_{\ell}^{(g)}$ for $g=-2,-1,0,1,2$ and $q\le12$.
\begin{figure}[h!]
\centering
\includegraphics[scale=0.65]{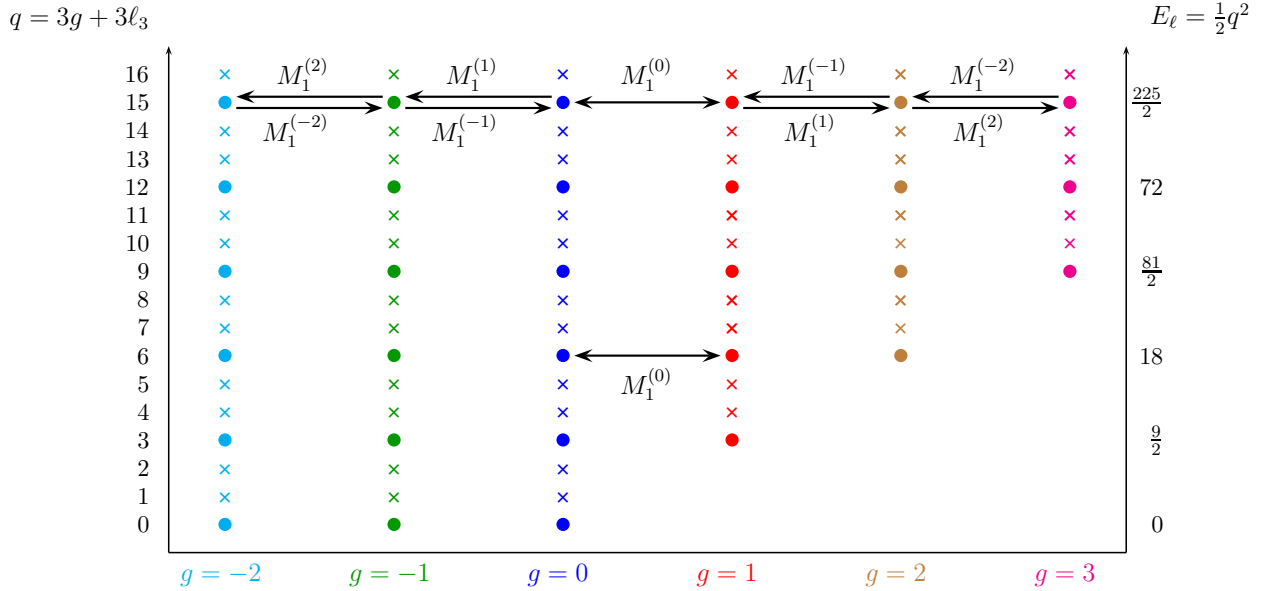}
\caption{Action of the intertwiner $M_1^{(g)}$ on the $A_2$ spectrum for small values of $g$.}
\label{fig1}
\end{figure}

Let us take a look at the possible ${\cal PT}$ involutions and the compatible complex deformations
for the P\"oschl-Teller model.
The only order-2 elements in~$S_3$ are the Coxeter reflections about the lines perpendicular 
to the roots, so without loss of generality we may fix ${\cal P}$ as the action of~$s_0$,
which belongs to the root $\ga=\sqrt{2}e_1$ and is a reflection on the $x^2$-axis,
\beq
\hat\ga \= \bigl(\begin{smallmatrix} 1 \\ 0 \end{smallmatrix}\bigr) 
\qquad\Leftrightarrow\qquad
{\cal P} : \ s_0 \= \bigl(\!\!\begin{smallmatrix} -1 & 0 \\ \ph 0 & 1 \end{smallmatrix}\bigr)\ .
\eeq
Obviously, $P_-$ and $P_+$ project onto the $x^1$ axis and the $x^2$ axis, respectively.
As usual, ${\cal T}$ is complex conjugation, but please be aware that this does {\it not\/} 
swap $w$ with~$\wb$ because the complex linear combination of the real coordinates $x^1$ and $x^2$
is unaffected by~${\cal T}$.

The coset SO$(2,\C)/$SO$(2,\R)$ is one-dimensional and parametrized as
\beq
\Ga(\eps) \= \ep^{\eps G} \= 
\exp\bigl\{\eps\bigl(\begin{smallmatrix} 0 & -\im \\ \im & \ph 0 \end{smallmatrix}\bigr)\bigr\} \=
\begin{pmatrix} \cosh(\eps) & -\im\sinh(\eps) \\[4pt] \im\sinh(\eps) & \ph\cosh(\eps) \end{pmatrix}
\= \cosh(\eps)\,\unity\ +\ \sinh(\eps)\,G\ .
\eeq
Since there is just one plane, necessarily $\hat\eta=e_2$ and $P_{\ga\wedge\eta}=\unity$.
Clearly, $s_0$ and $G$ anticommute, and so all such complex deformations 
\beq
\bigl(x^1 ,\, x^2\bigr) \ \mapsto \ \bigl(x^1(\eps)\, ,\, x^2(\eps) \bigr) \=
\bigl( \cosh(\eps)x^1-\im\sinh(\eps)x^2\, ,\, \cosh(\eps)x^2+\im\sinh(\eps)x^1 \bigr)
\eeq
are ${\cal PT}$ symmetric.  
In polar coordinates, this deformation takes a particularly simple form,
\beq
\bigl(r,\phi\bigr) \ \mapsto \ \bigl(r(\eps)\, ,\, \phi(\eps) \bigr) \=
\bigl( r\, ,\, \phi+\im\eps \bigr) \ ,
\eeq
but for the complex combinations $(w,\wb)$ one has to keep in mind that ${\cal T}$ does not conjugate 
\beq
w(\eps)\= \ep^{-\eps}\,w \qquad\textrm{or}\qquad \wb(\eps) \= \ep^\eps\,\wb
\eeq
but only flips the sign of~$\eps$.
For any root~$\al$ contributing to the potential, 
the singular locus of~$U(\eps)$ for $\eps{\neq}0$ lies at
\beq
\textrm{sing}(\al) \= \bigl\{\,x\,\bigm| \ \al\cdot x=0 \quad\&\quad \al\cdot G\,x=0 \;\bigr\}
\= \emptyset\ \ \forall\al\ ,
\eeq
since $\im G$ is a $\pi/2$ rotation in our plane. Hence, the deformed potential
\beq
U(\eps,\phi) \= 9\,g(g{-}1)\,\frac{1+\cosh(6\eps)\cos(6\phi)+\im\sinh(6\eps)\sin(6\phi)}
{\bigl(\cosh(6\eps)+\cos(6\phi)\bigr)^2}
\eeq
\begin{figure}[h!]
\centering
\includegraphics[scale=0.75]{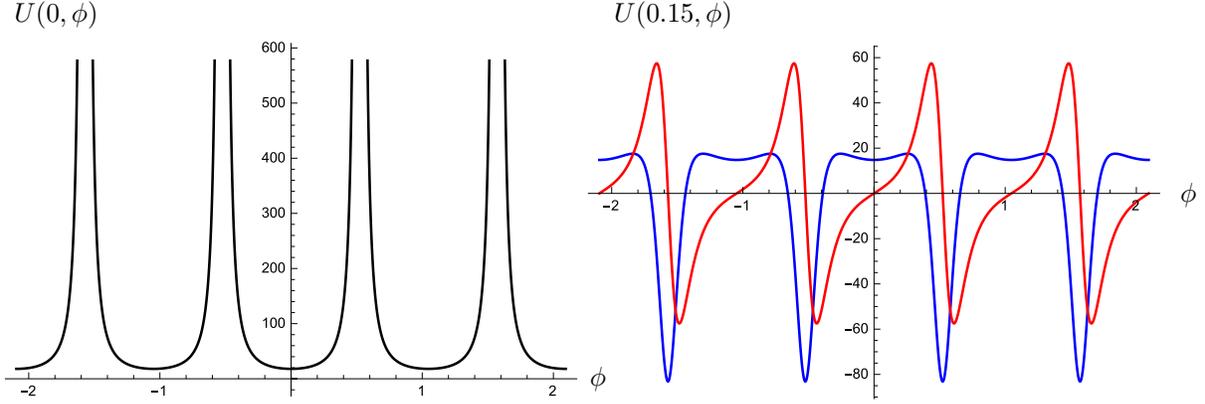}
\caption{Singular ($\eps{=}0$) and regularized ($\eps{=}0.15$) $A_2$ potential~$U(\eps,\phi)$ 
for $g{=}2$.\hfill\break
The blue curve displays $\textrm{Re}\,U$, the red one shows $\textrm{Im}\,U$.}
\label{fig2}
\end{figure}
as well as the deformed wave functions (see (\ref{hanalytic}))
\beq
v_{\ell\,\eps}^{(g)} (w,\wb) \= r^{-q}\,\De_\eps^g\,h_{\ell}^{(g)} (\ep^{-\eps}w,\ep^\eps\wb)
\eeq
for $g<0$ are free of singularities because
\beq
\De_\eps \ \sim\ \ep^{-3\eps} w^3 + \ep^{3\eps} \wb^3 \ \sim\
r^3\bigl( \cosh(3\eps)\cos(3\phi)-\im\sinh(3\eps)\sin(3\phi)\bigr)
\eeq
is regular everywhere.
Because the complex deformation is merely a constant shift of the polar angle,
the angular momentum and the potential-free intertwiner exceptionally remain undeformed,
\beq
\widetilde{M}_{1\eps} \= L_\eps \= \im(w\pa_w-\wb\pa_\wb) \= \pa_\phi\ .
\eeq

Our intertwiner $\widetilde{M}_1$ has a simple kernel. Since 
\beq
h_0^{(g>0)} = 1  \und h_{6|g|}^{(g\le0)} = (w\wb)^{3|g|} = r^{6|g|} \ ,
\eeq
$\widetilde{M}_1$ at any fixed~$g$ annihilates this one state but no other one. 
Our ${\cal PT}$ deformation leads to a rough doubling of the energy eigenstates,
because the spectrum of $H^{(g)}$ now has to be joined with that of $H^{(1-g)}$. 
So, for a given $g{>}\frac12$, we encounter two towers of states with $E=\half q^2$, for
\beq
q \= \ell + 3g \und q \= \ell + 3(1{-}g) \quad\with \ell=3\ell_3=0,3,6,\ldots \ ,
\eeq
where the second tower yields negative $q$ for $\ell<3(g{-}1)$.
When $g$ is integral or half-integral, the two towers meet, so the degeneracy doubles.
However, it turns out that flipping the sign of~$q$ yields the same state again, 
and so for positive integral~$g$ the level degeneracy becomes
\beq
\textrm{deg}(E_q^{(g)}) \= 
\begin{cases} 1 & \for q=0,3,6,\ldots,3(g{-}1) \\ 2 & \for q=3g,3(g{+}1),3(g{+}2),\ldots \end{cases}\ .
\eeq
The new states are again given by (\ref{hanalytic}), where in the limit of negative integral~$g$
the zeros of the Jacobi polynomial are cancelled by poles of the prefactor,
so a careful limit has to be taken.
Such a state structure is common for systems possessing a hidden supersymmetric structure~\cite{bososusy},
which is indeed the case here and revealed by the additional `odd' conserved charge
\beq
Q^{(g)}_\eps\= M_{1\,\eps}^{(g-1)} M_{1\,\eps}^{(g-2)} \cdots M_{1\,\eps}^{(2-g)} M_{1\,\eps}^{(1-g)}
\= \De_\eps^g \bigl( \De_\eps^{-1} \partial_\phi \bigr)^{2g-1}  \De_\eps^{g-1}=\De_\eps^{g} \widetilde{Q}^{(g)}_\eps \De_\eps^{g-1}\ .
\eeq
In the potential-free frame, it simplifies to
\beq
\widetilde{Q}^{(g)}_\eps \= \bigl( \De_\eps^{-1} \partial_\phi \bigr)^{2g-1} 
\= \bigl( (\ep^{-3\eps} w^3 {+} \ep^{3\eps} \wb^3)^{-1}  \im(w\pa_w{-}\wb\pa_\wb)\bigr)^{2g-1}  : \
h^{(1-g)}_{\ell\,\eps} \ \mapsto\ h^{(g)}_{\ell-3(2g-1)\,\eps}
\eeq
and clearly obeys the intertwining relation
\beq \label{intq}
\widetilde{Q}^{(g)}_\eps \widetilde{H}^{(1-g)}_\eps \= \widetilde{H}^{(g)}_\eps \widetilde{Q}^{(g)}_\eps\ ,
\eeq
relating the deformed harmonic polynomials at couplings $1{-}g$ and~$g$.
Since the transition from $h^{(g)}_{\ell}$ to $v^{(g)}_{\ell}$ involves the ($g$-dependent) factor of $\De^g$
and $\widetilde{H}^{(1-g)}\neq\widetilde{H}^{(g)}$, 
only in the potential frame this intertwining relation becomes a commutation relation,
\beq \label{comuq}
[ Q^{(g)}_\eps ,H^{(g)}_\eps] \= [ Q^{(g)}_\eps ,H^{(1-g)}_\eps] \=0\ .
\eeq
The $g$ singlet states (for $q<3g$) are annihilated by $Q^{(g)}_\eps$,
\beq
Q^{(g)}_\eps\,v^{(1-g)}_{\ell\,\eps} = 0 \quad\for \ell_3=g{-}1,g,g{+}1,\ldots,2g{-}2\ ,
\eeq
at energies
\beq
E_q \= \sfrac12 q^2 \= \sfrac{9}{2}(\ell_3+1{-}g)^2\=\sfrac{9}{2}\,j^2 \quad\for j=0,1,\ldots,g{-}1\ .
\eeq
For all other states, $Q^{(g)}_\eps$ maps the doublet partners to each other.
The square of $Q^{(g)}_\eps$ is a polynomial in the Hamiltonian,
\beq
\bigl( Q^{(g)}_\eps \bigr)^2 \ \propto\ \prod_{j=1-g}^{g-1} 
\left( H^{(g)}_\eps-\sfrac{9}{2}j^2\right)
\= H^{(g)}_\eps\,\prod_{j=1}^{g-1} \left( H^{(g)}_\eps-\sfrac{9}{2}j^2 \right)^2\ ,
\eeq
which also reveals the properties of the combined spectrum.
\begin{figure}[h!]
\centering
\includegraphics[scale=0.7]{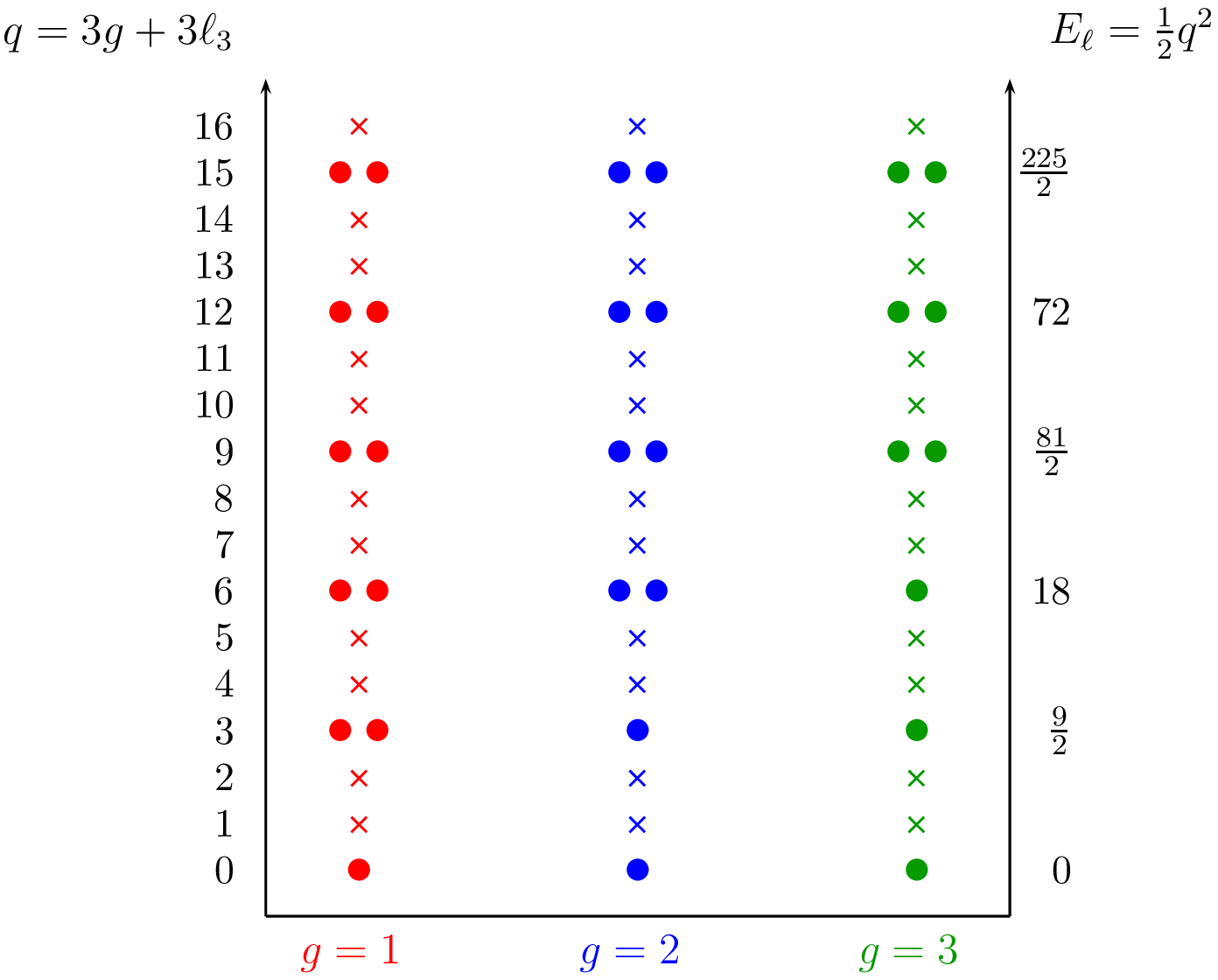}
\caption{Joint spectrum of $H^{(g)}_\eps$ and $H^{(1-g)}_\eps$ 
after the ${\cal PT}$ deformation for the $A_2$ model.}
\label{fig3}
\end{figure}

\section{$G_2$ model}

\noindent 
The $A_2$ model is the first of an infinite list of dihedral $I_2(p)$ models, with
\beq
I_2(2) = A_1\oplus A_1\ ,\quad
I_2(3) = A_2\ ,\quad
I_2(4) = BC_2\ ,\quad
I_2(6) = G_2\ ,
\eeq
and where for odd~$p$ all couplings must coincide while for even~$p$ the root system decomposes
into two $I_2(\frac{p}{2})$ subsystems with two couplings $g_S$ and $g_L$.
Let us illustrate the latter situation on the $G_2$ example, since it can be obtained by
a superposition of two $A_2$ systems (with a $\pi/2$ rotation), 
\beq
{\cal R}_+ \= \bigl\{ e_1\,,\,\sfrac12(e_1+\sqrt{3}e_2)\,,\,\sfrac12(-e_1+\sqrt{3}e_2) \,,\,
\sqrt{3}e_2\,,\,\sfrac12(3e_1+\sqrt{3}e_2)\,,\,\sfrac12(-3e_1+\sqrt{3}e_2)\bigr\}\ ,
\eeq
presented in the previous section.
The corresponding Coxeter reflections read
\beq
\Bigl(\begin{smallmatrix} -1 & \ph 0 \\[4pt] \ph 0 & \ph 1 & \end{smallmatrix}\Bigr) \ ,\
\sfrac12 \Bigl(\begin{smallmatrix} \ph 1 & -\sqrt{3} \\[2pt] -\sqrt{3} & -1 \end{smallmatrix}\Bigr) \ ,\
\sfrac12 \Bigl(\begin{smallmatrix} 1 & \sqrt{3} \\[2pt] \sqrt{3} & -1 \end{smallmatrix}\Bigr) \and
\Bigl(\begin{smallmatrix}  1 & \ph 0 \\[4pt] 0 & -1 & \end{smallmatrix}\Bigr) \ ,\
\sfrac12 \Bigl(\begin{smallmatrix} -1 & -\sqrt{3} \\[2pt] -\sqrt{3} & \ph 1 \end{smallmatrix}\Bigr) \ ,\
\sfrac12 \Bigl(\begin{smallmatrix} -1 & \sqrt{3} \\[2pt] \sqrt{3} & 1 \end{smallmatrix}\Bigr)\ .
\eeq
The potential is easily derived,
\beq
\begin{aligned}
U &\= \sfrac92\,g_S(g_S{-}1)\,r^6\,\bigl(x^1\bigr)^{-2}\bigl((x^1)^2-3(x^2)^2\bigr)^{-2}
\ +\ \sfrac92\,g_L(g_L{-}1)\,r^6\,\bigl(x^2\bigr)^{-2}\bigl((x^2)^2-3(x^1)^2\bigr)^{-2} \\[4pt]
&\= {g_S(g_S{-}1)}\sfrac{18\,(w\wb)^3}{(w^3+\wb^3)^2}
\ -\ {g_L(g_L{-}1)}\sfrac{18\,(w\wb)^3}{(w^3-\wb^3)^2} \\[4pt]
&\= \sfrac92 {g_S(g_S{-}1)}\,\cos^{-2}(3\phi) \ +\ \ \sfrac92 {g_L(g_L{-}1)}\,\sin^{-2}(3\phi)\ ,
\end{aligned}
\eeq
and exhibits the two subsystems.
The Coxeter group is the dihedral group ${\cal D}_6$ with 12 elements, 
which maps short roots to short roots and long roots to long roots.
The two basic ${\cal D}_6$-invariant homogeneous polynomials are
\beq
\si_2= w\,\wb = r^2 \und
\si_6 \sim w^6{+}\wb^6 \sim r^6\cos(6\phi)\ .
\eeq
Hence, $d_3=6$, $\{\ell\}=\ell_3$ and $\ell=6\ell_3$, and we have
the ${\cal D}_6$-invariant spectrum
\beq \label{G2energy}
E_\ell = \half\,q^2 \quad\with\quad q=\ell+3g_S+3g_L=3(2\ell_3{+}g_S{+}g_L)
\und \textrm{deg}(E_\ell)=1\ .
\eeq
For $g_L{=}0$ or $g_S{=}0$, we fall back to the P\"oschl-Teller model, but only its `even' states
survive the more restrictive Weyl invariance requirement, as $\ell$ must be a multiple of~6 now.
Compared to the P\"oschl-Teller model, the density of energy eigenstates is cut in half.
The Vandermonde factorizes,
\beq
\De = \De_S \De_L \quad\with\quad
\De_S \sim w^3+\wb^3 \sim r^3\cos(3\phi) \and
\De_L \sim w^3-\wb^3 \sim r^3\sin(3\phi)\ ,
\eeq
and the (potential-free) Dunkl operator reads
\beq
\begin{aligned}
\widetilde{\cD}_w \= \pa_w\ &+\ \frac{3\,g_S\,w^2}{w^3+\wb^3} \ -\ g_S\,\Bigl\{
\frac{1}{w+\wb}\,s_0 + \frac{\rho}{\rho w+\rb\wb}\,s_+ + \frac{\rb}{\rb w+\rho\wb}\,s_- \Bigr\} \\
&+\ \frac{3\,g_L\,w^2}{w^3-\wb^3} \ -\ g_L\,\Bigl\{
\frac{1}{w-\wb}\,\bar{s}_0 + \frac{\rho}{\rho w-\rb\wb}\,\bar{s}_+ + \frac{\rb}{\rb w-\rho\wb}\,\bar{s}_- \Bigr\}
\end{aligned}
\eeq
with the additional Coxeter reflections
\beq
\bar{s}_0: \ w\ \mapsto\ +\wb\quad,\qquad
\bar{s}_+: \ w\ \mapsto\ +\rho\wb\quad,\qquad
\bar{s}_-: \ w\ \mapsto\ +\bar{\rho}\wb\quad.
\eeq
With these ingredients, the wave functions in the potential-free frame can be constructed,
\beq \label{hganalytic}
\begin{aligned}
h_\ell^{(g_S,g_L)} &\=r^{2\ell+6g_S+6g_L}\bigl(\widetilde{\cD}_w^6+\widetilde{\cD}_\wb^6\bigr)^{\ell_3} r^{-6g_S-6g_L} \\[4pt]
&\ \sim\ P_{\ell_3}^{(g_S-\frac12,g_L-\frac12)}\bigl(\sfrac12(\sfrac{w}{\wb})^3{+}\sfrac12(\sfrac{\wb}{w})^3\bigr)
\ (w\,\wb)^\ell \quad\with \ell=6\ell_3\ .
\end{aligned}
\eeq
Since only even powers of $w$ or $\wb$ occur, its form is a bit simpler than (\ref{hanalytic}). 
Some low-lying wave functions are given explicitly in Appendix~A.2.

The Dunklized angular momentum is given by
\beq
\cL \= \im\bigl(w\cD_w-\wb\cD_\wb\bigr) \quad\with\quad
\cD_w=\widetilde{\cD}_w-\frac{3\,g_S\,w^2}{w^3+\wb^3}-\frac{3\,g_L\,w^2}{w^3-\wb^3}
\eeq
and essentially squares to the Hamiltonian,
\beq
{\cal C}_2 \= \cL^2 \= -2\,\cH\ +\ \bigl[g_S(s_0{+}s_+{+}s_-)+g_L(\bar{s}_0{+}\bar{s}_+{+}\bar{s}_-)\bigr]^2\ ,
\eeq
via $H=\textrm{res}({\cal H})$.
Again, for generic~$g$ this is the only conserved charge (case~D).
Like before, $\cL$ is Weyl antiinvariant (case~A), thus providing the basic intertwiner
\beq
\begin{aligned}
M_1\ \equiv\ \textrm{res}(\cL) &\= \im \bigl( w\pa_w-\wb\pa_\wb \bigr) 
- 3\im g_S\,\frac{w^3{-}\wb^3}{w^3{+}\wb^3} - 3\im g_L\,\frac{w^3{+}\wb^3}{w^3{-}\wb^3} \\[4pt]
&\= \pa_\phi + 3g_S\,\tan(3\phi) - 3g_L\,\cot(3\phi)\ .
\end{aligned}
\eeq
The intertwining relations read
\beq
\begin{aligned}
M_1^{(g_S,g_L)}\,H^{(g_S,g_L)} &\= H^{(g_S+1,g_L+1)}\,M_1^{(g_S,g_L)} \ , \\[4pt]
M_1^{(1-g_S,1-g_L)}\,H^{(g_S,g_L)} &\= H^{(g_S-1,g_L-1)}\,M_1^{(1-g_S,1-g_L)} \ , \\[4pt]
M_1^{(1-g_S,g_L)}\,H^{(g_S,g_L)} &\= H^{(g_S-1,g_L+1)}\,M_1^{(1-g_S,g_L)} \ , \\[4pt]
M_1^{(g_S,1-g_L)}\,H^{(g_S,g_L)}&\= H^{(g_S+1,g_L-1)}\,M_1^{(g_S,1-g_L)} \ .
\end{aligned}
\eeq
and again the potential-free intertwiner trivializes,
\beq 
\widetilde{M}_1\ \equiv\ \De_L^{-g_L}\De_S^{-g_S}M_1\,\De_S^{g_S}\De_L^{g_L} \=
L \= \im \bigl( w\pa_w-\wb\pa_\wb \bigr) \= \pa_\phi\ .
\eeq
The corresponding ladder relations for the wave functions are
\beq \label{G2ladders}
\begin{aligned}
\partial_\phi\ h_\ell^{(g_S,g_L)} &\= \De_S\De_L\,h_{\ell-6}^{(g_S+1,g_L+1)}\ ,\\[4pt]
\partial_\phi\ \De_S^{2g_S-1}\,h_\ell^{(g_S,g_L)} &\= \De_S^{2g_S-2}\De_L\,h_{\ell}^{(g_S-1,g_L+1)}\ ,\\[4pt]
\partial_\phi\ \De_L^{2g_L-1}\,h_\ell^{(g_S,g_L)} &\= \De_S\De_L^{2g_L-2}\,h_{\ell}^{(g_S+1,g_L-1)}\ ,\\[4pt]
\partial_\phi\ \De_S^{2g_S-1}\De_L^{2g_L-1}\,h_\ell^{(g_S,g_L)} &\= \De_S^{2g_S-2}\De_L^{2g_L-2}\,h_{\ell+6}^{(g_S-1,g_L-1)}\ ,
\end{aligned}
\eeq
with special relations for the vanishing of one of the couplings,
\beq \label{G2ladders0}
\begin{aligned}
\partial_\phi\ h_\ell^{(g_S,0)} \= \De_S\,h_{\ell-3}^{(g_S+1,0)} &\und
\partial_\phi\ \De_S^{2g_S-1}\,h_\ell^{(g_S,0)} \= \De_S^{2g_S-2}\,h_{\ell+3}^{(g_S-1,0)}\ ,\\[4pt]
\partial_\phi\ h_\ell^{(0,g_L)} \= \De_L\,h_{\ell-3}^{(0,g_L+1)} &\und
\partial_\phi\ \De_L^{2g_L-1}\,h_\ell^{(0,g_L)} \= \De_L^{2g_L-2}\,h_{\ell+3}^{(0,g_L-1)}\ ,
\end{aligned}
\eeq
where we intermediately allow Weyl `half-invariant' states at $\ell=3,9,12,\ldots$.
For integral couplings the above relations may be iterated for the alternative wave function reconstruction
\beq \label{G2iterate1}
h_\ell^{(g_S,g_L)} \= \begin{cases}
\bigl(\De_S^{-1}\De_L^{-1}\partial_\phi\bigr)^{g_L}\,\bigl(\De_S^{-1}\partial_\phi\bigr)^{g_S-g_L}\,
h^{(0,0)}_{\ell+3g_S+3g_L} & \for g_S\ge g_L\ge0 \\[4pt]
\bigl(\De_S^{-1}\De_L^{-1}\partial_\phi\bigr)^{g_S}\,\bigl(\De_L^{-1}\partial_\phi\bigr)^{g_L-g_S}\,
h^{(0,0)}_{\ell+3g_S+3g_L} & \for g_L\ge g_S\ge0 \end{cases}\ ,
\eeq
\beq \label{G2iterate2}
h_\ell^{(g_S,g_L)} \= \begin{cases}
\De_S^{-2g_S}\bigl(\De_L^{-1}\partial_\phi\De_S^{-1}\big)^{g_L} \bigl(\partial_\phi\De_S^{-1}\bigr)^{-g_S-g_L}\,
h^{(0,0)}_{\ell+3g_S+3g_L} & \for {-}g_S\ge g_L\ge0 \\[4pt]
\De_S^{-2g_S}\bigl(\De_L^{-1}\partial_\phi\De_S^{-1}\big)^{-g_S} \bigl(\De_L^{-1}\partial_\phi\bigr)^{g_S+g_L}\,
h^{(0,0)}_{\ell+3g_S+3g_L} & \for g_L\ge {-}g_S\ge0 \end{cases}
\eeq
and similarly for the four other domains of $(g_S,g_L)$, starting from 
\beq
h_{\ell+3g_S+3g_L}^{(0,0)} \ \sim\ w^{\ell+3g_S+3g_L}+(-\wb)^{\ell+3g_S+3g_L} \quad\with \ell=0,6,12,\ldots \ .
\eeq
When $g_S$ and $g_L$ are non-negative, the wave functions are normalizable. 
For integral couplings, the ${\cal D}_6$-invariant energy spectrum $E_\ell=\frac12 q^2$ is non-empty only for
\beq \label{G2qvalues}
q \= \begin{cases} \ph 0 \quad\textrm{mod} \ 6 & \quad\textrm{if}\quad g_S{+}g_L \ \textrm{is even} \ph \\
\ph 3 \quad\textrm{mod} \ 6 & \quad\textrm{if}\quad g_S{+}g_L \ \textrm{is odd} \end{cases} \Biggr\} \und
q\ \ge\ 3(g_S+g_L)\ .
\eeq
When a coupling turns negative, the zeros of the corresponding Vandermonde factor render the full wave function
$v_\ell^{(g_S,g_L)}$ non-normalizable. In order to make these states physical, we turn to the ${\cal PT}$ deformation.

The order-2 elements in ${\cal D}_6$ are precisely the 6 root reflections, 
so there are only two inequivalent cases, corresponding to 
$\hat\ga=\bigl(\begin{smallmatrix}1\\0\end{smallmatrix}\bigr)$ and to 
$\hat\ga=\bigl(\begin{smallmatrix}0\\1\end{smallmatrix}\bigr)$,
\beq
{\cal P} : \ s_0 \= \bigl(\!\!\begin{smallmatrix} -1 & 0 \\ \ph 0 & 1 \end{smallmatrix}\bigr) \und
{\cal P} : \ \bar{s}_0 \= \bigl(\begin{smallmatrix} 1 & \ph 0 \\ 0 & -1 \end{smallmatrix}\bigr)\ .
\eeq
However, since all linear complex coordinate deformations are admissible in two dimensions, 
the discussion is identical to the $A_2$ case, and the potential and wave functions lose all their singularities,
\beq
\begin{aligned}
U(\eps) &\= 9\,g_S(g_S{-}1)\,\frac{1+\cosh(6\eps)\cos(6\phi)+\im\sinh(6\eps)\sin(6\phi)}
{\bigl(\cosh(6\eps)+\cos(6\phi)\bigr)^2} \\
&\ +\ 9\,g_L(g_L{+}1)\,\frac{1-\cosh(6\eps)\cos(6\phi)-\im\sinh(6\eps)\sin(6\phi)}
{\bigl(\cosh(6\eps)-\cos(6\phi)\bigr)^2} \ ,
\end{aligned}
\eeq
\beq
\begin{aligned}
\De_{S\,\eps} \ &\sim\ \ep^{-3\eps} w^3 + \ep^{3\eps} \wb^3 \ \sim\
r^3\bigl( \cosh(3\eps)\cos(3\phi)-\im\sinh(3\eps)\sin(3\phi)\bigr)\ ,\\
\De_{L\,\eps} \ &\sim\ \ep^{-3\eps} w^3 - \ep^{3\eps} \wb^3 \ \sim\
r^3\bigl( \cosh(3\eps)\sin(3\phi)+\im\sinh(3\eps)\cos(3\phi)\bigr)\ .
\end{aligned}
\eeq
The ${\cal PT}$ deformation now leads to an approximate quadrupling of the eigenstates because
\beq \label{4towers}
H^{(g_S,g_L)}_\eps \= H^{(1-g_S,g_L)}_\eps \= H^{(g_S,1-g_L)}_\eps \= H^{(1-g_S,1-g_L)}_\eps
\eeq
tells us to join four towers of states.
Let us look at positive integral couplings $(g_S,g_L)$. 
Then, (\ref{G2energy}) implies that the first and fourth tower from (\ref{4towers}) coincide,
and likewise do the second and third tower. Depending on whether $g_S{+}g_L$ is even or odd, 
one pair of towers sits at $q=0,6,12,\ldots$ and the other one at $q=3,9,15,\ldots$.
Therefore, the density of energy eigenstates is about the same as in the $A_2$~model.
Like in the latter though, some states are missing for small values of~$q$, since the towers
do not reach all the way down to zero (see~(\ref{G2qvalues})).
\begin{figure}[h!]
\centering
\includegraphics[scale=0.7]{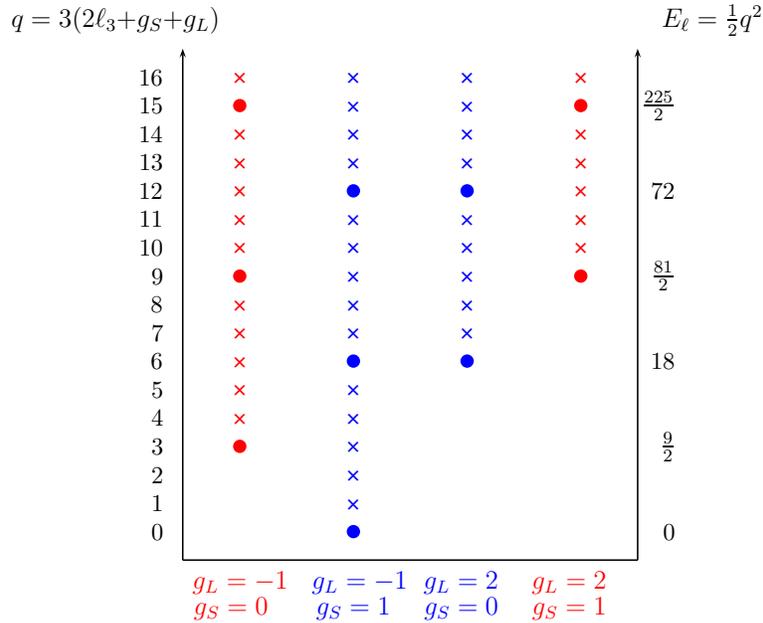}
\caption{The spectra of the $G_2$ Hamiltonians $H^{(g_S,g_L)}_\eps$ for the four towers 
should be joined for the full ${\cal PT}$-symmetric extension.
The blue and red towers are distinguished by $q$ taking odd and even integer values, respectively.}
\label{fig4}
\end{figure}

\begin{figure}[h!]
\centering
\includegraphics[scale=0.65]{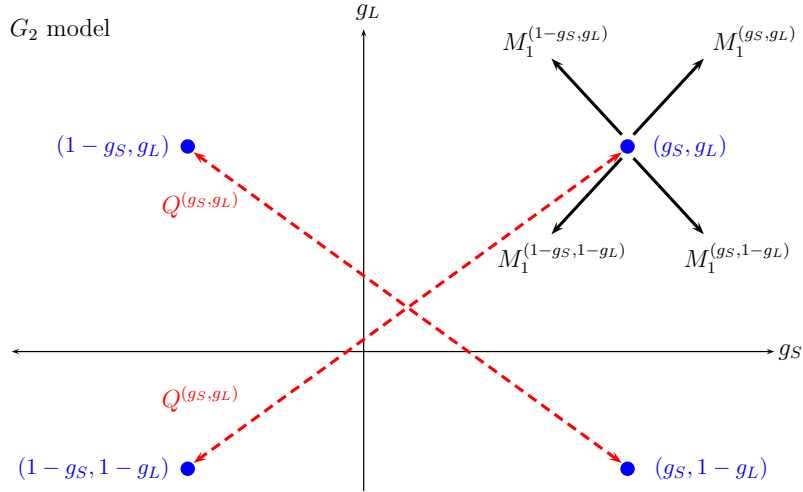}
\caption{Action of the intertwining operators and `odd' conserved charges in the $G_2$ model.}
\label{fig5}
\end{figure}

When $g_S$ and $g_L$ are positive integers, we can write down an additional `odd' conserved charge
\beq
\begin{aligned}
Q^{(g_S,g_L)}_{\eps} &: \ 
v^{(1-g_S,1-g_L)}_{\ell\ \eps} \ \mapsto\ v^{(g_S,g_L)}_{\ell-6(g_S+g_L-1)\ \eps}\ , 
\end{aligned}
\eeq
whose explicit form reads
\begin{align}\label{qope}
Q^{(g_S,g_L)}_{\eps}\=
\begin{cases}
\displaystyle
\left( \, \prod_{j=g_S+g_L-1}^{2g_S-2} M_{1\,\eps}^{(1-g_S-j,2 - g_L - 2 g_S+j)} \right)
\left( \prod_{i=0}^{g_S+g_L-2} M_{1\,\eps}^{(1-g_S-i,1-g_L-i)} \right)  \for g_S \ge g_L  \\ 
\displaystyle
\left( \, \prod_{j=g_S+g_L-1}^{2g_L-2} M_{1\,\eps}^{(2 - g_S - 2 g_L+j,1-g_L-j)} \right)
\left( \prod_{i=0}^{g_S+g_L-2} M_{1\,\eps}^{(1-g_S-i,1-g_L-i)} \right) \for g_L \ge g_S   
\end{cases}
\end{align}
where the product order must be assumed from right to left due to noncommuting action 
of the intertwining operators. 
The potential-free form reads
\beq\label{gopefree}
\widetilde{Q}^{(g_S,g_L)}_{\eps} \= \begin{cases}
\bigl(\De_{S\,\eps}^{-1}\De_{L\,\eps}^{-1}\pa_\phi\bigr)^{g_L}
\bigl(\De_{S\,\eps}^{-1}\pa_\phi\bigr)^{2(g_S-g_L)}
\bigl(\De_{S\,\eps}^{-1}\De_{L\,\eps}^{-1}\pa_\phi\bigr)^{g_L-1} 
& \for g_S\ge g_L \\[4pt]
\bigl(\De_{S\,\eps}^{-1}\De_{L\,\eps}^{-1}\pa_\phi\bigr)^{g_S}
\bigl(\De_{L\,\eps}^{-1}\pa_\phi\bigr)^{2(g_L-g_S)}
\bigl(\De_{S\,\eps}^{-1}\De_{L\,\eps}^{-1}\pa_\phi\bigr)^{g_S-1}
& \for g_L\ge g_S \end{cases} \ .
\eeq
The form (\ref{qope}) or (\ref{gopefree}) represents an action 
$(g_S,g_L)\mapsto(1{-}g_S,1{-}g_L)$ on the couplings. 
Analogously to~(\ref{intq}), $\widetilde{Q}^{(g_S,g_L)}_{\eps}$ obeys an intertwining relation,
while ${Q}^{(g_S,g_L)}_{\eps}$ commutes with the potential-frame Hamiltonian as in~(\ref{comuq}).
There exist other admissible actions like 
$(1{-}g_S,g_L)\mapsto(g_S,1{-}g_L)$ which only produce different factorizations of the same 
operator~(\ref{qope}) but no new conserved charges, see Fig.~\ref{fig5}.
For $g_S \geq g_L$, $Q^{(g_S,g_L)}_{\eps}$ annihilates the singlet states with energies
\begin{align}
E(g_S,g_L;j)\=  \begin{cases}
\frac{9}{2} j^2
& \for j-g_S+g_L <0 \\[4pt]
\frac{9}{2}( g_L{-}g_S {+}j)^2
& \for j-g_S+g_L \geq 0 \end{cases}, \quad  j=1,\ldots, g_S-1\ .
\end{align}
For $g_L \geq g_S$, the roles of $g_L$ and $g_S$ are reversed.
In analogy with the $A_2$ case, 
(\ref{qope})  squares to a polynomial in the Hamiltonian \cite{spectral}, 
\beq
\begin{aligned}
\bigl(Q_{\eps}^{(g_S,g_L)}\bigr)^2 \ \propto\ \begin{cases} \displaystyle
 H^{(g_S,g_L)}_\eps\prod_{j=1}^{g_S-1} \bigl( H^{(g_S,g_L)}_\eps-E(g_S,g_L;j) \bigr)^2 
 & \for g_S\ge g_L\ge0 \\ \displaystyle
 H^{(g_S,g_L)}_\eps\prod_{j=1}^{g_L-1} \bigl( H^{(g_S,g_L)}_\eps-E(g_L,g_S;j) \bigr)^2 
 & \for g_L\ge g_S\ge0
\end{cases}\ .
\end{aligned}
\eeq

The structure presented in the last two sections are easily generalized to all dihedral $I_2(p)$ models.
Essentially, $w^3$ is replaced by $w^p$ or $w^{p/2}$, $\ell=p\,\ell_3$, $\rho$ becomes a $p$th root of unity,
and the intertwiner shifts $(g_S,g_L,\ell)\to(g_S{+}1,g_L{+}1,\ell{-}p)$.
The wave-function formul\ae\ (\ref{hanalytic}) and (\ref{hganalytic}) generalize without any change after the first line.

\section{$AD_3$ model}

\noindent
A much richer and less trivial situation appears one dimension higher, i.e.~at rank~3. 
To reduce index cluttering, we redenote the coordinates as
\beq
x \= \bigl( x^1,x^2,x^3\bigr) \ =:\ \bigl(x,y,z\bigr) \ .
\eeq
The set of positive roots can be chosen as
\beq
{\cal R}_+ \= \bigl\{ e_x{+}e_y\,,\,e_x{-}e_y\,,\,e_x{+}e_z\,,\,e_x{-}e_z\,,\,e_y{+}e_z\,,\,e_y{-}e_z \bigr\}\ .
\eeq
We consider the $A_3$ (or $D_3$) Calogero model spherically reduced to what we have named
the tetrahexahedric model. Here, a particle moves on the 2-sphere with the potential
\beq
U \= 2\,g(g{-}1)\,(x^2{+}y^2{+}z^2)\, \biggl( 
\frac{x^2+y^2}{(x^2-y^2)^2} +
\frac{y^2+z^2}{(y^2-z^2)^2} +
\frac{z^2+x^2}{(z^2-x^2)^2} \biggr)\ .
\eeq
Since $W=S_4$ has just one orbit on the root system, again we have a single coupling, $g_\al=g$.
It can be generated by 
\beq
s_{x-y}= \Bigl(\begin{smallmatrix} 0 & \ph 1 & \ph 0 \\ 1 & \ph 0 & \ph 0 \\ 0 & \ph 0 & \ph 1 \end{smallmatrix}\Bigr) \ ,\
s_{y-z}= \Bigl(\begin{smallmatrix} 1 & \ph 0 & \ph 0 \\ 0 & \ph 0 & \ph 1 \\ 0 & \ph 1 & \ph0 \end{smallmatrix}\Bigr) \ ,\
s_{y+z}= \Bigl(\begin{smallmatrix} 1 & \ph 0 & \ph 0 \\ 0 & \ph 0 & -1 \\ 0 & -1 & \ph 0 \end{smallmatrix}\Bigr)\ .
\eeq
The three basic $S_4$-invariant homogeneous polynomials read
\beq
\si_2=x^2+y^2+z^2=: r^2 \quad,\qquad\
\si_3= x\,y\,z \quad,\qquad
\si_4=x^4+y^4+z^4\quad,
\eeq
and therefore $\{\ell\}=(\ell_3,\ell_4)$ and $\ell=3\ell_3+4\ell_4$, and the energy levels take the form
\beq
E_\ell \= \half\,q\,(q+1) \quad\with\quad
q\=\ell+6g\=3\ell_3+4\ell_4+6g
\eeq
for $S_4$-invariant states, with a degeneracy given by
\beq \label{A3deg}
\textrm{deg}(E_\ell) \= \Bigl\lfloor\frac{\ell}{12}\Bigr\rfloor\ +\ 
\begin{cases} 0 & \for \ \ell=1,2,5\ \ \textrm{mod}\ 12 \\
              1 & \for \ \ell=\textrm{else}\ \ \textrm{mod}\ 12 \end{cases}\ .
\eeq
The Vandermonde reads
\beq
\De \= (x^2-y^2)(y^2-z^2)(z^2-x^2) \ ,
\eeq
and the Dunkl operators in the potential-free frame are given by
\beq \label{D3Dunkl}
\begin{aligned}
\widetilde{\cD}_x &\= \pa_x\ +\ 
\frac{g}{x{+}y}(1{-}s_{x+y}) + \frac{g}{x{-}y}(1{-}s_{x-y}) +
\frac{g}{x{+}z}(1{-}s_{x+z}) + \frac{g}{x{-}z}(1{-}s_{x-z}) \ ,\\
\widetilde{\cD}_y &\= \pa_y\ +\ 
\frac{g}{y{+}z}(1{-}s_{y+z}) + \frac{g}{y{-}z}(1{-}s_{y-z}) +
\frac{g}{y{+}x}(1{-}s_{x+y}) + \frac{g}{y{-}x}(1{-}s_{x-y}) \ ,\\
\widetilde{\cD}_z &\= \pa_z\ +\ 
\frac{g}{z{+}x}(1{-}s_{x+z}) + \frac{g}{z{-}x}(1{-}s_{x-z}) +
\frac{g}{z{+}y}(1{-}s_{y+z}) + \frac{g}{z{-}y}(1{-}s_{y-z}) \ ,
\end{aligned}
\eeq
with the Coxeter reflections
\beq \label{D3reflections}
\begin{aligned} 
s_{x+y} : \ (x,y,z) \mapsto (-y,-x,+z) \quad,\qquad & 
s_{x-y} : \ (x,y,z) \mapsto (+y,+x,+z) \quad, \\
s_{x+z} : \ (x,y,z) \mapsto (-z,+y,-x) \quad,\qquad &
s_{x-z} : \ (x,y,z) \mapsto (+z,+y,+x) \quad,\\
s_{y+z} : \ (x,y,z) \mapsto (+x,-z,+y) \quad,\qquad & 
s_{y-z} : \ (x,y,z) \mapsto (+x,+z,+y) \quad.
\end{aligned}
\eeq
These ingredients enter the $S_4$-invariant energy eigenfunctions 
$v^{(g)}_{\{\ell\}}=r^{-q}\De^g h^{(g)}_{\{\ell\}}(x)$ in
\beq \label{AD3states}
h^{(g)}_{\{\ell\}}(x,y,z) \= r^{2\ell+12g+1}\,
\bigl(\widetilde{\cD}_x\widetilde{\cD}_y\widetilde{\cD}_z\bigr)^{\ell_3}\,
\bigl(\widetilde{\cD}_x^4{+}\widetilde{\cD}_y^4{+}\widetilde{\cD}_z^4\bigr)^{\ell_4}\,
r^{-1-12g}\ ,
\eeq
for which we cannot offer a more explicit expression.
The lowest-energy wave functions are given in the table of Appendix~A.3. 
Their degeneracies and corresponding quantum numbers $(\ell_3,\ell_4)$ are listed below,
where the notation $(\ell_3,\ell_4)^*$ identifies the $q{<}0$ states.
{\small
\beq\notag
\begin{array}{||c|c|c||c|c|c||c|c|c||}
\hline
g{=}{-}2  & \! \text{deg} \!\! & (\ell_3,\ell_4) &  
g{=}{-}1 &  \! \text{deg} \!\! &  \!\! (\ell_3,\ell_4) \!\! & 
g\geq0 &  \! \text{deg} \!\! & \!\! (\ell_3,\ell_4) \!\!  \\ \hline \hline \vphantom{\Big|}
E=0 & 3 & (4,0),(0,3),(1,2)^* &
E=0 & 1 & (2,0)  &
E=\frac{1}{2}6g(6g{+}1)\qquad{} & 1 & (0,0) \\ \hline\vphantom{\Big|}
E=1 & 2 & (3,1), (2,1)^* &
E=1 & 2 & (1,1),(0,1)^* &
E=\frac{1}{2}(6g{+}3)(6g{+}4) & 1 & (1,0) \\ \hline\vphantom{\Big|}
E=3 & 2 & (2,2),(3,0)^* &
E=3 & 2 & (0,2) ,(1,0)^* &
E=\frac{1}{2}(6g{+}4)(6g{+}5)& 1 & (0,1)  \\ \hline\vphantom{\Big|}
E=6 & 3 & (5,0),(1,3),(0,2)^* &
E=6 & 1 & (3,0)  &
E=\frac{1}{2}(6g{+}6)(6g{+}7)& 1 & (2,0)  \\ \hline\vphantom{\Big|}
E{=}10 & 3 & (4,1),(0,4),(1,1)^* &
E{=}10 & 1 & (2,1)  &
E=\frac{1}{2}(6g{+}7)(6g{+}8) & 1 & (1,1) \\ \hline
\end{array}
\eeq
}

The Dunkl-deformed angular momenta,
\beq
\cL_x \equiv \cL_{yz} = y\cD_z{-}z\cD_y \ ,\quad
\cL_y \equiv \cL_{zx} = z\cD_x{-}x\cD_z \ ,\quad
\cL_z \equiv \cL_{xy} = x\cD_y{-}y\cD_x 
\eeq
with $\cD_i=\widetilde{\cD}_i-g\,\pa_i\ln\De$ (amounting to dropping the `1's in (\ref{D3Dunkl})),
get permuted under the action of~$S_4$, with an odd number of sign flips thrown in.
The ring of Weyl invariant polynomials in $\{\cL_x,\cL_y,\cL_z\}$ (case~D) is generated by
\beq
{\cal C}_k \= \cL_x^k + \cL_y^k + \cL_z^k \quad\for k=0,2,4,6\ ,
\eeq
where
\beq
{\cal C}_2 \= -2\,\cH\ +\ S(S{+}1) \quad\with\quad S\=g\sum_\al s_\al\ ,
\eeq
giving rise to three algebraically independent conserved quantities, 
$C_k=\textrm{res}({\cal C}_k)$ for $k=2,4,6$, see also \cite{CoLe15}.
Their algebra seems to be freely generated, modulo the center spanned by~$C_2$.

The basic Weyl antiinvariants built from $\{\cL_x,\cL_y,\cL_z\}$ (case~A) are
\beq \label{AD3intertwiners}
\begin{aligned}
\cM_3 &\= 
\cL_x\cL_y\cL_z+\cL_x\cL_z\cL_y+\cL_y\cL_z\cL_x
+\cL_y\cL_x\cL_z+\cL_z\cL_x\cL_y+\cL_z\cL_y\cL_x\ ,\\[4pt]
\cM_6 &\=
\{\cL_x^4,\cL_y^2\}-\{\cL_y^4,\cL_x^2\}+\{\cL_y^4,\cL_z^2\}
-\{\cL_z^4,\cL_y^2\}+\{\cL_z^4,\cL_x^2\}-\{\cL_x^4,\cL_z^2\}\ ,
\end{aligned}
\eeq
and all higher ones are words in these and the ${\cal C}_k$.
Their restriction to $S_4$-symmetric functions produces two independent intertwiners,
$M_3$ and $M_6$, which obey the same relations~(\ref{intertwine}).
Their potential-free version\footnote{
Note that $\widetilde{M}_s$ are {\it not\/} the restrictions of the corresponding
polynomials $\widetilde{\cM}_s$ in the potential-free frame.}
\beq
\widetilde{M}_s \= \De^{-g} M_s\,\De^g \quad\for s=3,6
\eeq
can be employed to step up the energy eigenfunctions in the coupling,
\beq
\widetilde{M}_s\,h_{\{\ell\}}^{(g)} \= 
\smash{\sum_{\{\ell'\}\atop\ell'=\ell-6}}
c^{s\,\{\ell\}}_{\{\ell'\}} \ \De\,h_{\{\ell'\}}^{(g+1)}\ .
\eeq
$\ph$\\
In this way, eventually all states with positive integer coupling can be reached.
This may not be true for the (more numerous) negative integer coupling states, 
some of which can be found by applying the adjoint intertwiner.
In contrast to the previous section, $\widetilde{M}_s$ now depend on the value of~$g$,
which prevents a nice closed formula like (\ref{hanalytic}) for the polynomials $h_{\{\ell\}}^{(g)}$.
\begin{figure}[h!]
\centering
\includegraphics[scale=0.65]{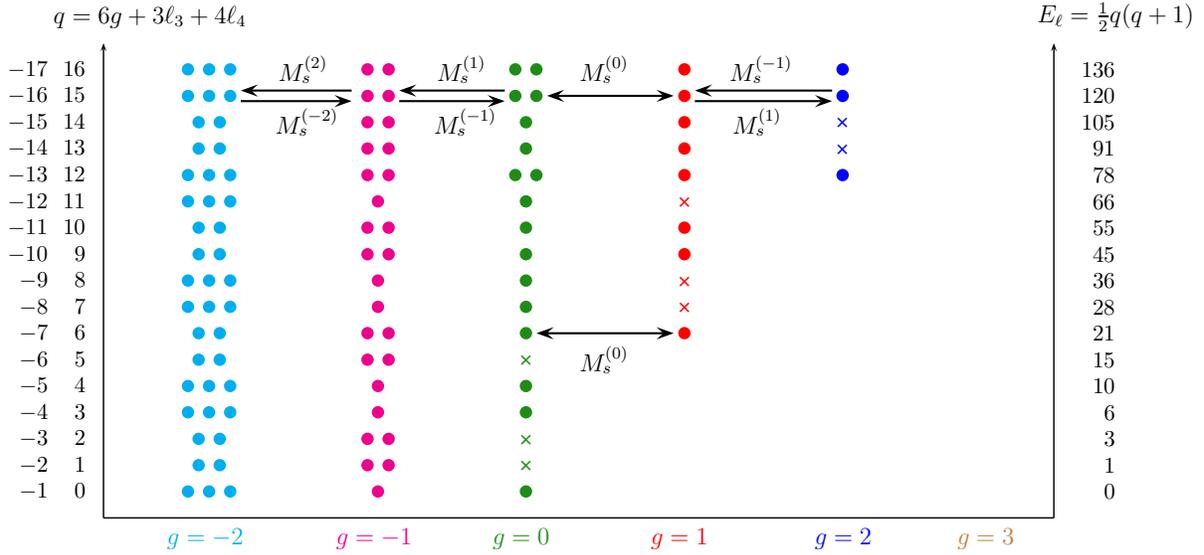}
\caption{Spectrum of $H^{(g)}_\eps$ and action of the interwiners for the $AD_3$ model.}
\label{fig6}
\end{figure}

What are the possibilities for a linear realization of ${\cal PT}$ transformations?
The Coxeter group $W=S_4$ contains one rank-zero involution (the identity),
6 rank-one involutions (the Coexeter reflections), and 3 rank-two involutions 
($\pi$ rotations on one of the three basic planes). The unique rank-three involution
(the negative identity) is the outer automorphism of~$A_3$, hence it is not in $S_4$ 
but generates its double cover. Vanishing rank or co-rank of $P_-$ does not admit 
a compatible complex deformation.
The three-dimensional coset $\textrm{SO}(3,\C)/\textrm{SO}(3,\R)$ is parametrized as
\beq \label{defA3}
\Gamma(\eps,e)\=\exp\Bigl\{-\im\eps\Bigl(\!\!\begin{smallmatrix}
\ph 0 & \ph w & -v \\ -w & \ph 0 & \ph u \\ \ph v & -u & \ph 0 \end{smallmatrix}\Bigr)\Bigr\} \=
\biggl( \begin{smallmatrix}
\c+(1{-}\c)u^2 & \ \ (1{-}\c)uv-\im\s w \ \ & (1{-}\c)uw+\im\s v \\[2pt]
(1{-}\c)vu+\im\s w & \c+(1{-}\c)v^2 & (1{-}\c)vw-\im\s u \\[2pt]
(1{-}\c)wu-\im\s v & (1{-}\c)wv+\im\s u & \c+(1{-}\c)w^2 
\end{smallmatrix} \biggr)
\eeq
where
\beq
e \= \Bigl(\begin{smallmatrix} u \\ v \\ w \end{smallmatrix}\Bigr) \ ,\quad u^2+v^2+w^2=1
\und \c\equiv\cosh\eps\ ,\quad \s\equiv\sinh\eps\ .
\eeq
Clearly, any nonvanishing~$G$ is of rank two.
Degeneracy in the singular locus $\al\cdot x(\eps)=0$ occurs only when $e$ is parallel to some root~$\al$.

For rank$(P_-)=1$, without loss of generality we choose ${\cal P}$ to permute $x$ and~$y$, i.e.
\beq \label{rank-one}
{\cal P}=s_{x-y}=
\Bigl(\begin{smallmatrix} 0 & 1 & 0 \\ 1 & 0 & 0 \\ 0 & 0 & 1 \end{smallmatrix}\Bigr) \quad,\qquad
\ga=\Bigl(\!\!\begin{smallmatrix} \ph 1 \\ -1 \\ \ph 0 \end{smallmatrix}\Bigr) \quad,\qquad
\eta=\Bigl(\begin{smallmatrix} w/2 \\ w/2 \\ -v \end{smallmatrix}\Bigr) \quad,
\eeq
with free real parameters $u$ and~$v$.
Compatibility of (\ref{defA3}) with the rank-one involution (\ref{rank-one}) requires merely $u=v$.
The simplest option is $(u,v,w)=(0,0,1)$, which copies the $n{=}2$ case into the $xy$~plane,
\beq \label{xyboost}
\Bigl(\begin{smallmatrix} x \\[2pt] y \\[2pt] z \end{smallmatrix}\Bigr)(\eps) \=
\Bigl(\begin{smallmatrix} \c\,x\ -\ \im\s\,y \\[2pt] \c\,y\ +\ \im\s\,x \\[2pt] z \end{smallmatrix}\Bigr)\ .
\eeq
Since no root is orthogonal to this plane, our option is generic, 
and each singular locus has a nontrivial imaginary part.
This is not the case for another option, $(u,v,w)=\pm(1,1,0)/\sqrt{2}$, 
since this unit vector is parallel to a root.

For rank$(P_-)=2$, we may take ${\cal P}$ to rotate by $\pi$ in the $yz$~plane, so effectively
\beq \label{rank-two}
{\cal P}=s_{y+z}s_{y-z}=
\Bigl(\begin{smallmatrix} 1 & \ph 0 & \ph 0 \\ 0 & -1 & \ph 0 \\ 0 & \ph 0 & -1 \end{smallmatrix}\Bigr) \quad,\qquad
\ga=\Bigl(\begin{smallmatrix} 1 \\ 0 \\ 0 \end{smallmatrix}\Bigr) \quad,\qquad
\eta=\Bigl(\!\!\begin{smallmatrix} \ph 0 \\ \ph w \\ -v \end{smallmatrix}\Bigr) \quad,
\eeq
The deformation~(\ref{defA3}) is consistent with (\ref{rank-two}) precisely if $u=0$. 
Specializing once more to $(u,v,w)=(0,0,1)$, without loss of generality, 
we again arrive at the boost~(\ref{xyboost}).
Also in this case, there are some degenerate options, namely $(u,v,w)=\pm(0,1,1)/\sqrt{2}$
and $(u,v,w)=\pm(0,1,-1)/\sqrt{2}$.

The singular set of the deformed potential 
\beq
\frac{U(\eps)}{2{g(g{-}1)}} \=
\frac1{\sin^2\!\th\,\cos^2 2(\phi{+}\im\eps)}+
\frac{\cos^2\!\th+\sin^2\!\th\cos^2(\phi{+}\im\eps)}{(\cos^2\!\th-\sin^2\!\th\cos^2(\phi{+}\im\eps))^2}+
\frac{\cos^2\!\th+\sin^2\!\th\sin^2(\phi{+}\im\eps)}{(\cos^2\!\th-\sin^2\!\th\sin^2(\phi{+}\im\eps))^2}
\eeq
consists of 6 antipodal pairs $(x_\al,-x_\al)$ of points,
\beq
\textrm{sing}\=\pm \Bigl\{ \ 
\Bigl(\begin{smallmatrix} 0 \\ 0 \\ 1 \end{smallmatrix}\Bigr)\ ,\
\Bigl(\begin{smallmatrix} 0 \\ 0 \\ 1 \end{smallmatrix}\Bigr)\ ,\
\Bigl(\begin{smallmatrix} 0 \\ \nu \\ \c\,\nu \end{smallmatrix}\Bigr)\ ,\
\Bigl(\begin{smallmatrix} 0 \\ -\nu \\ \c\,\nu \end{smallmatrix}\Bigr)\ ,\
\Bigl(\begin{smallmatrix} \nu \\ 0 \\ \c\,\nu \end{smallmatrix}\Bigr)\ ,\
\Bigl(\begin{smallmatrix} -\nu \\ 0 \\ \c\,\nu \end{smallmatrix}\Bigr) \ \Bigr\}
\quad\with \nu=1/\sqrt{1{+}\c^2}\ ,
\eeq
where the first two pairs coincide. With increasing deformation parameter~$\eps$, 
the other four pairs move from the location of the roots (outside the $xy$~plane) 
to the north and south poles.
Clearly, the singular Vandermonde factor $\De_\eps^g$ keeps the energy eigenstates
unphysical for negative values of~$g$. 
Hence, only the free state spaces at $g{=}0$ and $g{=}1$ should be combined, so that
its degeneracy becomes
\beq
\textrm{deg}(E_\ell^{(1)})\=\Bigl\lfloor\frac{\ell}{6}\Bigr\rfloor\ +\ 
\begin{cases} 0 & \for \ell=1,2,5\quad\textrm{mod}\ 6 \\ 1 & \for \ell=0,3,4\quad\textrm{mod}\ 6 \end{cases}\ .
\eeq
When the potential is turned on, the linear ${\cal PT}$ deformation hence does not alter the 
degeneracy of the energy spectrum but smoothly modifies the states.

With a nonlinear ${\cal PT}$ deformation of the type (\ref{nonlinearPT}) we may, however,
completely remove the wave-function and potential singularities. For the case at hand, it reads
\beq \label{nonlinearA3}
\begin{pmatrix} x \\[2pt] y \\[2pt] z \end{pmatrix}(\eps_1,\eps_2) \=
r\,\begin{pmatrix} 
\sin(\th{+}\im\eps_1)\cos(\phi{+}\im\eps_2) \\[2pt]
\sin(\th{+}\im\eps_1)\sin(\phi{+}\im\eps_2) \\[2pt]
\cos(\th{+}\im\eps_1) 
\end{pmatrix} \=
r\,\begin{pmatrix}
c_1 c_2\,x - \im c_1 s_2\,y + s_1 s_2\,\sfrac{z\,y}{\rho} + \im s_1 c_2\,\sfrac{z\,x}{\rho} \\[2pt]
c_1 c_2\,y + \im c_1 s_2\,x - s_1 s_2\,\sfrac{z\,x}{\rho} + \im s_1 c_2\,\sfrac{z\,y}{\rho} \\[2pt]
c_1\,z\ -\ \im s_1\,\rho 
\end{pmatrix}
\eeq
\beq
\with\quad c_i=\cosh(\eps_i)\ ,\quad s_i=\sinh(\eps_i) \und \rho=\sqrt{x^2+y^2}\ .
\eeq
For $\eps_1=0$ we come back to the linear complex boost in the $xy$~plane.
The ${\cal P}$ involution is chosen as the outer automorphism
\beq
{\cal P} : \ (\th,\phi)\mapsto(-\th,-\phi) \qquad\Leftrightarrow\qquad (x,y,z)\mapsto(x,-y,z)\ ,
\eeq
and it is easy to see that the deformed Hamiltonian $H(\eps)$ is ${\cal PT}$ symmetric.
We should note, however, that the above deformation modifies the kinetic term,
\beq
\begin{aligned}
L^2_\eps &\= -\pa_\th^2 
- \frac{c_1\cos(\th)-\im s_1\sin(\th)}{c_1\sin(\th)+\im s_1\cos(\th)}\,\pa_\th
- \frac{1}{(c_1\sin(\th)+\im s_1\cos(\th))^2}\,\pa^2_\phi \\[4pt]
&\= -\pa_\th^2  
+ \frac{\sin(2\th)-\im\sinh(2\eps_1)}{\cos(2\th)-\cosh(2\eps_1)}\,\pa_\th
- 2\frac{1-\cosh(2\eps_1)\cos(2\th)-\im\sinh(2\eps_1)\sin(2\th)}{(\cos(2\th)-\cosh(2\eps_1))^2}\,\pa_\phi^2\ .
\end{aligned}
\eeq

Because with this deformation the Vandermonde is nowhere vanishing,
\beq
\begin{aligned}
\De_\eps \ \sim\ & r^6\,\sin^2(\th{+}\im\eps_1)\,\cos^4(\th{+}\im\eps_1)\,\cos^2(2\phi{+}2\im\eps_2) \\
&\times\,\bigl( \tan^2(\th{+}\im\eps_1)\cos^2(\phi{+}\im\eps_2)-1 \bigr)
\bigl( \tan^2(\th{+}\im\eps_1)\sin^2(\phi{+}\im\eps_2)-1 \bigr)\ ,
\end{aligned}
\eeq
all state spaces at $g<0$ become physical, and so we should combine the tower for any $g{>}\frac12$ 
with the one at $1{-}g$.
In contrast to the $A_2$ model, both branches for $1{-}g<0$ contribute, 
and for positive integral~$g$ one finds (demanding $S_4$ invariance) that
\beq
\textrm{deg}(E_\ell) \= 
\begin{cases}
\ph g{-}1 \ +\ \begin{cases}
0 &\textrm{for}\quad q+6g=0,3,4,7,8,11\ \textrm{mod} \ 12 \\
1 &\textrm{for}\quad q+6g=1,2,5,6,9,10\ \textrm{mod} \ 12
\end{cases}
\Biggr\} &\textrm{if}\quad q<6g{-}6 \\[4pt]
\ph \Bigl\lfloor\frac{q}{6}\Bigr\rfloor \ \ + \ \begin{cases}
0 &\textrm{for}\quad q=1,2,5\ \textrm{mod} \ 6 \\
1 &\textrm{for}\quad q=0,3,4\ \textrm{mod} \ 6
\end{cases}
\Biggr\} &\textrm{if}\quad q\ge6g{-}6
\end{cases}
\eeq
which demonstrates the doubling (for large enough energy) compared to~(\ref{A3deg}).
\begin{figure}[h!]
\centering
\includegraphics[scale=0.6]{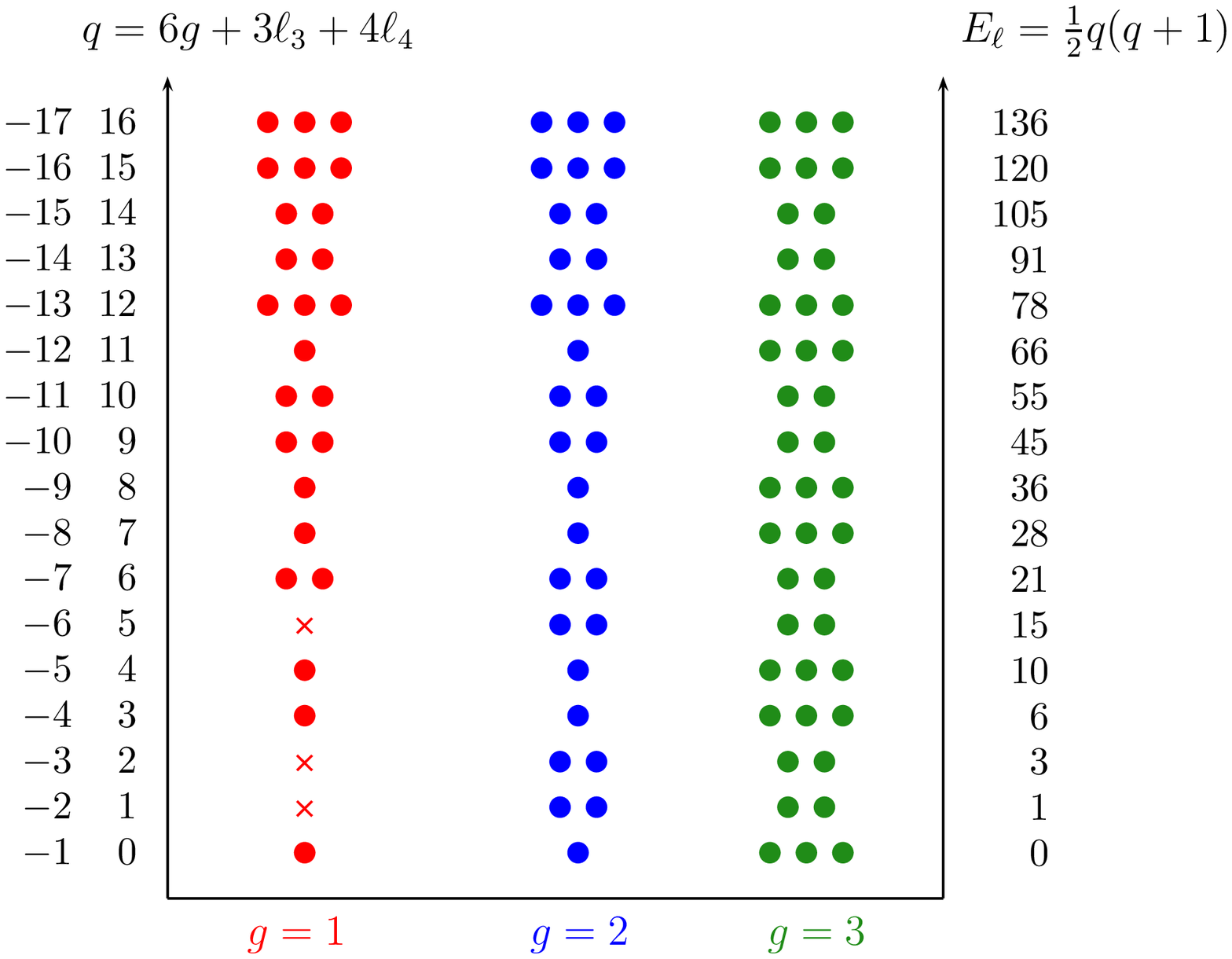}
\caption{Joint spectrum of $H^{(g)}_\eps$ and $H^{(1-g)}_\eps$ after the ${\cal PT}$ 
deformation in the $AD_3$ model.}
\label{fig7}
\end{figure}

In this situation we encounter additional `odd' conserved charges (suppressing~$\eps$)
\beq
Q^{(g)}_{\{s\}} \= 
M_{s_{g-1}}^{(g-1)} M_{s_{g-2}}^{(g-2)} \cdots M_{s_{2-g}}^{(2-g)} M_{s_{1-g}}^{(1-g)}
\quad\with \{s\}=\{s_i\} \and s_i\in\{3,6\}\ .
\eeq
They square to polynomials in the even charges $C_2$, $C_4$ and $C_6$, e.g.
\begin{align}\notag
\bigl(Q^{(2)}_{333}\bigr)^2 \ \propto\ & \ 
8 C_6^3-36 C_2 C_4 C_6^2+12 C_2^3 C_6^2+54 C_2^2 C_4^2 C_6
-36 C_2^4 C_4 C_6+6 C_2^6 C_6 \\[2pt]
&-27 C_2^3 C_4^3+27 C_2^5 C_4^2-9 C_2^7 C_4+ C_2^9
+\text{lower-order terms}\ .
\end{align}

\section{$BC_3$ model} 

\noindent
To understand the non-simply-laced situation at rank-three,
we study the model based on the $BC_3$ Coxeter system. 
It is obtained by extending the $AD_3$~root system to
\beq
{\cal R}_+ \= \bigl\{ 
e_x{+}e_y\,,\,e_x{-}e_y\,,\,e_x{+}e_z\,,\,e_x{-}e_z\,,\,e_y{+}e_z\,,\,e_y{-}e_z\,,\,
e_x\,,\,e_y\,,\,e_z \bigr\}\ ,
\eeq
yielding the potential
\beq
U \= 2\,g_L(g_L{-}1)\,r^2 \biggl(
\frac{x^2+y^2}{(x^2-y^2)^2} +
\frac{y^2+z^2}{(y^2-z^2)^2} +
\frac{z^2+x^2}{(z^2-x^2)^2} \biggr) 
+ \sfrac12 g_S(g_S{-}1)\,r^2 \biggl(
\frac1{x^2} + \frac1{y^2} + \frac1{z^2} \biggr) \ .
\eeq
The Coxeter group $W=S_4\ltimes\Z_2$ enlarges the previous $S_4$ by reflections on the basic coordinate planes,
and it may be generated by
\beq
s_{x-y}= \Bigl(\begin{smallmatrix} 0 & \ph 1 & \ph 0 \\ 1 & \ph 0 & \ph 0 \\ 0 & \ph 0 & \ph 1 \end{smallmatrix}\Bigr) \ ,\
s_{y-z}= \Bigl(\begin{smallmatrix} 1 & \ph 0 & \ph 0 \\ 0 & \ph 0 & \ph 1 \\ 0 & \ph 1 & \ph0 \end{smallmatrix}\Bigr) \ ,\
s_{z}= \Bigl(\begin{smallmatrix} 1 & \ph 0 & \ph 0 \\ 0 & \ph 1 & \ph 0 \\ 0 & \ph 0 & -1 \end{smallmatrix}\Bigr)\ .
\eeq
The basic invariant polynomials are\footnote{
The choice of $\si_3$ is not unique. 
Other options are \ $x^6{+}y^6{+}z^6$ \ or \ $\si_2\si_4-(x^6{+}y^6{+}z^6)$.}
\beq
\si_2=x^2+y^2+z^2=:r^2 \quad,\qquad
\si_3=x^2\,y^2\,z^2\quad,\qquad
\si_4=x^4+y^4+z^4 \quad,
\eeq
which leads to $\ell=6\ell_3+4\ell_4$ and $W$-invariant energy levels
\beq \label{B3energy}
E_\ell\= \half\,q\,(q+1) \quad\with\quad q\=\ell+6g_L+3g_S\=6\ell_3+4\ell_4+6g_L+3g_S
\eeq
and a degeneracy deg$(E_\ell)=0$ when $\ell$ is odd and 
\beq \label{B3deg}
\textrm{deg}(E_\ell) \= \Bigl\lfloor\frac{\ell}{12}\Bigr\rfloor\ +\
\begin{cases} 0 & \for \ \ell=2\ \ \textrm{mod}\ 12 \\
              1 & \for \ \ell=\textrm{else}\ \ \textrm{mod}\ 12 \end{cases}
\eeq
when $\ell$ is even.
Putting $g_S=0$, we are back to the $AD_3$ case, 
but its states with odd $\ell_3$ and thus odd $\ell$ are absent here.
The Vandermonde splits,
\beq
\De=\De_L\De_S \quad\with\quad
\De_L=(x^2-y^2)(y^2-z^2)(z^2-x^2) \and
\De_S=x\,y\,z\ .
\eeq
The Dunkl operators $\widetilde{\cD}_i$ can be obtained from (\ref{D3Dunkl}) by
specifying $g\to g_L$ and adding a term $\frac{g_S}{x^i}(1{-}s_i)$ with the
additional Coxeter reflections
\beq
s_x : (x,y,z)\mapsto (-x,y,z)\ ,\quad 
s_y : (x,y,z)\mapsto (x,-y,z)\ ,\quad 
s_z : (x,y,z)\mapsto (x,y,-z)
\eeq
complementing (\ref{D3reflections}).
For the $W$-invariant energy eigenfunctions 
$v^{(g)}_{\{\ell\}}=r^{-q}\De^g h^{(g)}_{\{\ell\}}(x)$
we must construct the degree-$\ell$ homogeneous polynomials
\beq \label{BC3states}
h^{(g)}_{\{\ell\}}(x,y,z) \= r^{2\ell+12g_L+6g_S+1}\,
\bigl(\widetilde{\cD}_x\widetilde{\cD}_y\widetilde{\cD}_z\bigr)^{2\ell_3}\,
\bigl(\widetilde{\cD}_x^4{+}\widetilde{\cD}_y^4{+}\widetilde{\cD}_z^4\bigr)^{\ell_4}\,
r^{-1-12g_L-6g_S}\ .
\eeq
Comparing with the $AD_3$ case, apart from the extended Dunkl operators this formula 
is very similar to~(\ref{AD3states}), but all odd-$\ell$ states have disappeared.  
The following tables show the states and degeneracy at small values of the energy 
for a few values of $g_S$ and $g_L$, where again a $*$ denotes the $q{<}0$ states. 
We see that the latter appear even when only one of the couplings is negative.
Some of the wave functions can be calculated explicity from the table in Appendix~A.4. 
{\small
\beq\notag
\begin{array}{||c|c|c||c|c|c||c|c|c||}
\hline
g_S{=}{-}1  & \! \text{deg} \!\! & (\ell_3,\ell_4) & 
g_S{=}{-}1 &  \! \text{deg} \!\! &  \!\! (\ell_3,\ell_4) \!\! &
g_S{=}{-}1 &  \! \text{deg} \!\! & \!\! (\ell_3,\ell_4) \!\!   \\ 
g_L{=}{-}2  &  &   &
g_L{=}{-}1 &   &    &
g_L{=}0 &    & \\  \hline \vphantom{\Big|}
E=0 & 1 &  (1,2)^* &
E=0 & 1 &  \,\,(0,2)^* &
E=0 & 0 & \\ \hline\vphantom{\Big|}
E=1 & 2 &  (2,1), (0,4)  &
E=1 & 1 & (1,1)   &
E=1 & 1 & (0,1) \\ \hline\vphantom{\Big|}
E=3 & 2 &  (2,0)^*,(0,3)^* &
E=3 & 1 &  \,\,(1,0)^* &
E=3 & 1 &    \,\,(0,0)^* \\ \hline\vphantom{\Big|}
E=6 & 2 & (3,0),(1,3)   &
E=6 & 2 & (2,0),(0,3)  &
E=6 & 1 & (1,0)  \\ \hline\vphantom{\Big|}
E{=}10 & 1 &   (1,1)^* &
E{=}10 & 1 &  \,\,(0,1)^* &
E=10  & 0 &  \\ \hline
\end{array}
\eeq
}
{\small
\beq\notag
\begin{array}{||c|c|c||c|c|c||c|c|c||}
\hline
g_L{=}{-}1  & \! \text{deg} \!\! & (\ell_3,\ell_4) & 
g_L{=}{-}1 &  \! \text{deg} \!\! &  \!\! (\ell_3,\ell_4) \!\! & 
g_L{=}{-}1 &  \! \text{deg} \!\! & \!\! (\ell_3,\ell_4) \!\!   \\ 
g_S{=}{-}2  & & &  
g_S{=}{-}1 &   &  &  
g_S{=}0 &   &  \\  \hline \vphantom{\Big|}
E=0 & 2 &  (2,0),(0,3)   &
E=0 & 1 &   \,\, (0,2)^* &
E=0 & 1 & (1,0) \\ \hline\vphantom{\Big|}
E=1 & 1 &   (1,1)^* &
E=1 & 1 & (1,1) &  
E=1 & 1 &  \,\,(0,1)^* \\ \hline\vphantom{\Big|}
E=3 & 1 &   (1,2)^* &
E=3 & 1 &   \,\,(1,0)^* &
E=3 & 1 &  (0,2)   \\ \hline\vphantom{\Big|}
E=6 & 1 &  (0,2)^* &
E=6 & 2 & (2,0),(0,3)  &
E=6 & 0  & \\ \hline\vphantom{\Big|}
E{=}10 & 2 & (2,1),(0,4)   &
E{=}10 & 1 &   \,\,(0,1)^* &
E=10  & 1 &  (1,1) \\ \hline
\end{array}
\eeq
}

The Dunkl-deformed angular momenta 
\beq
\cL_i\=\eps_{ijk}x^j\cD_k \quad\with\quad
\cD_i=\widetilde{\cD}_i-g_L\,\pa_i\ln\De_L-g_S\,\pa_i\ln\De_S
\eeq
do not differ much from those of the $AD_3$ model. The Coxeter reflections permute
them and can flip the sign of any number of them. Therefore, the Weyl invariant polynomials
in $\{\cL_x,\cL_y,\cL_z\}$ are the same as in the $AD_3$ case, generated by
$\{C_0,C_2,C_4,C_6\}$, and the conserved charges agree with the previous ones,
except that the constituting Dunkl operators have been extended by the short-root terms.
What about Weyl antiinvariants, corresponding to cases A, B or C in Section~2?
Unfortunately, because 
\beq \label{triplereflection}
s_x s_y s_z :\ \bigl(\cL_x,\cL_y,\cL_z\bigr)\ \mapsto\ \bigl(\cL_x,\cL_y,\cL_z\bigr)\ ,
\eeq
there do not exist $\cL_i$ polynomials which are antiinvariant under the short-root reflections.
Besides, an intertwiner shifting $g_S$ by unity would connect states with an even value of~$q$
to states with an odd one, which is incompatible with~(\ref{B3energy}).
Therefore, besides case~D (the invariants) we can only realize case~B, 
which copies the $AD_3$ intertwining situation.
As a result, the two basic $AD_3$ intertwiners $M_3$ and $M_6$, based on~(\ref{AD3intertwiners})
with the $\cL_i$ pertaining to the $BC_3$ system, will obey the relations
\beq
\begin{aligned}
M_s^{(g_L,g_S)}\,H^{(g_L,g_S)} &\= H^{(g_L+1,g_S)}\,M_s^{(g_L,g_S)} \ , \\[4pt]
M_s^{(1-g_L,g_S)}\,H^{(g_L,g_S)} &\= H^{(g_L-1,g_S)}\,M_s^{(1-g_L,g_S)} 
\end{aligned}
\eeq
but do not shift the $g_S$ value. Therefore, iterating the $\widetilde{M}_s$ action, 
we can produce the polynomials $h^{(g_L,g_S)}_{\{\ell\}}$ from $h^{(0,g_S)}_{\{\ell'\}}$.

The discussion of ${\cal PT}$ deformations may be completely borrowed from the previous section.
The additional $\textrm{rank}(P_-){=}1$ option of ${\cal P}=s_x$ does not produce anything new.
Under the nonlinear deformation~(\ref{nonlinearA3}), again the Vandermonde loses its zeros,
and the negative-$g$ state spaces become physical. 
So for positive integral values of $g_L$ and $g_S$, we must combine two state towers at
\beq \label{towers}
(g_L,g_S)\ \&\ (1{-}g_L,g_S) \qquad\textrm{as well as}\qquad
(g_L,1{-}g_S)\ \&\ (1{-}g_L,1{-}g_S) \ ,
\eeq
where one pair has states only at even~$q$ and the other pair only at odd~$q$.
For $q\ge6(g_L{-}1)+3(g_S{-}1)$, the irregularities due to missing low-energy states disappear,
and the degeneracy grows approximately like $\frac{\ell}{6}$ both for even and odd $q$~values.
For $g_L\in\Z$ there appear `odd' conserved charges $Q^{(g_L,g_S)}_{\{s\}}$ mapping
$(1{-}g_L,g_S)\mapsto(g_L,g_S)$. They are formally identical to those of the $AD_3$ model. 
Analogous odd operators connecting the states at $1{-}g_S$ and $g_S$ do not exist
since the two pairs of towers have disjoint spectra.
\begin{figure}[h!]
\centering
\includegraphics[scale=0.6]{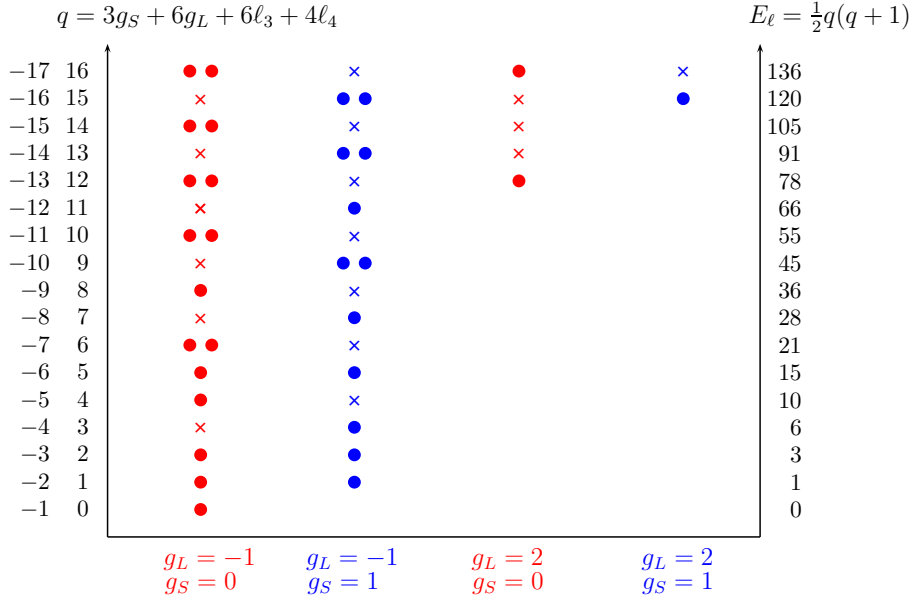}
\caption{Spectrum for $(g_L,g_S)=(2,1)$ comprising four towers for the ${\cal PT}$-extended
$BC_3$ model. The blue and red towers carry odd and even integer values of $q$, respectively.}
\label{fig8}
\end{figure}
\begin{figure}[h!]
\centering
\includegraphics[scale=0.7]{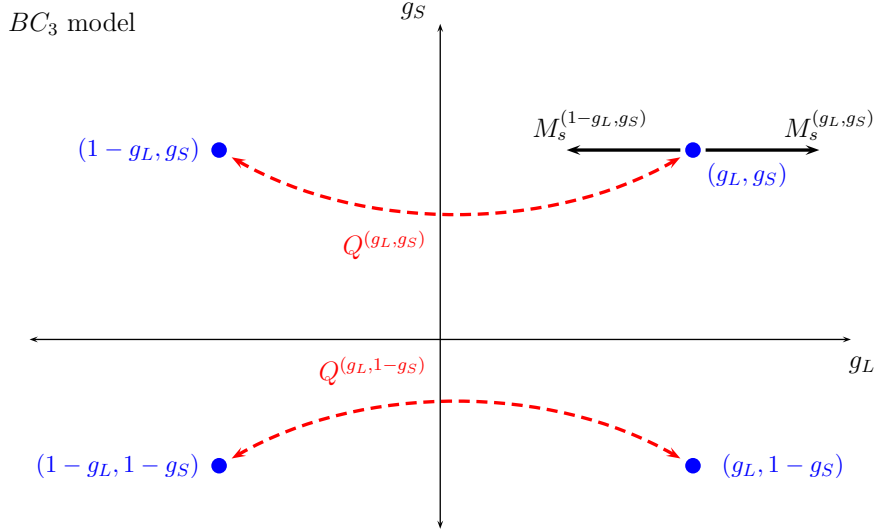}
\caption{Action of the intertwining operators and `odd' conserved charges in the $BC_3$ model.}
\label{fig9}
\end{figure}

\section{$A_1^{\oplus 3}$ model}

\noindent
The previous section reduced the $AD_3$ system to the $A_1^{\oplus 3}$ system of short roots, 
\beq
{\cal R}_+ \= \bigl\{ e_x\,,\,e_y\,,\,e_z \bigr\}\ .
\eeq
When the radial excitations are included, this model is reducible and decomposes 
into three copies of the rank-one system with inverse-square potential and coinciding couplings $g_s=g$.
However, the spherical reduction couples the three subsystems to a potential
\beq
U \= \sfrac12 g(g{-}1)\,r^2 \biggl( \frac1{x^2} + \frac1{y^2} + \frac1{z^2} \biggr)
\= \sfrac12 g(g{-}1)\,\biggl( 3 + \frac{x^2{+}y^2}{z^2} + \frac{y^2{+}z^2}{x^2} + \frac{z^2{+}x^2}{y^2} \biggr)\ .
\eeq
The Coxeter group $W=\Z_2^3$ consists merely of the 3 reflections about the elementary coordinate planes,
\beq
s_x:\ (x,y,z)\mapsto(-x,y,z)\ ,\
s_y:\ (x,y,z)\mapsto(x,-y,z)\ ,\
s_z:\ (x,y,z)\mapsto(x,y,-z)\ ,
\eeq
and the basic invariant polynomials can be taken as\footnote{
The choice of $\si_3$ and $\si_4$ is ambiguous; 
other possibilities are $\si_3=x^2{-}y^2$ and $\si_4=x^2{+}y^2{-}2 z^2 $
or $\si_3=x^2$ and $\si_4=x^2{+}y^2$.}
\beq
\si_2 = x^2+y^2+z^2=:r^2 \quad,\qquad
\si_3 = x^2 \quad,\qquad
\si_4 = y^2\quad,
\eeq
and thus 
\beq
E_\ell\= \half\,q\,(q+1) \quad\with\quad q\=\ell+3g\=2(\ell_3{+}\ell_4)+3g
\eeq
for the $W$-invariant states, with a degeneracy
\beq
\textrm{deg}(E_\ell) \= \sfrac{\ell}{2}+1\ .
\eeq
This is consistent with the fact that only the spherical-harmonic combinations
\beq
Y_{\ell,0} \and Y_{\ell,m}+Y_{\ell,-m} \quad\for \ell,m=0,2,4,\ldots
\eeq
are Weyl invariant.
The Vandermonde is simply $\De=x\,y\,z$, and the potential-free wave functions arise from
\beq \label{A13states}
h^{(g)}_{\{\ell\}}(x,y,z) \= r^{2\ell+6g+1}\,
\widetilde{\cD}_x^{2\ell_3}\,\widetilde{\cD}_y^{2\ell_4}\,r^{-1-6g}
\eeq
with
\beq
\widetilde{\cD}_x \= \pa_x + \frac{g}{x}(1{-}s_x) \quad,\qquad
\widetilde{\cD}_y \= \pa_y + \frac{g}{y}(1{-}s_y) \quad,\qquad
\widetilde{\cD}_z \= \pa_z + \frac{g}{z}(1{-}s_z) \quad.
\eeq
With the above choice of symmetric polynomials we could find the following formul\ae\ 
for the states with $\ell_4=0$,
\begin{align}\notag
h_{(\ell_3,0)}^{(g)}(x,y,z)&\=
\sum_{i=0}^{\ell_3}\frac{ 2^{\ell_3} (-1)^{-i{+}\ell_3} \Gamma (\ell_3{+}1) 
\Gamma(g{+}\ell_3{+}\frac{1}{2}) \Gamma (2 g{+}\ell_3{+}1)}
{\Gamma (i{+}1) \Gamma (2 g{+}i{+}1) \Gamma (-i{+}\ell_3{+}1) 
\Gamma(g{-}i{+}\ell_3{+}\frac{1}{2})}\,x^{2 (\ell_3-i)} \left(y^2{+}z^2\right)^i \\
&\=x^{2 \ell_3} \, _2F_1\bigl(-\ell_3,-g{-}\ell_3{+}\sfrac{1}{2};2 g{+}1;-\sfrac{y^2+z^2}{x^2}\bigr)\ ,
\end{align}
and due to the symmetry $\ell_3\leftrightarrow\ell_4$ plus $x\leftrightarrow y$
we can obtain the $\ell_3=0$ states,
\beq
h_{(0,\ell_4)}^{(g)}(x,y,z)\=h_{(\ell_4,0)}^{(g)}(y,x,z) \ .
\eeq
Below we present the low-lying degeneracies and quantum numbers at $g\ge-2$. 
Their explicit form can be found in Appendix~A.5, where without loss of generality
we restrict to $\ell_3\ge\ell_4$.
{\tiny
\beq\notag
\begin{array}{||c|c|c||c|c|c||c|c|c|c|c||}
\hline
g{=}{-}2  & \!\! \text{deg} \!\! & (\ell_3,\ell_4) &  
g{=}{-}1 & \!\! \text{deg} \!\! & (\ell_3,\ell_4) &
g\geq0  & \!\! \text{deg} \!\! & (\ell_3,\ell_4) \\ \hline \hline \vphantom{\Big|}
E=0 & 4 & (3,0),(2,1),(1,2),(0,3) &
E=0 & 2 & (1,0)^*,(0,1)^* &
E=\frac{1}{2}3g(3g{+}1)\qquad{} & 1 & (0,0) \\ \hline\vphantom{\Big|}
E=1 & 3 & (2,0)^*,(1,1)^*,(0,2)^* &
E=1 & 3 & (2,0),(1,1),(0,2) &
E=\frac{1}{2}(3g{+}2)(3g{+}3) & 2 & (1,0),(0,1) \\ \hline\vphantom{\Big|}
E=3 & 5 & (4,0),(3,1),\ldots,(1,3),(0,4) &
E=3 & 1 & (0,0)^* &
E=\frac{1}{2}(3g{+}4)(3g{+}5)& 3 & (2,0),(1,1),(0,2)  \\ \hline\vphantom{\Big|}
E=6 & 2 & (1,0)^*,(0,1)^* &
E=6 & 4 & (3,0),(2,1),(1,2),(0,3) &
E=\frac{1}{2}(3g{+}6)(3g{+}7)& 4 & (3,0),(2,1),(1,2),(0,3)  \\ \hline\vphantom{\Big|}
E{=}10 & 6 & \!\! (5,0),(4,1),\ldots,(1,4),(0,5)\!\!  &
E{=}10 & 0 &  &
E=\frac{1}{2}(3g{+}8)(3g{+}9) & 5 & \!\! (4,0),(3,1),\ldots,(1,3),(0,4) \!\! \\ \hline\vphantom{\Big|}
E{=}15 & 1 & (0,0)^* &
E{=}15 & 5 & \!\! (4,0),(3,1),\ldots,(1,3),(0,4) \!\! &
\!\! E=\frac{1}{2}(3g{+}10)(3g{+}11) \!\! & 6 & \!\! (5,0),(4,1),\ldots,(1,4),(0,5) \!\! \\ \hline 
\end{array}
\eeq
}

The Dunklized angular momenta have the simple form
\beq
\cL_x = y\pa_z{-}z\pa_y-g\bigl(\sfrac{y}{z}s_z{-}\sfrac{z}{y}s_y\bigr) \ ,\
\cL_y = z\pa_x{-}z\pa_z-g\bigl(\sfrac{z}{x}s_x{-}\sfrac{x}{z}s_z\bigr) \ ,\
\cL_z = x\pa_y{-}z\pa_x-g\bigl(\sfrac{x}{y}s_y{-}\sfrac{y}{x}s_x\bigr) \ ,
\eeq
and any word in $\cL_i^2$ and $\cL_x\cL_y\cL_z$ (and permutations) will restrict
to a conserved quantity. 
As was argued in the previous section, 
there exist neither Weyl antiinvariant polynomials in $\cL_i$ nor intertwiners
shifting $g$ by unity.\footnote{
The intertwiners proposed in~\cite{Feigin03} are not $W$ invariant.}
As a consequence, an `odd' conserved charge for integral~$g$ cannot be constructed in this way.

A linear ${\cal PT}$ deformation of the type~(\ref{xyboost}) (but with a non-coordinate plane) 
still leaves three pairs of singular points in the potential $U_\eps$, while the nonlinear
deformation~(\ref{nonlinearA3}) yields the fully regularized potential
\beq
U_{\eps_1,\eps_2} \= \sfrac12 g(g{-}1) 
\biggl( \frac{4}{\sin^2(\theta{+}\im\eps_2)\sin^2(2\phi{+}2\im\eps_2)} + \frac{1}{\cos^2(\theta{+}\im\eps_1)} \biggr)\ .
\eeq
This revives the negative-$g$ state spaces and lets us combine the towers at $1{-}g$ and $g$.
The result is a linearly (with $q$) growing $W$-invariant spectrum both for even and odd values of~$q$,
\beq
\textrm{deg}(E_\ell) \= \begin{cases}
\ph\sfrac12(q-3g+2) &\for\quad q{+}g\ \textrm{even} \\[4pt]
\ph\sfrac12(q+3g-1) &\for\quad q{+}g\ \textrm{odd}
\end{cases} \Biggl\} \qquad\textrm{when}\quad q\ge3(g{-}1)\ .
\eeq
\begin{figure}[h!]
\centering
\includegraphics[scale=0.6]{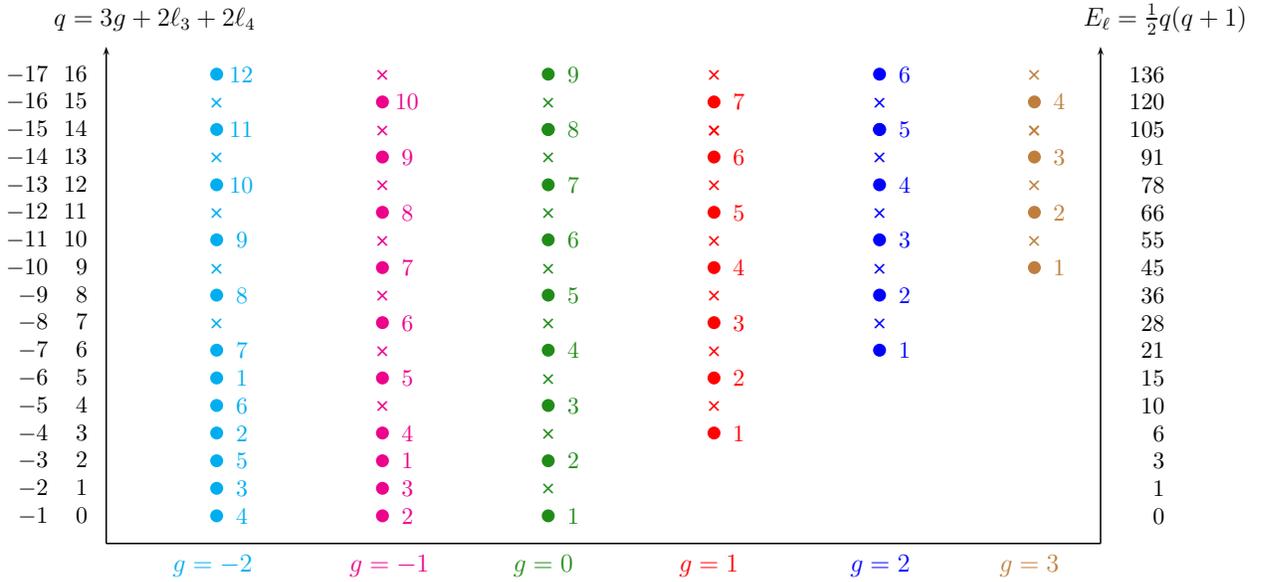}
\caption{Low-lying energy spectrum for the $A_1^{\oplus3}$ model. 
The levels are labeled with their degeneracy. 
States at $g{<}0$ become physical only under a ${\cal PT}$ deformation,
which adds them to the tower at $1{-}g$.}
\label{fig10}
\end{figure}

The $A_1^{\oplus3}$ model is the simplest of an infinite reducible series, based on $A_1\oplus I_2(p)$.
We leave it to the reader to work out the details for $p>2$.

\newpage

\section{$H_3$ model}

\begin{figure}[h!]
\centering
\includegraphics[scale=0.5]{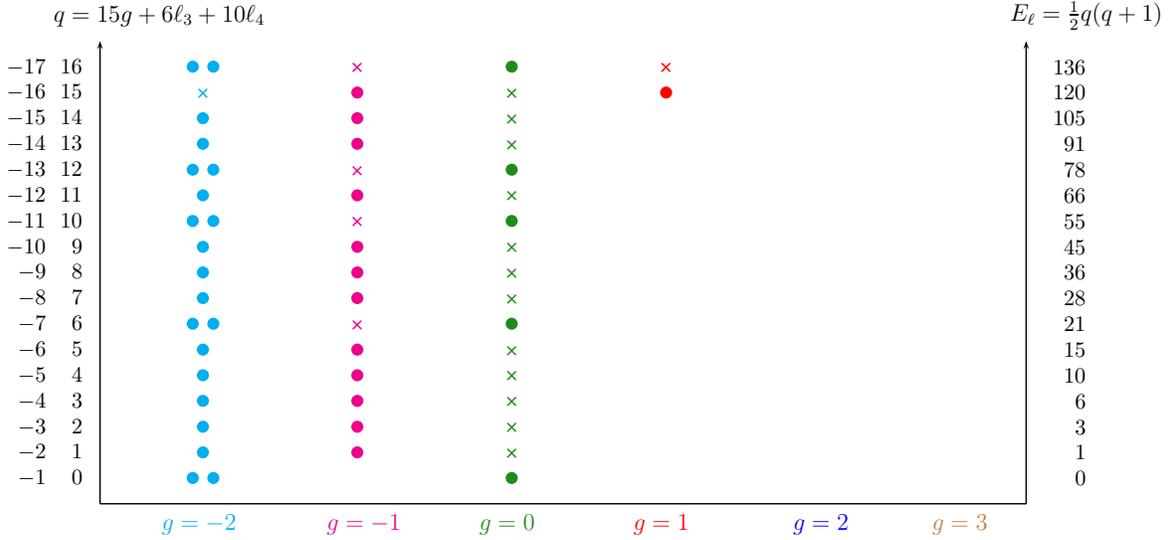}
\caption{Low-lying energy spectrum for the $H_3$ model. 
The towers at $g{=}2$ and $g{=}3$ are invisible 
because their spectrum begins at $E_\ell=465$ and $E_\ell=1035$ respectively.
States at $g{<}0$ become physical only under a ${\cal PT}$ deformation,
which adds them to the tower at $1{-}g$.}
\label{fig11}
\end{figure}

\noindent
Finally, to fully cover the rank-3 landscape, let us turn to the non-crystallographic case of~$H_3$.
Abbreviating the golden ratio and its algebraic conjugate,
\beq
\tau \= \sfrac12(1{+}\sqrt{5}) \und \bar\tau \= \sfrac12(1{-}\sqrt{5})\ ,
\eeq
the set of 15 positive roots,\footnote{
All four sign combinations appear. These roots do not lie in a half-space, but this is irrelevant here.}
\beq
{\cal R}_+ \= \bigl\{ 
e_x{\pm}\tau e_y{\pm}\bar\tau e_z\,,\,e_y{\pm}\tau e_z{\pm}\bar\tau e_x\,,\,e_z{\pm}\tau e_x{\pm}\bar\tau e_y\,,\,
e_x\,,\,e_y\,,\,e_z \bigr\}\ .
\eeq
Accordingly, the potential takes the form
\beq
\begin{aligned}
U &\= \sfrac12 g(g{-}1)\,r^2 \biggl( \frac1{x^2} + \frac1{y^2} + \frac1{z^2} \biggr) \\[4pt]
&\ +\ 2 g(g{-}1)\,r^2 \biggl( 
\frac1{(x{+}\tau y{+}\bar\tau z)^2} + \frac1{(x{+}\tau y{-}\bar\tau z)^2} +
\frac1{(x{-}\tau y{+}\bar\tau z)^2} + \frac1{(x{-}\tau y{-}\bar\tau z)^2} + \textrm{cyclic} \biggr)
\end{aligned}
\eeq
with 15 double poles.
The Coxeter group is the icosahedral group~$I$ of 120 elements, 
and it may be generated by the elements
\beq
\Bigl(\begin{smallmatrix} -1 & \ph 0 & \ph 0 \\ \ph 0 & -1    & \ph 0 \\ \ph 0 & \ph 0 & \ph 1 \end{smallmatrix}\Bigr)\ ,\
\Bigl(\begin{smallmatrix}  0 & \ph 0 & \ph 1 \\     1 & \ph 0 & \ph 0 \\     0 & \ph 1 & \ph 0 \end{smallmatrix}\Bigr)\ ,\
\sfrac12 \Bigl(\begin{smallmatrix} \ph 1 & -\tau & -\bar\tau \\ \ph\tau & -\bar\tau & -1 \\ -\bar\tau & \ph 1 & \tau \end{smallmatrix}\Bigr)\ .
\eeq
We can choose three basic invariant polynomials of degrees~2, 6 and 10, for instance
\beq
\begin{aligned}
\si_2(x,y,z) &\=x^2{+}y^2{+}z^2\ =:\ r^2\ ,\\ 
\si_3(x,y,z) &\=(\tau{-}\bar\tau)(x^6{+}y^6{+}z^6)-15\,\bar{\tau} (x^2 y^4{+}y^2 z^4{+}z^2 x^4)
+30\,x^2 y^2 z^2 \ ,\\
\si_4(x,y,z) &\= x^{10}{+}y^{10}{+}z^{10}+30\,x^2 y^2 z^2 (x^2 y^2{+}y^2 z^2{+}z^2 x^2)
+15 (\tau{+}1) (x^8 y^2{+}y^8 z^2{+}z^8 x^2)\\
&\quad+30 (\tau{+}1) (x^6 y^4{+}y^6 z^4{+}z^6x^4 ) -60\,\tau\,x^2 y^2 z^2 (x^4{+}y^4{+}z^4)\ .
\end{aligned}
\eeq
Hence, $\ell=6\ell_3+10\ell_4$, and the $I$-invariant energy levels are given by
\beq
E_\ell\=\sfrac12\,q\,(q{+}1) \quad\with\quad
q\= \ell+15 g \= 6\ell_3+10\ell_4+15 g 
\eeq
and a degeneracy deg$(E_\ell)=0$ when $\ell$ is odd and 
\beq \label{H3deg}
\textrm{deg}(E_\ell) \= \Bigl\lfloor\frac{\ell}{30}\Bigr\rfloor\ +\
\begin{cases} 0 & \for \ \ell=2,4,8,14\ \ \textrm{mod}\ 30 \\
              1 & \for \ \ell=\textrm{else}\ \ \textrm{mod}\ 30 \end{cases}
\eeq
when $\ell$ is even.
The Vandermonde factor $\De=\De_1 \De_2$ is split in terms of \ $\De_1 = x\,y\,z$ \ and
\beq
\begin{aligned}
\De_{2} &\= \prod_{\epsilon_{1,2}=0,1} \left(x{+}(-1)^{\epsilon_1} \tau  y{+}(-1)^{\epsilon_2} \bar\tau z\right) 
(\tau x{+} (-1)^{\epsilon_1}\bar\tau y{+}(-1)^{\epsilon_2}z) (\bar\tau x{+}  (-1)^{\epsilon_1}  y{+}(-1)^{\epsilon_2}\tau z) \\
 &\= x^{12}-(13{-}\rf)x^{10}y^2-(13{+}\rf)x^{10}z^2+\sfrac12(113{-}11\rf)x^8y^4+\sfrac12(113{+}11\rf)x^8z^4\\
&\quad+50\,x^8y^2z^2-84\,x^6y^6-(90{-}66\rf)x^6y^4z^2-(90{+}66\rf)x^6y^2z^4+126\,x^4y^4z^4\\[2pt]
&\quad+\,\textrm{cyclic permutations}\ .
\end{aligned}
\eeq
The analytical computation of the energy eigenfunctions
$v^{(g)}_{\{\ell\}}=r^{-q}\De^g h^{(g)}_{\{\ell\}}(x)$ in
\beq \label{h3states}
h^{(g)}_{\{\ell\}}(x,y,z) \= r^{2\ell+15g+1}\, 
\sigma_3(\widetilde{\cD}_x,\widetilde{\cD}_y,\widetilde{\cD}_z\bigr)^{\ell_3}\,
\sigma_4(\widetilde{\cD}_x,\widetilde{\cD}_y,\widetilde{\cD}_z\bigr)^{\ell_4}\,
r^{-1-15g}
\eeq
becomes quite more complicated in contrast with the previous cases.  Because of the sum over the 15 positive roots, 
the Dunkl operators yield quite tedious expressions considering also that the invariant polynomials are given 
in terms of powers of differential operators of order 6 and~10. We present here the simplest wave functions of order 6, 
with $(\ell_3,\ell_4)=(1,0)$,
\begin{align}\notag
h^{(g)}_{\{\ell=6\}} &\=(1+2g) {[600]}-3 \tau  (7 \tau +8\bar{\tau}+(30  \tau +32 \bar{\tau})g ){[420]}\\ 
& \quad -3 \bar{\tau}  (7 \bar{\tau} +8{\tau}+(30  \bar{\tau} +32 {\tau})g ){[240]}+2(15 + 62 g) {[222]}\ ,
\end{align}
and of order 10, with $(\ell_3,\ell_4)=(0,1)$,
\begin{align}\notag
h^{(g)}_{\{\ell=10\}} &\=(1+2g)^2 {[10 00]}+8 \left(63+285 g+310 g^2\right){[622]}+10 \left(63+278 g+288 g^2\right){[442]}\\ 
& \quad +\kappa(\tau,\bar\tau){[640]}+\kappa(\bar\tau,\tau){[460]} + \lambda(\tau,\bar\tau) {[820]} + \lambda(\bar\tau,\tau) {[280]}\ ,
\end{align}
where we defined \ ${[rst]}:= x^r y^s z^t+x^ty^rz^s+x^sy^tz^r$ \ and abbreviated
\begin{align}
\kappa(\tau,\bar\tau)  &\= -\sfrac{126}{5} {\tau} +\sfrac{336}{5} \bar{\tau} + (-144 {\tau} + 284 \bar{\tau}) g + (-200 {\tau} + 280 \bar{\tau}) g^2\ ,\\
\lambda(\tau,\bar\tau) &\= -\left(\sfrac{53}{5}{\tau}+\sfrac{162}{5} \bar{\tau} +(72 {\tau}+148 \bar{\tau}) g+(100 {\tau}+160 \bar{\tau}) g^2\right)\ .
\end{align}
It is possible to check that they are symmetric under the simultaneous transposition of variables plus $\tau \rightarrow \bar\tau$, concretely
\beq
(x,y,z,\tau)\rightarrow(y,x,z, \bar\tau)\ ,\
(x,y,z,\tau) \rightarrow(z,y,x,\bar\tau)\ ,\
(x,y,z,\tau) \rightarrow(x,z,y,\bar\tau)\ .
\eeq
The low-lying degeneracies and quantum numbers at $g\ge-2$ are presented in the following table and also illustrated in Fig.~\ref{fig11}. 
{\small
\beq\notag
\begin{array}{||c|c|c||c|c|c||c|c|c||}
\hline
g{=}{-}2  &  \text{deg}  & (\ell_3,\ell_4) & 
g{=}{-}1 &  \text{deg}  & (\ell_3,\ell_4) & 
g\geq0  &  \text{deg}  & (\ell_3,\ell_4) \\ \hline \hline \vphantom{\Big|}
E=0 & 2 &(5,0),(0,3) &  
E=0 & 0 &  & 
E=\frac{1}{2}15g(15g{+}1)\qquad{} & 1 & (0,0) \\ \hline\vphantom{\Big|}
E=1 & 1 & \,\, (3,1)^*& 
E=1 & 1 & (1,1) & 
E=\frac{1}{2}(15g{+}6)(15g{+}7) & 1 & (1,0) \\ \hline\vphantom{\Big|}
E=3 & 1 & (2,2) & 
E=3 & 1 & \,\, (2,0)^* & 
E=\frac{1}{2}(15g{+}10)(15g{+}11)& 1 & (0,1)  \\ \hline\vphantom{\Big|}
E=6 & 1 & \,\, (1,2)^* & 
E=6 & 1 & (3,0) & 
E=\frac{1}{2}(15g{+}12)(15g{+}13)& 1 & (2,0)  \\ \hline\vphantom{\Big|}
E{=}10 & 1 & (4,1)  & 
E{=}10 & 1 & \,\, (0,1)^* & 
E=\frac{1}{2}(15g{+}16)(15g{+}17) & 1 &  (1,1)  \\ \hline\vphantom{\Big|}
E{=}15 & 1 & \,\, (4,0)^* & 
E{=}15 & 1 & (0,2) &  
\!\! E=\frac{1}{2}(15g{+}18)(15g{+}19) \!\! & 1 & (3,0) 
\\ \hline\vphantom{\Big|}
E{=}21 & 2 & (6,0),(1,3) & 
E{=}21 & 0 &  &  
\!\! E=\frac{1}{2}(15g{+}20)(15g{+}21) \!\! & 1 & (0,2)
\\ \hline \hline
\end{array}
\eeq
}

\section{Outlook}

\noindent
We have investigated the ${\cal PT}$ deformation of the angular Calogero model 
firstly in general and secondly in detail for rank-two and rank-three systems. 
Among the different ways to introduce an antilinear symmetry like ${\cal PT}$, 
nonlinear complex deformations of the coordinates seem to be more effective 
for removing the singularities of the potential than linear ones. 
As a result of such a `${\cal PT}$  regularization', the energy spectrum gets enlarged
due to the $g\mapsto1{-}g$ invariance of the (potential-frame) Hamiltonian: 
The previously non-normalizable eigenstates at $g{<}0$ become physical and have to be included. 
In non-simply-laced cases this holds separately for the short- and long-root couplings. 
For integer (or half-integer) values of~$g$, the energy levels at $1{-}g$ concide with those at~$g$, 
increasing the degeneracy of the latter. 
In this situation, a suitable product of intertwiners produces conserved charges, 
which act in a regular way thanks to the ${\cal PT}$  regularization.
When $g$ is an integer, these charges represent `square roots' of conserved charges 
defined for any $g$-value, which extends their algebra to a nonlinear $\Z_2$-graded one. 
In the light of our results it is interesting to investigate how the energy spectra get modified 
for ${\cal PT}$-deformed trigonometric, hyperbolic or elliptic Calogero models. 
We plan to address these problems in the future. 

\begin{figure}[h!]
\centering
\includegraphics[scale=0.2]{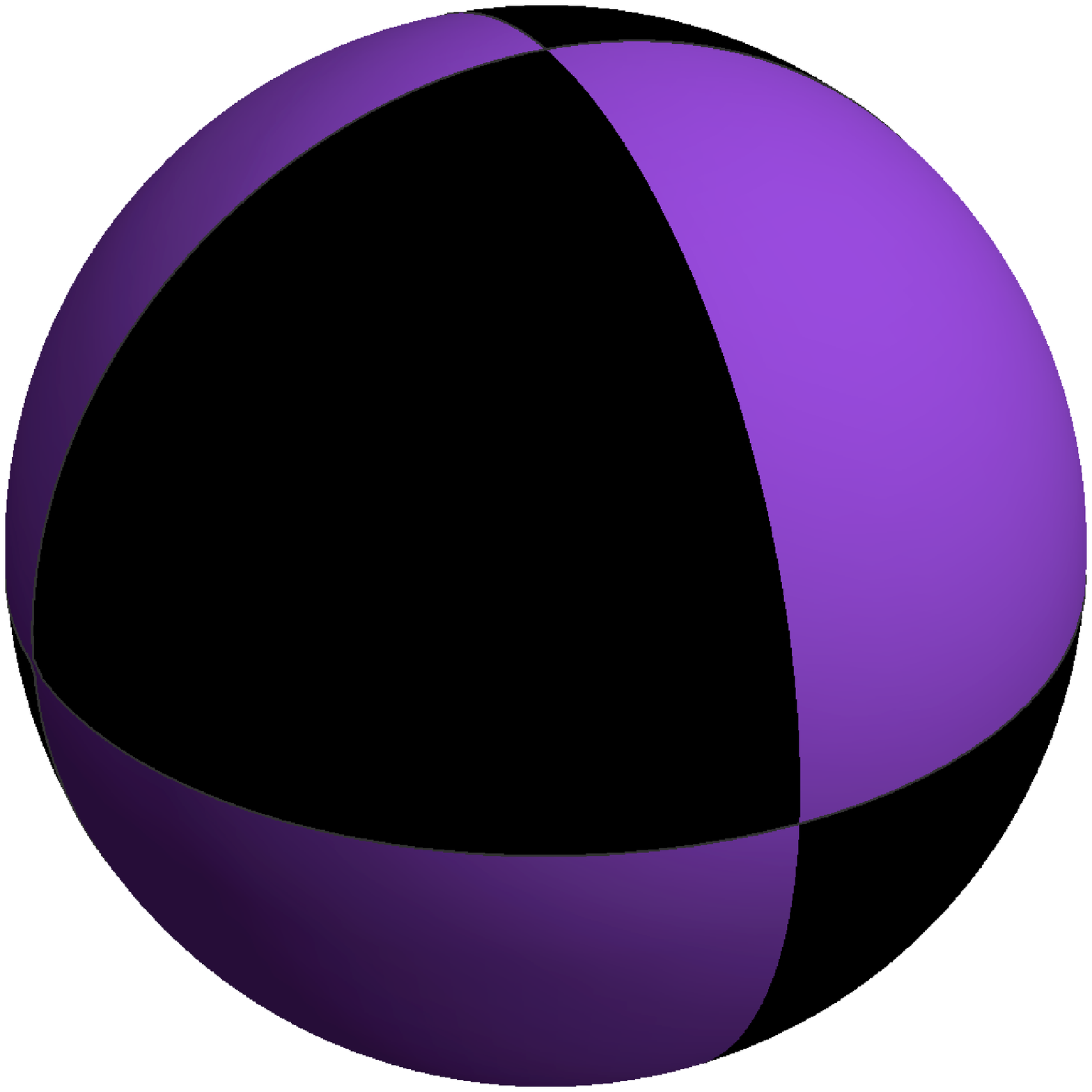}
\qquad
\includegraphics[scale=0.7]{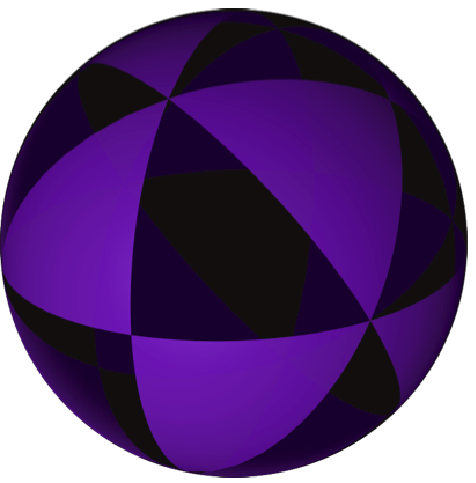}
\qquad
\includegraphics[scale=0.75]{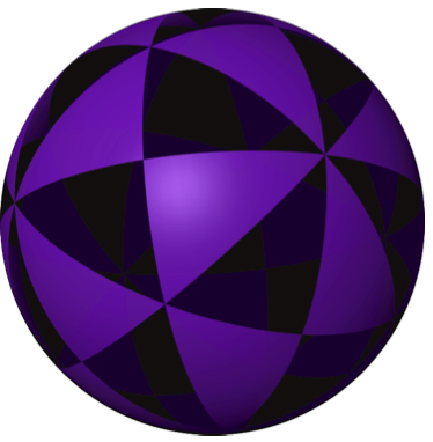}
\qquad
\includegraphics[scale=0.6]{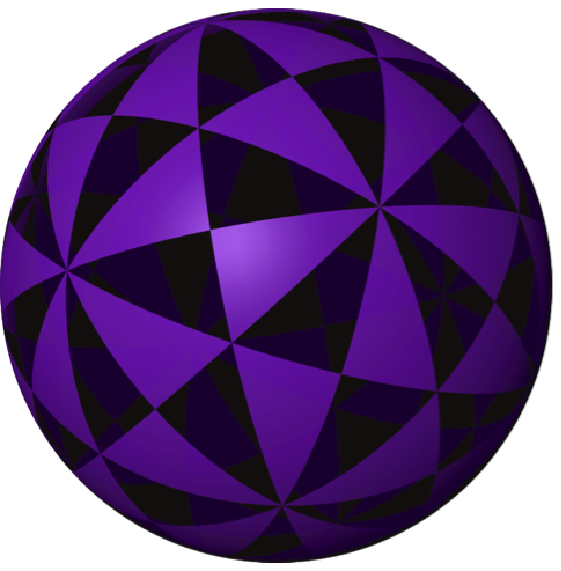}
\caption{We close with the a visualisation of the Coxeter groups~$W$ for the 
$A_1^{\oplus 3}$, $AD_3$, $BC_3$ and $H_3$ models,
given by the Coxeter complexes for three orthogonal lines, the tetrahedron, 
the hexahedron/octahedron and the dodecahedron/icosahedron, respectively.
This illustrates the close relation of irreducible rank-three Coxeter systems
and platonic solids.}
\label{fig12}
\end{figure}


\subsection*{Acknowledgments}

\noindent
This work was partially supported by
the Alexander von Humboldt Foundation under grant CHL~1153844~STP
and by the Deutsche Forschungsgemeinschaft under grant LE 838/12.
This article is based upon work from COST Action MP1405 QSPACE, 
supported by COST (European Cooperation in Science and Technology). 
F.C.~is grateful for the warm hospitality at Leibniz Universit\"at Hannover, 
where the main part of this work was done. 

\bigskip

{\small
\phantom{.}
}

\begin{landscape}

\appendix
\section{Appendix}

\subsection{$A_2$ model}

\renewcommand{\arraystretch}{1.7} 

{\scriptsize
\begin{align}\notag
\begin{array}{|c|c||c|}
\hline
\ell  & \ell_3 & h^{(g)}_\ell  \\ \hline \hline 
0  & 0 &1  \\ \hline 
1  & 3 & w^3-\bar w^3 \\ \hline 
2  & 6 & (g+1) \left(w^6+ \bar w^6\right)-2 g w^3 \bar w^3   \\ \hline 
3 & 9 & (w^3-\bar w^3) \left[(2 + g) \left(w^6+\bar w^6 \right) - 
   2 (g-1) w^3 \bar w^3 \right] \\ \hline
   4 & 12 & (g+2) (g+3)(w^{12}+\bar w^{12})-4 g (g+2)\left( w^9\bar w^3+w^3 \bar w^9 \right)+6 g (1 + g) w^6 \bar w^6 \\ \hline
   5 & 15 &  (w^3-\bar w^3)\left[  (g+3) (g+4)(w^{12}+\bar w^{12})-4 (g-1) (g+3)\left( w^9\bar w^3+w^3 \bar w^9 \right)+2(3g^2 + g+6) w^6 \bar w^6 \right] 
   \\ \hline
   6 & 18 & (g+3)(g+4)(g+5)(w^{18}+\bar w^{18})-6g(g+3)(g+4)\left( w^{15}\bar w^3+w^3 \bar w^{15} \right)+15g(g+1)(g+3)\left( w^{12}\bar w^6+w^6 \bar w^{12} \right) -20 g (g+1)(g+2) w^9 \bar w^9 \\ \hline
      7 & 21 &  (w^3-\bar w^3)\left[ (g+4)(g+5)(g+6)(w^{18}+\bar w^{18})-6(g-1)(g+4)(g+5)\left( w^{15}\bar w^3+w^3 \bar w^{15} \right)+3(g+4)(5g^2-g+10)\left( w^{12}\bar w^6+w^6 \bar w^{12} \right) -4 (g-1) (5g^2+17g+30) w^9 \bar w^9 \right] \\ \hline
\end{array}
\end{align}
}

{\scriptsize
\begin{equation*} 
\begin{tabular}{|c|c|c|c|c|c|}
\hline
$\!q\!$ & 
$h_\ell^{({-2})}$ &
$h_\ell^{({-1})}$ &
$h_\ell^{({0})}$ & 
$h_\ell^{({1})}$ & 
$h_\ell^{({2})} \phantom{\Big|}$ \\ 
\hline
$\!\!\phantom{\Big|}0\!$ & $w^6-4w^3\bar w^3+\bar w^6$
& 
$w^3-\bar w^3$ & $1$ &  &  \\
$\!\!\phantom{\Big|}3\!$ &  $w^3-\bar w^3$
&
$w^3\bar w^3$ & $w^3-\bar w^3$ & $1$ &  \\
$\!\!\phantom{\Big|}6\!$ & $w^6\bar w^6$
&
$w^9+3w^6\bar w-3w\bar w^6-\bar w^9$ & $w^6+\bar w^6$ & $w^3-\bar w^3$ & $1$  \\
$\!\!\phantom{\Big|}9\!$ & $w^{15}+5  w^{12}\bar w^3+10 w^9 \bar w^6 -10 w^6\bar w^9 -5 w^3\bar w^{12} -\bar w^{15}$
&
$\! w^{12}+2w^9\bar w^3+2w^3\bar w^9+\bar w^{12}\!$ & $w^9-\bar w^9$ & $w^6-w^3\bar w^3+\bar w^6$ & $w^3-\bar w^3$ \\
$\!\!\phantom{\Big|}12\!$ & $w^{18}+4  w^{15}\bar w^3+5 w^{12}\bar w^6 +5 w^6\bar w^{12} +4 w^3 \bar w^{15} +\bar w^{18}$
& 
$\!3w^{15}+5w^{12}\bar w^3-5w^3\bar w^{12}-3\bar w^{15}\!$ & $w^{12}+\bar w^{12}$ & $\!w^9-w^6\bar w^3+w^3\bar w^6-w^9\!$ & $\!3w^6-4 w^3\bar w^3+3\bar w^6\!$ \\
$\phantom{\Big|}\vdots$ & $\vdots$ & $\vdots$ & $\vdots$ & $\vdots$ & $\vdots$ \\
\hline
\end{tabular}
\end{equation*}
}

Table 1: \ Low-lying wave functions $v_\ell^{(g)}=r^{-\ell-3g}\De^g h_\ell^{(g)}$ 
of the P\"oschl-Teller model with $E_\ell=\sfrac12 q^2$ and $q=\ell{+}3g$.

\subsection{$G_2$ model}

\renewcommand{\arraystretch}{1.7} 
{\scriptsize
\begin{align}\notag
\begin{array}{|c|c||c|}
\hline
\ell  & \ell_3 & h^{(g_S,g_L)}_{\ell}  \\ \hline \hline 
0  & 0 &1  \\ \hline 
6  & 1 & (1+g_S+g_L)(w^6+\bar w^6)+2(g_L-g_S) (w \bar w)^3 \\ \hline 
12  & 2 & (2+g_S+g_L)(3+g_S+g_L)(w^{12}+\bar w^{12})+4(g_L-g_S)(2+g_S+g_L)(w^{9}\bar w^{3}+w^{3}\bar w^{9})+2(3g_L^2+3 g_L-6 g_L g_S+3 g_S+3g_S^2) (w \bar w)^6 \\ \hline
18 & 3 & (3 + g_S+g_L) (4 + g_S+g_L) (5 + g_S+g_L)(w^{18}+\bar w^{18})+ 6 (g_L-g_S) (3 + g_S+g_L) (4 + g_S+g_L)(w^{15}\bar w^{3}+w^{3}\bar w^{15}) \\
&  & +3(3 + g_S+g_L)(5g_L^2+5 g_L-6 g_L g_S+5 g_S+5g_S^2)(w^{12}\bar w^{6}+w^{6}\bar w^{12})+4(g_L-g_S)(5g_L^2+15 g_L+2g_L g_S+10+15 g_S+5g_S^2)(w \bar w)^9 \\ \hline
24 & 4 & (4 + g_S+g_L) (5 + g_S+g_L)(6 + g_S+g_L) (7 + g_S+g_L)(w^{24}+\bar w^{24})+(g_L-g_S) (4 + g_S+g_L) (5 + g_S+g_L)(6 + g_S+g_L)(w^{21}\bar w^{3}+w^{3}\bar w^{21}) \\
& &+4(4 + g_S+g_L) (5 + g_S+g_L)(7g_L^2+7 g_L-10 g_L g_S+7 g_S+7g_S^2)(w^{18}\bar w^{6}+w^{6}\bar w^{18}) \\ 
& &+8(g_L-g_S)(4 + g_S+g_L)(7g_L^2+21 g_L-2 g_L g_S+14+21 g_S+7g_S^2)(w^{15}\bar w^{9}+w^{9}\bar w^{15}) \\ 
& & +(70 g_S^4+420 g_S^3+770 g_S^2+420 g_S-40 g_S^3 g_L+36 g_S^2 g_L^2-84 g_S^2 g_L-40 g_S g_L^3-84 g_S g_L^2-124 g_S g_L+420 g_L+770 g_L^2+420 g_L^3+70 g_L^4)(w \bar w)^{12} \\ \hline 
\end{array}
\end{align}
}

Table 2: \ Deformed harmonic polynomials for low-lying wave functions of the $G_2$ model at general couplings $g_L$ and $g_S$.   

\renewcommand{\arraystretch}{1.7} 

\subsection{$AD_3$ model}

{\scriptsize
\begin{align}\notag
\begin{array}{|c|c|c||c|c|c|c|c|c|c|c|}
\hline
\ell & \ell_3 &\ell_4 & h^{(g)}_{\ell} \\ \hline \hline
0 & 0 & 0 & \{000\}  \\ \hline
3 & 1 & 0 & \{111\}  \\ \hline
4 & 0 & 1 &(1{+}2 g) \{400\}{-}(3{+}8g)\{220\}  \\ \hline
6 & 2 & 0 &(1{+}2g)\{600\}{-}3(5{+}8 g)\{420\}{+}2(3{+}4g)(5{+}9g)\{222\}   \\ \hline
7 & 1 & 1 & (3{+}2g)\{511\}{-}(5{+}8g)\{331\}  \\ \hline
8 & 0 & 2 & (1{+}2g)(3{+}2g)\{800\}{-}4(7{+}8g)(3{+}2g)\{620\}{+}3(35{+}56g{+}24g^2)\{440\}
+12g(7{+}8g)\{422\}\   \\ \hline
9 & 3 & 0 & 3(3{+}2g)\{711\}{-}9(7{+}8g)\{531\}{+}2(35{+}69g{+}36g^2)\{333\}\ \\ \hline
10 & 2 & 1 & (1{+}2g)(3{+}2g)\{1000\}{-}5(9{+}8g)(3{+}2g)\{820\}{+}2(63{+}149g{+}76 g^2)\{640\} +4 
(3{+}g)(126{+}239 g{+}108 g^2)\{622\}{-}6(315{+}914 g{+}892g^2{+}288 g^3)\{442\}  \\ \hline
11 & 1 & 2 & (3{+}2g)(5{+}2g)\{911\}{-}4(9{+}8g)(5{+}2g)\{731\}{+}9(21{+}24 g{+}8g^2)\{551\}{+}
12g(9{+}8g)\{533\}  \\ \hline
12 & 4 & 0 & 4 (1 + 2 g) (3 + 2 g)^2\{1200\}- 24 (11 + 8 g)  (3 + 2 g)^2\{1020\}+3 (815 + 2244 g + 
1904 g^2 + 512 g^3) \{840\}{+}(4893 + 11868 g + 9296 g^2 + 2368 g^3)\{660\}{+}  \\ & & &
3 (15375 + 40696 g + 38928 g^2 + 15872 g^3 + 2304 g^4)\{822\}{-}6 (35875 + 114060 g + 135440 
g^2 + 70976 g^3+13824 g^4)\{642\}{+} \\ & & & (179375 + 658280 g + 972744 g^2 + 725280 g^3 + 273024 
g^4 +41472 g^5)\{444\}  \\ \hline
12 & 0 & 3 & -32 (1 + 2 g) (3 + 2 g) (707 + 1774 g + 1296 g^2 + 288 g^3)\{1200\}{+}192 (3 + 2 g) 
(11 + 8 g) (707 + 1774 g + 1296 g^2 + 
    288 g^3)\{1020\}{+} \\ & & & 
384 (82425 + 272562 g + 359094 g^2 + 235132 g^3 + 76320 g^4 + 
    9792 g^5)\{840\}{+}   32 (894789 + 2962624 g + 3859432 g^2 + 2483776 g^3 + 793728 g^4 + 
    101376 g^5)\{660\}{+}  \\ & & &
-96 (60795 + 331284 g + 626140 g^2 + 537952 g^3 + 214848 g^4 + 
   32256 g^5)\{822\}{+}192 (141855 + 471268 g + 675384 g^2 + 543616 g^3 + 243072 g^4 + 
   46080 g^5)\{642\}{+} \\& & & 96 (-236425 - 635390 g - 560488 g^2 - 113008 g^3 + 77184 g^4 + 
   29952 g^5)\{444\} \\ \hline
   \ell & \ell_3 &\ell_4 & h^{(-2)}_{\ell} \\ \hline
13 & 3 & 1 & 334 \{5 5 3\} + 176 \{7 3 3\} + 106 \{7 5 1\} + 25 \{9 3 1\} - 
  \{11 1 1\} \\ \hline
  14 & 2 & 2 & 1780 \{6 4 4\} + 880 \{6 6 2\} + 1010 \{8 4 2\} + 95 \{8 6 0\} + 
 64 \{10 2 2\} + 17 \{10 4 0\} - 7 \{12 2 0\} + \{14 0 0\} \\ \hline
15 &  5 & 0 &  5\{5 5 5\} + 18\{7 5 3\} + 3\{7 7 1\} + 3\{9 3 3\} + 
  3\{9 5 1\} \\ \hline
15 & 1 & 3 & 229 \{5 5 5\} + 826 \{7 5 3\} + 101 \{7 7 1\} + 151 \{9 3 3\} + 
 116 \{9 5 1\} + 18 \{11 3 1\} - \{13 1 1\} \\ \hline
   16 & 4 & 1 & 234 \{6 6 4\} + 153 \{8 4 4\} + 157 \{8 6 2\} + 6 \{8 8 0\} + 
 77 \{10 4 2\} + 7 \{10 6 0\} + 5 \{12 2 2\} + \{12 4 0\}  \\ \hline
\end{array}
\end{align}}
Table 3: Low-lying polynomials for the $AD_3$ model. The notation is \ 
$ \{rst\} \ :=\ x^ry^sz^t+x^ry^tz^s+x^sy^tz^r+x^sy^rz^t+x^ty^rz^s+x^ty^sz^r$.

\renewcommand{\arraystretch}{1.7} 

\subsection{$BC_3$ model}

{\tiny
\begin{align}\notag
\begin{array}{|c|c|c||c|c|c|c|c|c|c|c|}
\hline
\ell & \ell_3 &\ell_4 & h^{(g_L,g_S)}_{\ell} \\ \hline \hline
0 & 0 & 0 & \{000\}  \\ \hline
4 & 0 & 1 & (1{+}2 g_L{+}2 g_S) \{4 0 0\} - (3{+}8 g_L{+}2 g_S) \{2 2 0\}  \\ \hline
6 & 1 & 0 & (1{+}2 g_S) (1{+}2 g_L{+}2 g_S)  \{6 0 0\} - 
 3 (1{+}2 g_S) (5{+}8 g_L{+}2 g_S) \{4 2 0\}+ 
 2 (15{+}47 g_L{+}36 g_L^2{+}16 g_S{+}22 g_L g_S{+}4 g_S^2) \{222\}\\ \hline
 8 & 0& 2 & (1{+}2 g_L{+}2 g_S) (3{+}2 g_L{+}2 g_S)  \{800\}- 
 4 (3{+}2 g_L{+}2 g_S) (7{+}8 g_L{+}2 g_S)  \{620\}{+}
 3 (35{+}56 g_L{+}24 g_L^2{+}24 g_S{+}16 g_L g_S{+}4 g_S^2)  \{440\}{+}
 12 g_L (7{+}8 g_L{+}2 g_S)  \{422\}\\ \hline
 10  & 1 & 1 & (1{+}2 g_S) (1{+}2 g_L{+}2 g_S) (3{+}2 g_L{+}2 g_S) \{10 0 0\} - 
  5 (1{+}2 g_S) (3{+}2 g_L{+}2 g_S) (9{+}8 g_L{+}2 g_S) \{820\}{+}
  2 (1{+}2 g_S) (63{+}149 g_L{+}76 g_L^2{+}32 g_S{+}26 g_L g_S{+}4 g_S^2) \{640\}+ \\
  & & &
  4 (378{+}843 g_L{+}563 g_L^2{+}108 g_L^3{+}444 g_S{+}688 g_L g_S{+}
     262 g_L^2 g_S{+}152 g_S^2{+}116 g_L g_S^2{+}16 g_S^3) \{622\}{+}\\
     & & & - 
  6 (315{+}914 g_L{+}892 g_L^2{+}288 g_L^3{+}286 g_S{+}484 g_L g_S{+}
     200 g_L^2 g_S{+}84 g_S^2{+}64 g_L g_S^2{+}8 g_S^3) \{442\} \\ \hline
 12 & 2 & 0 &   
 4 (3{+}2 g_L{+}g_S) (1{+}2 g_S) (3{+}2 g_S) (1{+}2 g_L{+}2 g_S) (3{+}2 g_L{+}
    2 g_S) \{12 0 0\} - 
 24 (3{+}2 g_L{+}g_S) (1{+}2 g_S) (3{+}2 g_S) (3{+}2 g_L{+}2 g_S) (11{+}
    8 g_L{+}2 g_S) \{10 2 0\}+ \\
     & & &
 3 (1{+}2 g_S) (3{+}2 g_S) (815{+}2244 g_L{+}1904 g_L^2{+}512 g_L^3{+}
    354 g_S{+}720 g_L g_S{+}320 g_L^2 g_S{+}4 g_S^2{+}16 g_L g_S^2 - 
    8 g_S^3) \{84 0\}{+}\\
     & & &
 3 (3{+}2 g_S) (15375{+}40696 g_L{+}38928 g_L^2{+}15872 g_L^3{+}
    2304 g_L^4{+}22328 g_S{+}45240 g_L g_S{+}29824 g_L^2 g_S{+}
    6400 g_L^3 g_S{+}11160 g_S^2{+}15136 g_L g_S^2{+}5056 g_L^2 g_S^2{+}
    2336 g_S^3{+}1568 g_L g_S^3{+}176 g_S^4) \{8 2 2\}  \\
     & & &+ (1{+}2 g_S) (3{+}2 g_S) (4893{+}11868 g_L{+}9296 g_L^2{+}
      2368 g_L^3{+}4086 g_S{+}6432 g_L g_S{+}2464 g_L^2 g_S{+}1132 g_S^2{+}
      848 g_L g_S^2{+}104 g_S^3) \{6 6 0\} \\
     & & &- 
 6 (3{+}2 g_S) (35875{+}114060 g_L{+}135440 g_L^2{+}70976 g_L^3{+}
    13824 g_L^4{+}38432 g_S{+}88920 g_L g_S{+}68544 g_L^2 g_S{+}
    17536 g_L^3 g_S{+}15304 g_S^2{+}22864 g_L g_S^2{+}8512 g_L^2 g_S^2{+}
    2688 g_S^3{+}1952 g_L g_S^3{+}176 g_S^4) \{6 4 2\}\\
     & & &+ 
\beta_1 \{ 4 4 4\}
  \\ \hline
12 & 0 & 3 & (1{+}2 g_L{+}2 g_S) (3{+}2 g_L{+}2 g_S) (707{+}1774 g_L{+}1296 g_L^2{+}
     288 g_L^3{+}990 g_S{+}1584 g_L g_S{+}576 g_L^2 g_S{+}468 g_S^2{+}
     360 g_L g_S^2{+}72 g_S^3) \{1200\} +\\  & & & - 
  6 (3{+}2 g_L{+}2 g_S) (11{+}8 g_L{+}2 g_S) (707{+}1774 g_L{+}1296 g_L^2{+}
     288 g_L^3{+}990 g_S{+}1584 g_L g_S{+}576 g_L^2 g_S{+}468 g_S^2{+}
     360 g_L g_S^2{+}72 g_S^3) \{1020\} +\\  & & &+ \alpha_1 \{840\}+
  \alpha_2 \{820\}{+}\alpha_3 \{660\}+\alpha_4 \{642\}{+}\alpha_5 \{444\}  \\ \hline
\end{array}
\end{align}}
Table 4: Low-lying polynomials for the $BC_3$ model. The constants $\beta_1$ and $\alpha_m$ for $m=1,\ldots,5$ are given below.

{\scriptsize
\begin{center}
\begin{align*}\notag
 \beta_1=& 3 (179375 + 658280 g_L + 972744 g_L^2 + 725280 g_L^3 + 273024 g_L^4 + 
    41472 g_L^5 + 263910 g_S + 755080 g_L g_S + 813232 g_L^2 g_S + 
    391936 g_L^3 g_S + 71424 g_L^4 g_S \\&+ 153384 g_S^2 + 322608 g_L g_S^2 + 
    226400 g_L^2 g_S^2 + 53120 g_L^3 g_S^2 + 44048 g_S^3 + 60896 g_L g_S^3 + 
    21056 g_L^2 g_S^3 + 6256 g_S^4 + 4288 g_L g_S^4 + 352 g_S^5) \\ \notag
\alpha_1=&12 (82425 + 272562 g_L + 359094 g_L^2 + 235132 g_L^3 + 
       76320 g_L^4 + 9792 g_L^5 + 140802 g_S + 367196 g_L g_S + 
       358868 g_L^2 g_S + 155808 g_L^3 g_S + 25344 g_L^4 g_S + 
       \\ 
       +& 93944 g_S^2 + 
       180904 g_L g_S^2 + 116136 g_L^2 g_S^2 + 24912 g_L^3 g_S^2 + 
       30384 g_S^3 + 38160 g_L g_S^3 + 11952 g_L^2 g_S^3 + 4752 g_S^4 + 
       2880 g_L g_S^4 + 288 g_S^5)  \\ \notag
   \alpha_2=&    3 (60795 + 331284 g_L + 626140 g_L^2 + 537952 g_L^3 + 214848 g_L^4 + 
     32256 g_L^5 + 123666 g_S + 501240 g_L g_S + 685160 g_L^2 g_S + 
     386496 g_L^3 g_S + 77184 g_L^4 g_S +\\ &  93224 g_S^2 + 270256 g_L g_S^2 + 
     239760 g_L^2 g_S^2 + 66816 g_L^3 g_S^2 + 32112 g_S^3 + 
     60192 g_L g_S^3 + 26208 g_L^2 g_S^3 + 5040 g_S^4 + 4608 g_L g_S^4 + 
     288 g_S^5)\\ \notag
     \alpha_3=& + (-894789 - 2962624 g_L - 3859432 g_L^2 - 2483776 g_L^3 - 
     793728 g_L^4 - 101376 g_L^5 - 1379182 g_S - 3583344 g_L g_S - 
     3421424 g_L^2 g_S - 1425024 g_L^3 g_S - 218880 g_L^4 g_S +\\
     & - 
     828536 g_S^2 - 1581472 g_L g_S^2 - 982368 g_L^2 g_S^2 - 
     198144 g_L^3 g_S^2 - 243792 g_S^3 - 304704 g_L g_S^3 - 
     92736 g_L^2 g_S^3 - 35280 g_S^4 - 21888 g_L g_S^4 - 2016 g_S^5) 
    \\ \notag
    \alpha_4=& - 6 (141855{{{+}}}471268 g_L {+} 675384 g_L^2 {+} 543616 g_L^3 {+} 
     243072 g_L^4 {+} 46080 g_L^5 {+} 234514 g_S {+} 610784 g_L g_S {+} 
     631568 g_L^2 g_S {+} 319104 g_L^3 g_S {+} 66816 g_L^4 g_S \\ &{+} 143624 g_S^2 {+} 
     266656 g_L g_S^2 {+} 170784 g_L^2 g_S^2 {+} 39168 g_L^3 g_S^2 {+} 
     41328 g_S^3 {+} 47232 g_L g_S^3 {+} 13248 g_L^2 g_S^3 {+} 5616 g_S^4 {+} 
     2880 g_L g_S^4 {+} 288 g_S^5) \\ \notag
     \alpha_5=& -3 (-236425 - 635390 g_L - 560488 g_L^2 - 113008 g_L^3 + 
     77184 g_L^4 + 29952 g_L^5 - 327810 g_S - 644352 g_L g_S - 
     347696 g_L^2 g_S + 12672 g_L^3 g_S + 36864 g_L^4 g_S \\ & - 177176 g_S^2 - 
     236560 g_L g_S^2 - 63072 g_L^2 g_S^2 + 10944 g_L^3 g_S^2 - 
     46512 g_S^3 - 36864 g_L g_S^3 - 2880 g_L^2 g_S^3 - 5904 g_S^4 - 
     2016 g_L g_S^4 - 288 g_S^5) 
\end{align*}
\end{center}
}

\renewcommand{\arraystretch}{1.5} 

\subsection{$A_1^{\oplus 3}$ model}

{\scriptsize
\begin{align}\notag
\begin{array}{|c|c|c||c|c|c|c|c|c|c|c|}
\hline
\ell&\ell_3 & \ell_4 & h_\ell^{(g)}  \\ \hline \hline
0 & 0 & 0 & 1  \\ \hline
2 & 1 & 0 & -2 x^2 + y^2 + z^2  \\ \hline
4 & 1 & 1 & 4 (g+1) (2 g+1) x^4-(2 g+3) x^2 \left((10 g+9) y^2-(2 g+1) z^2\right)+(2 
g+1) \left(y^2+z^2\right) \left(4 (g+1) y^2-(2 g+1) z^2\right)  \\ \hline
4 & 2 & 0 & 8 (g+1) (2 g+1) x^4-8 (g+1) (2 g+3) x^2 \left(y^2+z^2\right)+(2 g+1) (2 
g+3) \left(y^2+z^2\right)^2 \\ \hline
6 & 2 & 1& 8 (g+1) (2 g+1) x^6-4 x^4 \left((4 g (3 g+10)+29) y^2+(2 g+1) z^2\right) 
+x^2 \left(y^2+z^2\right) \left((4 g (9 g+31)+101) y^2-(2 g+1) (6 g+11) z^2\right) \\ & &  &-(2 g
+1) \left(y^2+z^2\right)^2 \left((4 g+6) y^2-(2 g+1) z^2\right) \\ \hline
6 & 3 & 0 & -16 (g+1) (2 g+1) x^6+24 (g+1) (2 g+5) x^4 \left(y^2+z^2\right)-6 (2 g
+3) (2 g+5) x^2 \left(y^2+z^2\right)^2+(2 g+1) (2 g+5) \left(y^2+z^2\right)^3 \\ \hline
8 & 3 & 1 &
32 (g+1) (g+2) (2 g+1) x^8-8 (g+2) x^6 \left((4 g (7 g+30)+101) y^2+(2 g+1) (2 g
+11) z^2\right)+\\ & & & 12 (g+2) (2 g+5) x^4 \left(y^2+z^2\right) \left((10 g+29) y^2-(2 g+1) 
z^2\right) -(2 g+5) x^2 \left(y^2+z^2\right)^2 \left((4 g (13 g+58)+247) y^2-(2 g+1) (10 g
+23) z^2\right)\\ & & &+(2 g+1) (2 g+5) \left(y^2+z^2\right)^3 \left(4 (g+2) y^2-(2 g+1) z^2\right)  
\\ \hline
8 & 2 & 2& 8 (g+2) (2 g+1) (2 g+3) x^8-8 (g+2) (2 g+3) x^6 \left((10 g+29) y^2-(2 
g+1) z^2\right) \\ & & &+3 (2 g+5) x^4 \left((4 g (11 g+52)+237) y^4-2 (2 g+3) (2 g+5) y^2 
z^2-(4 g (g+2)+3) z^4\right)\\  & & &-2 (2 g+3) x^2 \left(y^2+z^2\right) \left(4 (g+2) (10 g+29) 
y^4-(4 g (7 g+34)+157) y^2 z^2+(2 g+1) (2 g+7) z^4\right)\\ & & &+(2 g+1) (2 g+3) 
\left(y^2+z^2\right)^2 \left(8 (g+2) y^4-8 (g+2) y^2 z^2+(2 g+1) z^4\right) \\ \hline
8 & 4 & 0& 64 (g+1) (g+2) (2 g+1) x^8-128 (g+1) (g+2) (2 g+7) x^6 
\left(y^2+z^2\right)+48 (g+2) (2 g+5) (2 g+7) x^4 \left(y^2+z^2\right)^2 \\ & & &-16 (g+2) (2 g
+5) (2 g+7) x^2 \left(y^2+z^2\right)^3+(2 g+1) (2 g+5) (2 g+7) \left(y^2+z^2\right)^4 \\ \hline
10 & 4 & 1 &
-64 (g+2) (2 g+3) x^8 \left(\left(8 g^2+42 g+39\right) y^2+(g+6) (2 g+1) z^2\right)+64 
(g+1) (g+2) (2 g+1) (2 g+3) x^{10}+\\ & & &16 (g+2) (2 g+7) x^6 \left(y^2+z^2\right) \left((4 
g (11 g+59)+267) y^2+(3-4 (g-1) g) z^2\right)\\ & & &-8 (2 g+5) (2 g+7) x^4 
\left(y^2+z^2\right)^2 \left((2 g+7) (14 g+27) y^2-(2 g+1) (4 g+9) z^2\right)+\\ & & & (2 g+3) (2 
g+7) x^2 \left(y^2+z^2\right)^3 \left((4 g (17 g+93)+489) y^2-(2 g+1) (14 g+39) z^2\right)\
\&-(2 g+1) (2 g+3) (2 g+7) \left(y^2+z^2\right)^4 \left(2 (2 g+5) y^2-(2 g+1) z^2\right) \\ 
\hline
10 & 3 & 2 &-(2 g+7) x^4 \left(y^2+z^2\right) \left(\left(2 g \left(292 g^2+2138 g
+5103\right)+7947\right) y^4-2 (2 g+3) (4 g (17 g+82)+387) y^2 z^2+(2 g+1) (2 g+3) (2 
g+9) z^4\right)\\ & & &+16 (g+2) (2 g+1) (2 g+3)^2 x^{10}-8 (g+2) (2 g+3) x^8 \left((4 g 
(11 g+59)+267) y^2+(3-4 (g-1) g) z^2\right)+\\ & & & 2 (2 g+7) x^6 \left((2 g (2 g (86 g
+615)+2813)+4161) y^4+2 (2 g+3) (4 g (g+16)+111) y^2 z^2-(2 g+1) (2 g+3) (10 g
+21) z^4\right)\\ & & &+2 (2 g+3) (2 g+7) x^2 \left(y^2+z^2\right)^2 \left(2 (2 g+5) (14 g+27) y^4-
(2 g+3) (22 g+51) y^2 z^2+2 (4 g (g+2)+3) z^4\right)\\ & & &-(4 g (g+2)+3) 
\left(y^2+z^2\right)^3 \left(8 (g+2) (2 g+5) y^4-4 (2 g+3) (2 g+5) y^2 z^2+(4 g (g+2)+3) 
z^4\right) \\ \hline
10 & 5 & 0 &-128 (g+1) (g+2) (2 g+1) (2 g+3) x^{10}+320 (g+1) (g+2) (2 g+3) (2 
g+9) x^8 \left(y^2+z^2\right)\\ & & &-160 (g+2) (2 g+3) (2 g+7) (2 g+9) x^6 
\left(y^2+z^2\right)^2+80 (g+2) (2 g+5) (2 g+7) (2 g+9) x^4 \left(y^2+z^2\right)^3\\ & & &-10 (2 
g+3) (2 g+5) (2 g+7) (2 g+9) x^2 \left(y^2+z^2\right)^4+(2 g+1) (2 g+3) (2 g+7) (2 
g+9) \left(y^2+z^2\right)^5 \\ \hline
\end{array}
\end{align}}
Table 5: Low-lying polynomials for the $A_1^{\oplus 3}$ model at coinciding couplings
for $\ell_3\ge\ell_4$.

\end{landscape}


\bigskip

\begin{thebibliography}{99}

\bibitem{Poly06-rev}
A.P. Polychronakos,\\
{\it Physics and mathematics of Calogero particles},\\
J. Phys. A: Math. Gen. {\bf 39} (2006) 12793 {\tt [arXiv:hep-th/0607033]}.

\bibitem{Feigin03}
M.V. Feigin, \\
{\it Intertwining relations for the spherical parts of generalized Calogero operators},\\
Theor. Math. Phys. {\bf 135} (2003) 497--509.

\bibitem{HNY}
T. Hakobyan, A. Nersessian, V. Yeghikyan,\\
{\it The cuboctahedric Higgs oscillator from the rational Calogero model},\\
J. Phys. A: Math. Theor. {\bf 42} (2009) 205206 {\tt [arXiv:0808.0430[hep-th]]}.

\bibitem{HKLN}
T. Hakobyan, S. Krivonos, O. Lechtenfeld, A. Nersessian,\\
{\it Hidden symmetries of integrable conformal mechanical systems},\\
Phys. Lett. A {\bf 374} (2010) 801--806 {\tt [arXiv:0908.3290[hep-th]]}.

\bibitem{LNY}
O. Lechtenfeld, A. Nersessian, V. Yeghikyan,\\
{\it Action-angle variables for dihedral systems on the circle},\\
Phys. Lett. A {\bf 374} (2010) 4647--4652 {\tt [arXiv:1005.0464[hep-th]]}.

\bibitem{HLNS}
T. Hakobyan, O. Lechtenfeld, A. Nersessian, A. Saghatelian,\\
{\it Invariants of the spherical sector in conformal mechanics},\\
J. Phys. A: Math. Theor. {\bf 44} (2011) 055205 {\tt [arXiv:1008.2912[hep-th]]}.

\bibitem{HLN}
T. Hakobyan, O. Lechtenfeld and A. Nersessian,\\
{\it The spherical sector of the Calogero model as a reduced matrix model},\\
Nucl. Phys. B {\bf 858} (2012) 250--266 {\tt [arXiv:1110.5352[hep-th]]}.

\bibitem{HLNSY}
T. Hakobyan, O. Lechtenfeld, A. Nersessian, A. Saghatelian, V. Yeghikyan,\\
{\it Action-angle variables and novel superintegrable systems},\\
Physics of Particles and Nuclei {\bf 43} (2012) 577--582.

\bibitem{FeLePo13}
M. Feigin, O. Lechtenfeld, A. Polychronakos,\\
{\it The quantum angular Calogero-Moser model},\\
JHEP {\bf 1307} (2013) 162 {\tt [arXiv:1305.5841[math-ph]]}.

\bibitem{CoLePl13}
F. Correa, O. Lechtenfeld, M. Plyushchay,\\
{\it Nonlinear supersymmetry in the quantum Calogero model},\\
JHEP {\bf 1404} (2014) 151 {\tt [arXiv:1312.5749[hep-th]]}.

\bibitem{FeHa14}
M. Feigin, T. Hakobyan,\\
{\it On Dunkl angular momenta algebra},\\
JHEP {\bf 1511} (2015) 107 {\tt [arXiv:1409.2480[math-ph]]}.

\bibitem{CoLe15}
F. Correa, O. Lechtenfeld,\\
{\it The tetrahexahedric angular Calogero model},\\
JHEP {\bf 1510} (2015) 191 {\tt [arXiv:1508.04925[hep-th]]}.

\bibitem{BeBo98}
C.M. Bender, S. Boettcher, \\ 
{\it Real spectra in non-hermitian Hamiltonians having ${\cal PT}$ symmetry},\\
Phys. Rev. Lett. {\bf 80} (1998) 5243--5246 {\tt [arXiv:physics/9712001]}.

\bibitem{FrZn08}
A. Fring, M. Znojil,\\  
{\it ${\cal PT}$-Symmetric deformations of Calogero models},\\
J. Phys. A {\bf 41} (2008) 194010 {\tt [arXiv:0802.0624[hep-th]]}.

\bibitem{fringsmith1} 
A. Fring, M. Smith, \\
{\it Antilinear deformations of Coxeter groups, an application to Calogero models},\\
J. Phys. A {\bf 43} (2010) 325201 {\tt [arXiv:1004.0916[hep-th]]}.
  
\bibitem{fringsmith2} 
A. Fring, M. Smith, \\
{\it ${\cal PT}$ invariant complex $E_8$ root spaces},\\
Int. J. Theor. Phys. {\bf 50} (2011) 974 {\tt [arXiv:1010.2218[math-ph]]}.
  
\bibitem{fringsmith3} 
A. Fring, M. Smith, \\
{\it Non-Hermitian multi-particle systems from complex root spaces},\\
J. Phys. A {\bf 45} (2012) 085203 {\tt [arXiv:1108.1719[hep-th]]}.

\bibitem{Fring12-rev}
A. Fring,\\
{\it ${\cal PT}$-symmetric deformations of integrable models},\\
Phil. Trans. Roy. Soc. Lond. A {\bf 371} (2013) 20120046 {\tt [arXiv:1204.2291[hep-th]]}.

\bibitem{Cal71}
F. Calogero,\\
{\it Solution of the one-dimensional N-body problem with quadratic
      and\slash or inversely quadratic pair potentials},\\
J. Math. Phys. {\bf 12} (1971) 419--436;
Erratum, {\sl ibidem\/} {\bf 37} (1996) 3646.

\bibitem{OlshaPere81-rev}
M.A. Olshanetsky, A.M. Perelomov,\\
{\it Classical integrable finite-dimensional systems related to Lie algebras},\\
Phys. Rept. {\bf 71} (1981) 313--400.

\bibitem{OlshaPere83-rev}
M.A. Olshanetsky, A.M. Perelomov,\\
{\it Quantum integrable systems related to Lie algebras},\\
Phys. Rept. {\bf 94} (1983) 313--404.

\bibitem{Woj}
S. Wojciechowski,\\
{\it Superintegrability of the Calogero--Moser system},\\
Phys. Lett. {\bf 95A} (1983) 279--281.

\bibitem{Dunkl89}
C.F. Dunkl,\\
{\it Differential-difference operators associated to reflection groups},\\
Trans. Amer. Math. Soc. {\bf 311} (1989) 167--183.

\bibitem{ChaVes90}
O.A. Chalykh, A.P. Veselov,\\
{\it Commutative rings of partial differential operators and Lie algebras},\\
Commun. Math. Phys. {\bf 126} (1990) 597--611.

\bibitem{Heckman}
G.J. Heckman,\\
{\it A remark on the Dunkl differential-difference operators},\\
in: W. Barker, P. Sally (eds.),
{\em Harmonic analysis on reductive groups},\\
Progr. Math. {\bf 101}, 181--191, Birkh\"auser, 1991.

\bibitem{higgs}
P.W. Higgs,\\
{\it Dynamical symmetries in a spherical geometry I},\\
J. Phys. A: Math. Gen. {\bf 12} (1979) 309--323.

\bibitem{leemon}
H.I. Leemon,\\
{\it Dynamical symmetries in a spherical geometry II},\\
J. Phys. A: Math. Gen. {\bf 12} (1979) 489--501.

\bibitem{DunklXu}
C.F. Dunkl, Y. Xu,\\
{\em Orthogonal polynomials of several variables},\\
Cambridge University Press, 2001.

\bibitem{bososusy}
M.S. Plyushchay,\\
{\it Deformed Heisenberg algebra, fractional spin fields and supersymmetry without fermions,}\\
Annals Phys.  {\bf 245} (1996) 339 {\tt [arXiv:hep-th/9601116]};\\
{\it Hidden nonlinear supersymmetries in pure parabosonic systems,}\\
Int. J. Mod. Phys. A {\bf 15} (2000) 3679 {\tt [arXiv:hep-th/9903130]}.

\bibitem{spectral} 
F. Correa, M.S. Plyushchay,\\
{\it Spectral singularities in ${\cal PT}$-symmetric periodic finite-gap systems},\\
Phys. Rev. D {\bf 86} (2012) 085028 {\tt [arXiv:1208.4448[hep-th]]}.

\end{thebibliography}
\end{document}